\documentclass[a4paper,fleqn]{elsarticle}

\usepackage{bm}
\usepackage{subcaption}
\usepackage{nicefrac}
\usepackage{amssymb}
\usepackage{latexsym}
\usepackage{amsmath}
\usepackage{graphicx}
\usepackage{natbib}
\usepackage{fleqn}
\usepackage{geometry}
\usepackage{url}
\usepackage{lineno,hyperref}
\newcommand*{\Scale}[2][4]{\scalebox{#1}{$#2$}}

\DeclareMathAlphabet{\mathpzc}{OT1}{pzc}{m}{it}
\begin{document}

\begin{frontmatter}

\title{Exclusive robustness of Gegenbauer method to truncated convolution errors}

\author[mymainaddress]{Ehsan Faghihifar}
\ead{ehsan.faghihifar@alum.sharif.edu}

\author[mymainaddress]{Mahmood Akbari}
\ead{makbari@sharif.edu}

\address[mymainaddress]{Sharif University of Technology, Azadi Ave., Tehran, 1136511155, Iran}

\begin{abstract}
Spectral reconstructions provide rigorous means to remove the Gibbs phenomenon and accelerate the convergence of spectral solutions in non-smooth differential equations. In this paper, we show the concurrent emergence of truncated convolution errors could entirely disrupt the performance of most reconstruction techniques in the vicinity of discontinuities. They arise when the Fourier coefficients of the product of two discontinuous functions, namely $\Scale[.9]{f=gh}$, are approximated via truncated convolution of the corresponding Fourier series, i.e. $\Scale[.9]{\hat{f}_k\approx \sum_{|\ell|\leqslant N}{\hat{g}_\ell\hat{h}_{k-\ell}}}$. Nonetheless, we numerically illustrate and rigorously prove that the classical Gegenbauer method remains exceptionally robust against this phenomenon, with the reconstruction error still diminishing proportional to $\Scale[.9]{\mathcal{O}(N^{-1})}$ for the Fourier order $\Scale[.9]{N}$, and exponentially fast regardless of a constant. Finally, as a case study and a problem of interest in grating analysis whence the phenomenon initially was noticed, we demonstrate the emergence and practical resolution of truncated convolution errors in grating modes, which constitute the basis of Fourier modal methods.
\end{abstract}


\end{frontmatter}

\section{Introduction}
\label{Sec: Intro}

\begin{figure}[!t]
\centering
\includegraphics[width=.55\linewidth]{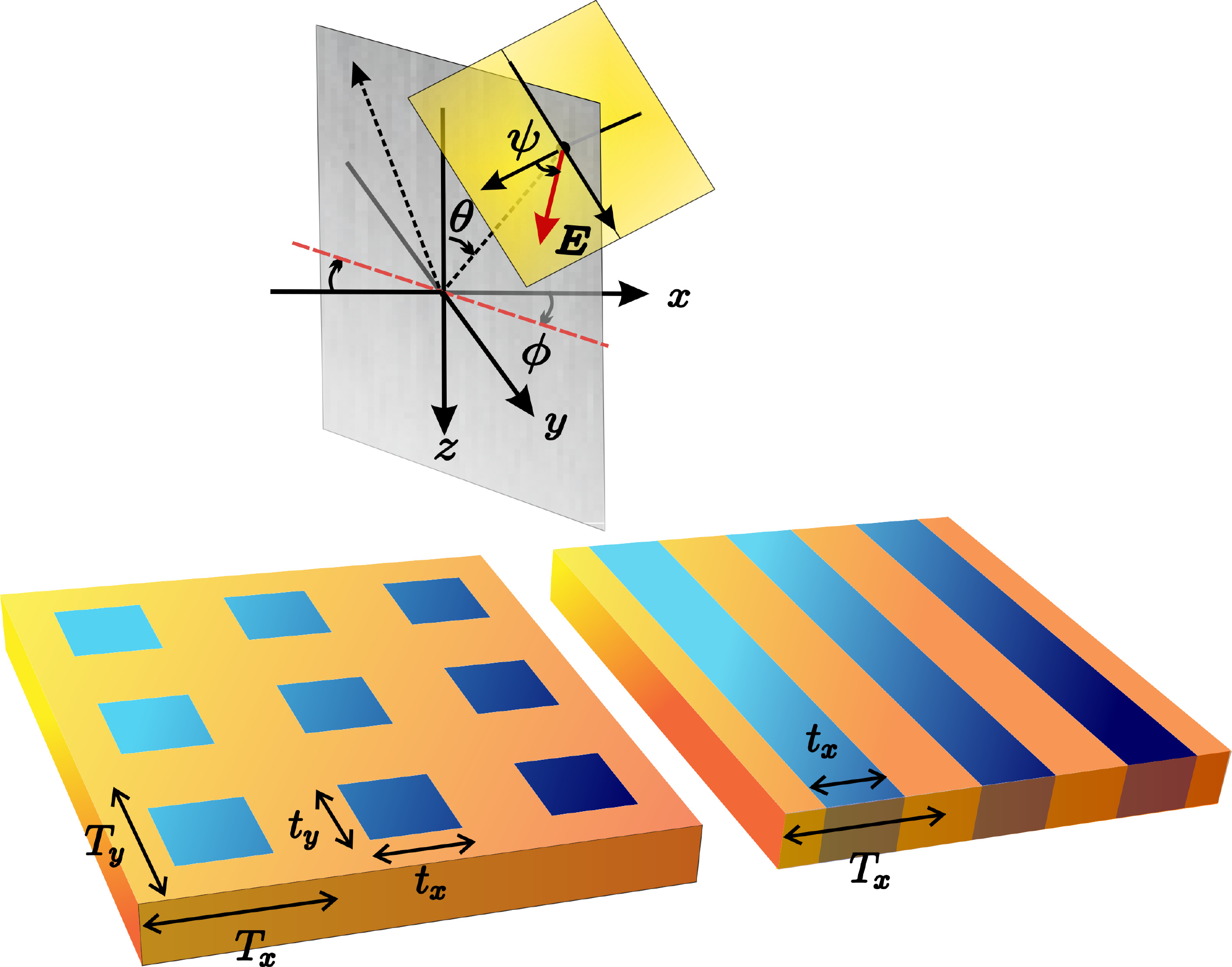}
\caption{Grating diffraction problem: the incidence of a plane wave on a lamellar/crossed gratings with normalized dimensions indicated.}
\label{fig_3d}
\end{figure}

Spectral methods emerged as an alternative to the finite-difference and finite-element methods for the numerical solution of partial differential equations. They involve representing the solution to a boundary value, eigenvalue, or time-dependent problem, as a truncated series of known basis functions with global (or at least extended) support on the domain \cite{boyd2013chebyshev,hesthaven2007spectral}. The coefficients of the basis functions can be considered as a spectrum of the solution, which explains the name of the method. Long before the first unifying mathematical assessment of the theory was provided by Gottlieb and Orszag in \cite{gottlieb1977numerical}, spectral methods had found applications in various problems, particularly as a large-scale computational tool in fluid dynamics \cite{canuto1988spectral}. The fascinating merit of spectral methods is the infinite-order convergence rate; i.e. under fortunate circumstances, the error can go to zero faster than any finite power of the number of retained modes \cite{gottlieb1977numerical}, while finite-difference and finite-element methods can not normally converge faster than algebraically \cite{boyd2013chebyshev, Babuska1982}. The superior accuracy, however, comes at the expense of less geometrical flexibility, which can be improved using multidomain techniques \cite{Yang1997, Hesthaven1999}.
The geometry also predominantly determines the choice of basis functions.
The use of the Fourier series, in particular, has been most dramatically successful in problems with periodic geometries. Thanks to the computational simplicity and availability of fast Fourier transform, Fourier spectral methods are now utilized in a wide variety of applications, particularly in photonics as a field of interest, wherein periodic structures play a major role \cite{Antoniades2003, Landy2008, Kildishev2013, Grady2013}.
In that regard, Fourier series are either utilized directly, e.g. in the derivation of band structures in photonic crystals\cite{Ho1990, Plihal1991, Meade1992, joannopoulos2011photonic, Johnson2001}, or as an intermediary tool in Fourier modal methods \cite{Moharam1981, Moharam1995, li2014chap}, providing an efficient technique for grating diffraction problems (figure \ref{fig_3d}); compared to the more computationally expensive alternatives, e.g. the finite elements \cite{Bao2005}, finite-difference time-domain \cite{Correia2004}, integral equation \cite{Dobson1991}, and boundary variation methods \cite{Bruno1993}.
Not to mention the noted techniques may also benefit nonperiodic structures via supercell approximations \cite{Payne1992} or perfectly matched layers \cite{Hugonin2005}.

However, despite the unequaled convenience of Fourier spectral methods, they have to deal with the inevitable emergence of the Gibbs phenomenon in the presence of discontinuities. In grating analysis in particular, not only does this phenomenon significantly deteriorate the convergence of near-field distributions, but it may also have indirect detrimental implications on the derivation of far-field parameters, e.g. the reflection coefficient, etc. Such far-field effects were initially recognized from the notoriously slow convergence of diffraction analyses in metallic lamellar gratings with the TM polarization \cite{Li1993}. Notwithstanding the prior discovery of an empirical treatment \cite{Lalanne1996, Granet1996} for this problem, it was Li's influential paper \cite{Li1996} that provided the first analytical explanation and led to further generalization and extensive applications \cite{Li1997, Popov2000, Popov2001, Li2003}. He attributed the problem to truncated convolution errors (TCE) which manifest when Fourier coefficients of the product two discontinuous functions are approximated through truncated convolution of the corresponding Fourier series. Though he reformulated the equations accordingly to speed up the convergence, the very presence of the Gibbs phenomenon could not be eliminated. Specifically, uniform resolution and fluctuations of the permittivity expansion can still deleteriously affect the convergence of far-field parameters \cite{Granet1999_2, Kim2012}, as well as the near-field distribution of the eigenmodes. The latter can particularly be significant in fast-Fourier-based mode solvers, where the computational algorithm does not leave room for the application of factorization rules \cite{Johnson2001, Silvestre2005, Lopez2005, Ortega2006, Ortega2007}. Overall, despite the scarcity of rigorous results regarding the convergence of Fourier modal methods such as \cite{Li1999}, it seems safe to conclude that the convergence of both near- and far-field parameters, could not exceed a modest polynomial rate---as illustrated in \cite{Walz2013, Morf1995}.

On the other side, circumvention of the Gibbs phenomenon using multi-domain techniques might still be possible, though typically to the detriment of simplicity and flexibility. For grating problems, elegant use of subsectional or non-smooth basis functions, e.g. the B-spline modal method \cite{Bouchon2010, Granet2014, Walz2013}, subsectional polynomial expansions \cite{Morf1995, Edee2011, Edee2015}, or pseudo-spectral techniques \cite{Chiou2009,Song2011}, can provide exponential or high-degree algebraic convergence rates \cite{Morf1995,  Walz2013}.
Nevertheless, the mathematical approach towards the resolution of the Gibbs phenomenon in spectral solutions would be to use spectral reconstructions. These post-processing methods are easily implemented and widely applicable as they merely depend on the solution, whereas problem-specific techniques like Fourier factorizations or adaptive spatial resolution \cite{Li1996, Granet1999_2, Weiss2009}, rely on reformulating the governing equation to mitigate manifestations of the Gibbs phenomenon. Despite high potentials, however, their application in photonic problems remains largely unexplored, save for sporadic research works \cite{Min2006, Perez2007, Piotrowska2019}.

Over the years, the Gibbs phenomenon has been subject to extensive scrutiny \cite{jerri1998gibbs}, resulting in the development of a variety of interesting techniques to deal with it; e.g. frequency-domain filters \cite{Cai1992, Vandeven1991}, physical-domain and adaptive filters \cite{Gottlieb1985, Tadmor2002, Tanner2006}, polynomial subtraction or Eckhoff method \cite{Eckhoff1993, Eckhoff1995, Eckhoff1998, Adcock2011}, singular Fourier-Pad\'e approximation \cite{Driscoll2001}, spectral reprojection using Gegenbauer and Freud polynomials \cite{Gottlieb1992, Gottlieb1994, Gottlieb1995, Gottlieb1997, Gelb1997, Gelb2006}, inverse polynomial reconstruction \cite{Shizgal2003, Pasquetti2004, Jung2004}, polynomial least-squares \cite{Hrycak2010, Adcock2012, Adcock2014}, Fourier extension \cite{Huybrechs2010, Adcock2012_2, Adcock2014_2, Adcock2014_3}, etc.
However, notwithstanding the rich literature of spectral post-processing techniques, their robustness against TCE---a phenomenon that may accompany the Gibbs phenomenon in spectral solutions--has not been studied yet, to the best of our knowledge. Here, we will attend to this problem analytically and numerically, only to reach this striking conclusion that TCE can disrupt almost all reconstruction techniques around discontinuities, save only the Gegenbauer method---a promising result in favor of the classical method, contrary to \cite{Boyd2005}. Moreover, our study will include a case study in grating modes reconstruction using the Gegenbauer method, though the idea might be later generalized to far-field problems as well.
This paper comprises three main sections. In section \ref{Sec: TCE}, TCE is explained and the performance of popular spectral reconstructions in its presence is numerically assessed. In section \ref{Sec: Gegen}, the Gegenbauer method is described and a theorem regarding its exclusive robustness to TCE proved. Eventually, as a case study in section \ref{Sec: Grat}, the emergence and resolution of TCE in grating modes are discussed.

\section{Truncated Convolution Error (TCE)}
\label{Sec: TCE}

The essence of all computational techniques for solving differential equations necessitates that somehow the information characterizing the coefficients and the solution be truncated into finite sets of numbers, i.e. vectors.
Here, we show how performing algebraic operations on Fourier series while keeping the number of retained orders constant causes the obtained solution to numerically deviate and result in disruption of most reconstruction techniques.

\subsection{Convolutional Fourier Series}
\label{Sub: CFS}

Without loss of generality, let all the functions and solutions henceforth be defined over $\Scale[.9]{[-1,1]}$. Letting $\Scale[.9]{\{\psi_k=e^{ik\pi x}\}}$ denote the Fourier basis functions, the partial Fourier expansion of the function $\Scale[.9]{f}$ is defined as follows:

\begin{eqnarray}\label{z1}
\Scale[.9]{
{f}_N\left( x \right) = \displaystyle\sum\nolimits_{\left| k \right| \leqslant N} {{{\hat f}_k}\psi_k} ,{\hspace{50pt}}{\hat f_k} = \frac{1}{2} \left\langle f,\psi_k \right\rangle= \frac{1}{2}\displaystyle\int_{ - 1}^1 {f\left( x \right)\psi_k^*(x) dx}.
}
\end{eqnarray}

According to Carlson's theorem, if $\Scale[.9]{f\in\mathcal{L}^2(-1,1)}$ then $\Scale[.9]{\lim\nolimits_{N\to\infty}f_N=f}$ almost everywhere \cite{Carleson1966}. It implies that point-wise disagreement between a function and its Fourier series could exceptionally occur even at continuities, as corroborated by specially fabricated examples. Nonetheless, the solutions we typically encounter are well-behaved partially smooth functions with at most, a finite number of discontinuities. Hence, we will presume that $\Scale[.9]{f=f_\infty}$ everywhere except at discontinuities.

Now, let us assume $\Scale[.9]{f}$ is the product of two given functions $\Scale[.9]{g}$ and $\Scale[.9]{h}$.
According to the popular relationship $\Scale[.9]{{\hat f_k} = \sum_{\ell=-\infty}^{\infty} {{{\hat g}_\ell}{\hat h}_{k-\ell}}}$ with $\Scale[.9]{{\hat g}_\ell}$ and $\Scale[.9]{{\hat h}_\ell}$ denoting Fourier coefficients of the multiplicands, exact characterization of $\Scale[.9]{f_N}$ is not possible within an $\Scale[.9]{N}$-th order spectral framework, as it also depends on the Fourier coefficients of $\Scale[.9]{g-g_N}$ and $\Scale[.9]{h-h_N}$ which are not retained in the process.
Hence, in practice, the truncated series $\Scale[.9]{f_N=\{g h\}_N}$ is estimated and replaced by an estimate, called the $\Scale[.9]{N}$-th order convolutional Fourier series, which we denote and define as follows:

\begin{eqnarray}\label{z2}
\Scale[.9]{
\tilde{f}_N(x)=\left\{{g}_N\star{h}_N\right\}\left( x \right) = \displaystyle\sum\nolimits_{\left| k \right| \leqslant N} {{{\hat c}_k}\psi_k} ,{\hspace{50pt}}{\hat c_k} = \displaystyle\sum\nolimits_{\left| \ell \right| \leqslant N} {{{\hat g}_\ell}{\hat h}_{k-\ell}}.
}
\end{eqnarray}

This truncation process naturally comes at the expense of a numerical deviation, which we call the truncated convolution error (TCE).
In explaining what makes the convolutional Fourier series of two functions $\Scale[.9]{g}$ and $\Scale[.9]{h}$ with concurrent jumps at $\Scale[.9]{x = x_1}$, effectively different from the Fourier series, we will invoke Li's important finding \cite{Li1996, li2001chap}. It follows that subtracting the convolutional Fourier series of a multiplicative function from its truncated Fourier series, gives rise to the emergence of an oscillatory function at the point of discontinuity, as follows:

 \begin{eqnarray}\label{z3}
 \Scale[0.9]{
\left\{ {g h} \right\}_N - {g}_N \star {h}_N = \left({d_{g}d_{h}}/{2\pi^2}\right){\Phi_N}\left( {x - {x_1}} \right) + o\left( 1 \right)
 }
 \end{eqnarray}

Wherein $\Scale[.9]{d_g}$ and $\Scale[.9]{d_h}$ are the corresponding jumps of $\Scale[.9]{g}$ and $\Scale[.9]{h}$, the term $\Scale[.9]{o(1)\to 0}$ uniformly as $\Scale[.9]{N\to\infty}$, and $\Scale[.9]{{\Phi_N}(x)}$ which will be called the Li's function henceforth, is defined as follows:

\begin{eqnarray}\label{z4}
\Scale[0.9]{
{\Phi_N}\left( x \right) = \displaystyle\sum\nolimits_{\left| n \right| \leqslant N} {{\alpha _n}{\psi_n}}, \hspace{25pt}
{\alpha _n} = \frac{{1 - {\delta _{n}}}}{{2\left| n \right|}}\displaystyle\sum\nolimits_{\left| k \right| > N} {\frac{1}{{k - \left| n \right|}} = \frac{{1 - {\delta _{n}}}}{{2\left| n \right|}}\left\{ H_{N+\left|n\right|} - H_{N-\left|n\right|} \right\}}.
}
\end{eqnarray}

In the above equation, $\Scale[.9]{H_{m>0}=\sum\nolimits_{i=1}^{m}{i^{-1}}}$ denote harmonic numbers, $\Scale[.9]{H_0=0}$, and $\Scale[.9]{\delta _{n}=\delta _{n,0}}$ is the Kronecker delta. A fundamental property of Li's function is that when $\Scale[.9]{N \to \infty}$, the function tends to zero at all points, except at $\Scale[.9]{x = 0} $ where it tends to $\Scale[.9]{\pi^2/4}$. Hence, from one point of view, TCE can basically be thought of as an approximation to a single point discontinuity in the limiting case.
Note that this interpretation readily contradicts the underlying assumption of all spectral reconstructions that the function should be comprised of smooth subintervals, and thus, it can be expected from these methods to lose their efficacy near discontinuities.

In the derivational process of equation \ref{z3} in \cite{li2001chap}, the partially smooth functions are decomposed as the sum of a continuous function and a number of linear terms (ramp functions).
It can be shown that the non-vanishing dominant term of TCE, i.e. the noted Li's function, stems from the multiplication of two periodically extended linear terms. Therefore, the process can be generalized to consider higher-order discontinuities as well, resulting in higher-order perturbations that are already incorporated in the $\Scale[.9]{o(1)}$ term of equation \ref{z3}. However, they have much less numerical significance, and hence, will be disregarded in the analysis of robustness against TCE in section \ref{Sec: Gegen}.

\subsection{Spectral Reconstructions in Presence of TCE}
\label{Sub: Perf}

In the literature, the efficacy of mathematical techniques for the removal or amelioration of the Gibbs phenomenon is numerically assessed through properly defined discontinuous test functions.
In order to consider TCE as well, we propose to define test functions as the product of two discontinuous functions and employ the associated convolutional Fourier series to assess the performance of post-processing methods. As the phenomenon was initially scrutinized within the context of grating analysis, the examples are selected particularly to resemble typical grating eigenmodes, though the comparison and conclusions remain generally valid.

\begin{figure}[!t]	
	\centering
	\begin{subfigure}{0.41\textwidth}
		\centering
		\includegraphics[width=\linewidth]{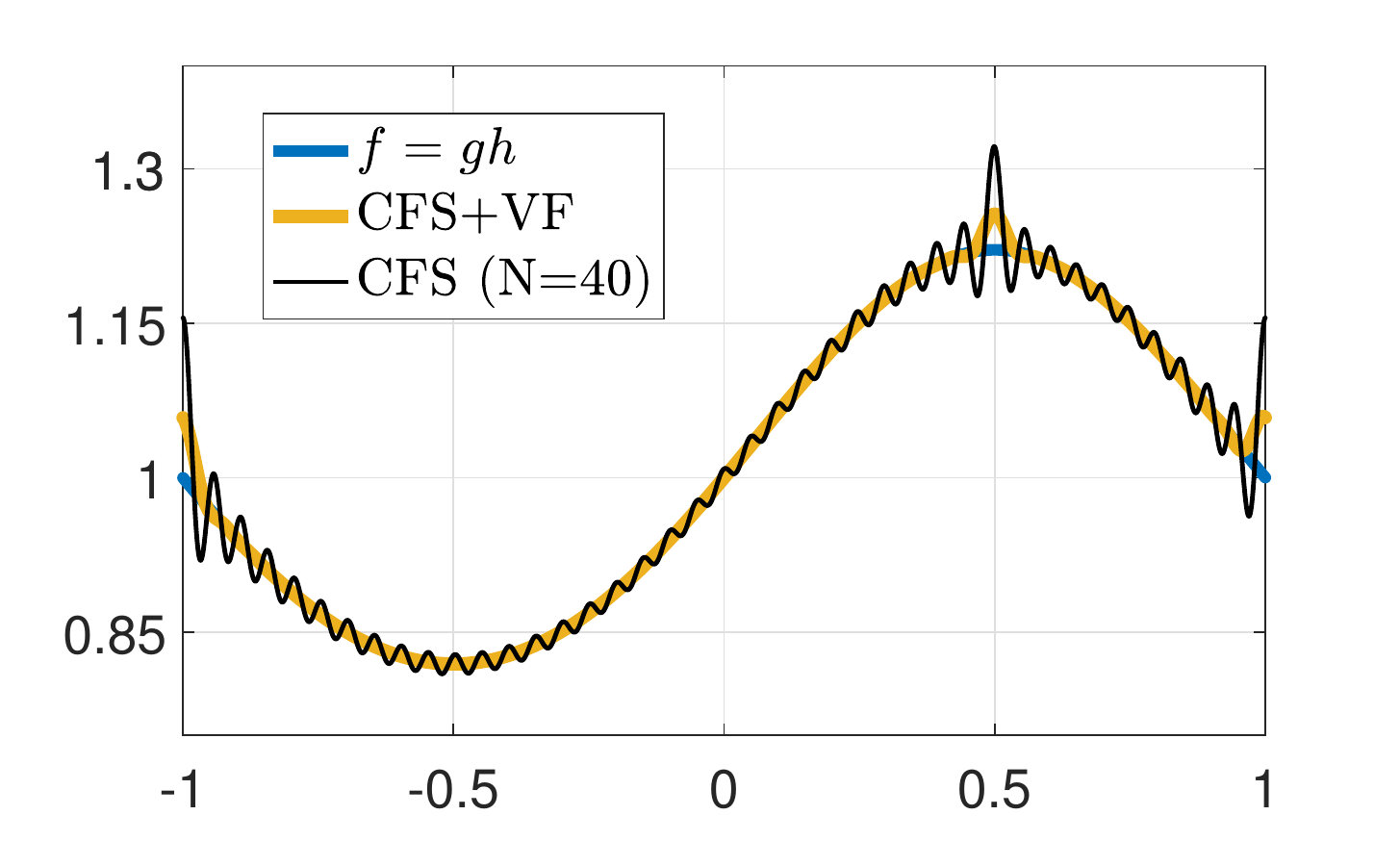}
		\caption{}\label{fil_con}		
	\end{subfigure}
~
	\begin{subfigure}{0.41\textwidth}
		\centering
		\includegraphics[width=\linewidth]{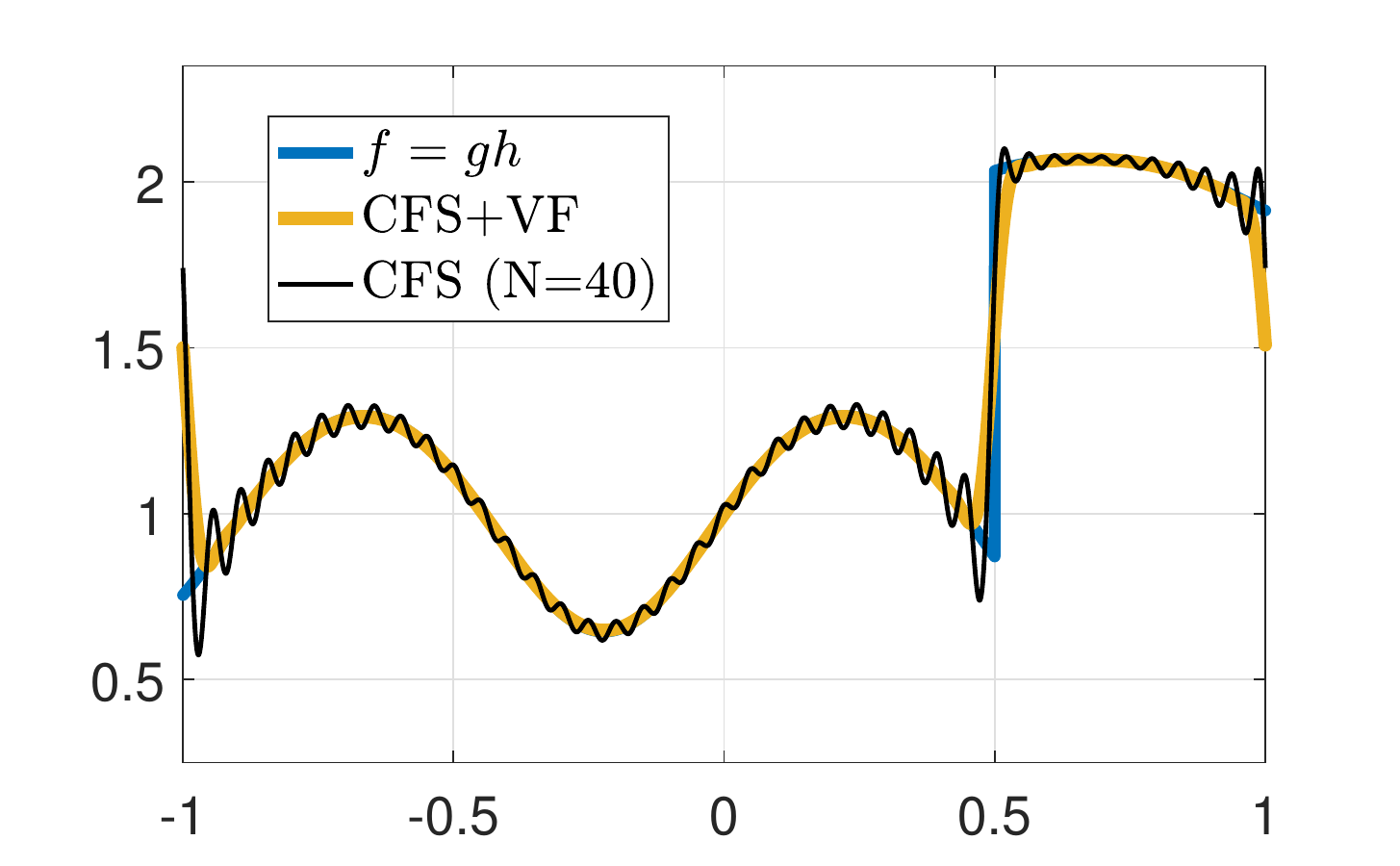}
		\caption{}\label{fil_unc}
	\end{subfigure}

	\caption{Depiction of test functions, their convolutional Fourier series, and reconstructions (via Vandeven filter) for: (\subref{fil_con}) continuous product \ref{z14}, and (\subref{fil_unc}) discontinuous product \ref{z15}.}
	\label{fig_ellip}
\end{figure}

Let $\Scale[.9]{f=g h}$ be the product of two piece-wise smooth functions $\Scale[.9]{g}$ and $\Scale[.9]{h}$ with concurrent jumps at $\Scale[.9]{x=0.5}$.
Two different sets of definitions for $\Scale[.9]{g}$ and $\Scale[.9]{h}$ are considered. The first set of functions is such that their product is smooth everywhere as follows, with $\Scale[.9]{g}$ denoting the Heaviside step function:

\begin{eqnarray}\label{z14}
\Scale[.9]{
\begin{array}{l}
\left\{ \begin{array}{l}
{g}(x) = 4 - 2u\left( {x - 0.5} \right) - 0.3\sin( {5\sqrt{2} x})\\
{h}(x) = \exp\{{0.2\sin(\pi x)}\}/g(x)\\
\end{array} \right.
\end{array}
}
\end{eqnarray}

As the product is continuous, the Gibbs phenomenon is not directly present in its convolutional Fourier series, and consequently, it would be TCE that give rise to the observable oscillations. We define the second set such that the product of discontinuous functions $\Scale[.9]{g}$ and $\Scale[.9]{h}$ is also discontinuous at $\Scale[.9]{x=0.5}$, as follows:

\begin{eqnarray}\label{z15}
\Scale[.9]{
\begin{array}{l}
\left\{ \begin{array}{l}
{g}(x) = 4 - 2u\left( {x - 0.5} \right) - 0.3{\sin( {5\sqrt{2} x})}\\
{h}(x) = 0.25 + 0.75u\left( {x - 0.5} \right) + 0.1\sin( {5\sqrt{2} x})\\
\end{array} \right.
\end{array}
}
\end{eqnarray}

The test functions $\Scale[.9]{f}$, and their convolutional Fourier series $\Scale[.9]{\tilde{f}_N = g_N \star h_N}$ are depicted in figure \ref{fil_con} and figure \ref{fil_unc}, respectively. For a comparison, the application of the Vandeven filter on both spectra is depicted as well.

Resolution of the Gibbs phenomenon predominantly depends on prior knowledge of the location of discontinuities, which otherwise should be estimated using edge detection techniques \cite{Gelb1999,Tadmor2007}. On the other hand, however, we assume the amplitudes of the jumps are unknown, hence, those techniques which require such information, e.g. the Eckhoff method \cite{Eckhoff1993}, are not studied here. The methods we evaluate their performance here are the Vandeven filter \cite{Vandeven1991}, polynomial least-squares \cite{Adcock2012}, Fourier extension \cite{Huybrechs2010}, singular Fourier-Pad\'e \cite{Driscoll2001}, Freud \cite{Gelb2006}, and Gegenbauer methods \cite{Gottlieb1992}, as some of the most prominent spectral reconstruction techniques. While the formulation of the latter method is discussed in section \ref{Sec: Gegen}, a very concise explanation of the rest and their associated parameters could be found in \ref{Apx_B}. Still, for elaborate theoretical discussions and generalizations refer to the references specified here and throughout the text.

\begin{figure*}[!t]	
	\centering

	\begin{subfigure}{0.31\textwidth}
		\centering
		\includegraphics[width=\linewidth]{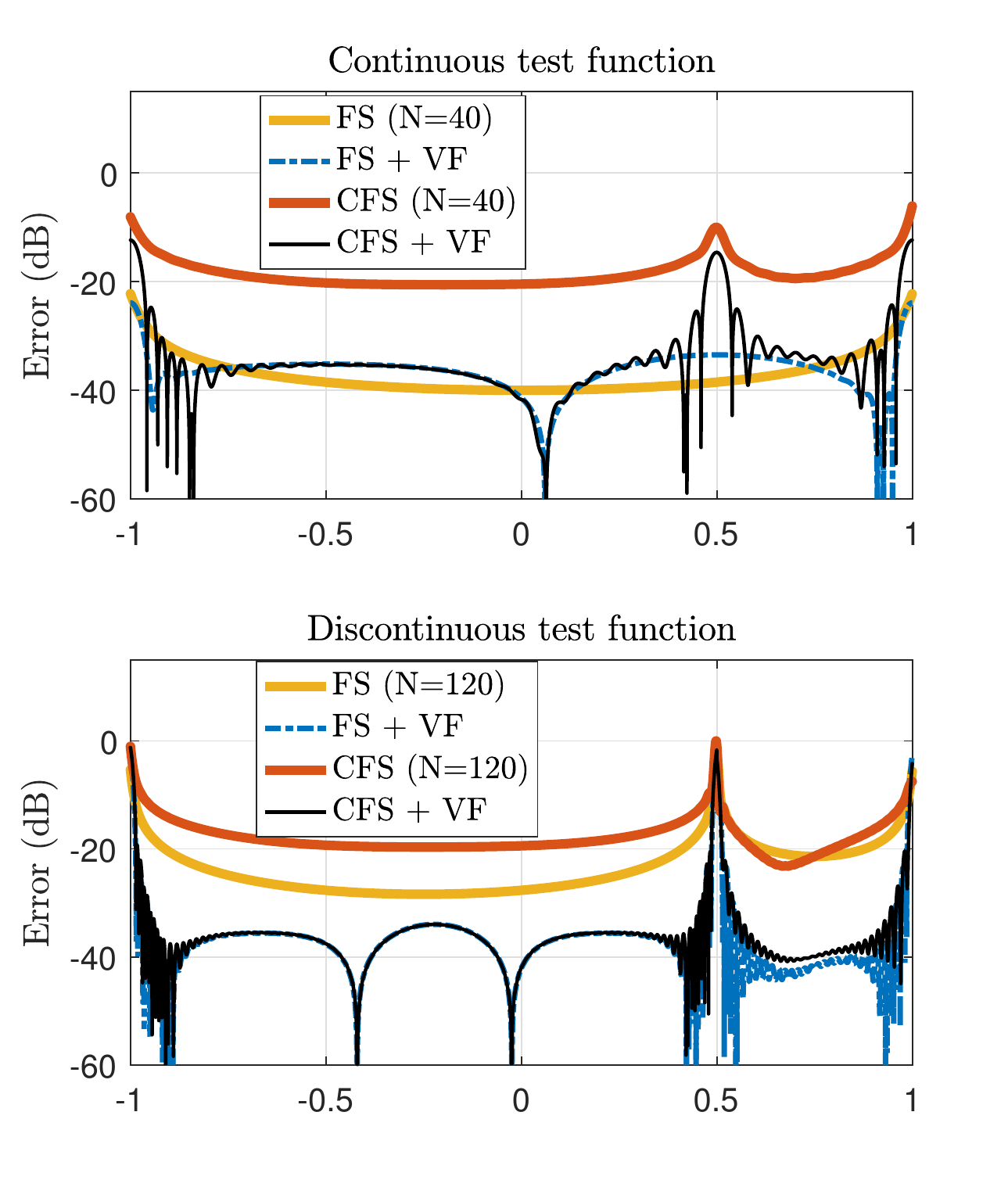}
		\caption{}\label{error_vf}
	\end{subfigure}	
~
	\begin{subfigure}{0.31\textwidth}
		\centering
		\includegraphics[width=\linewidth]{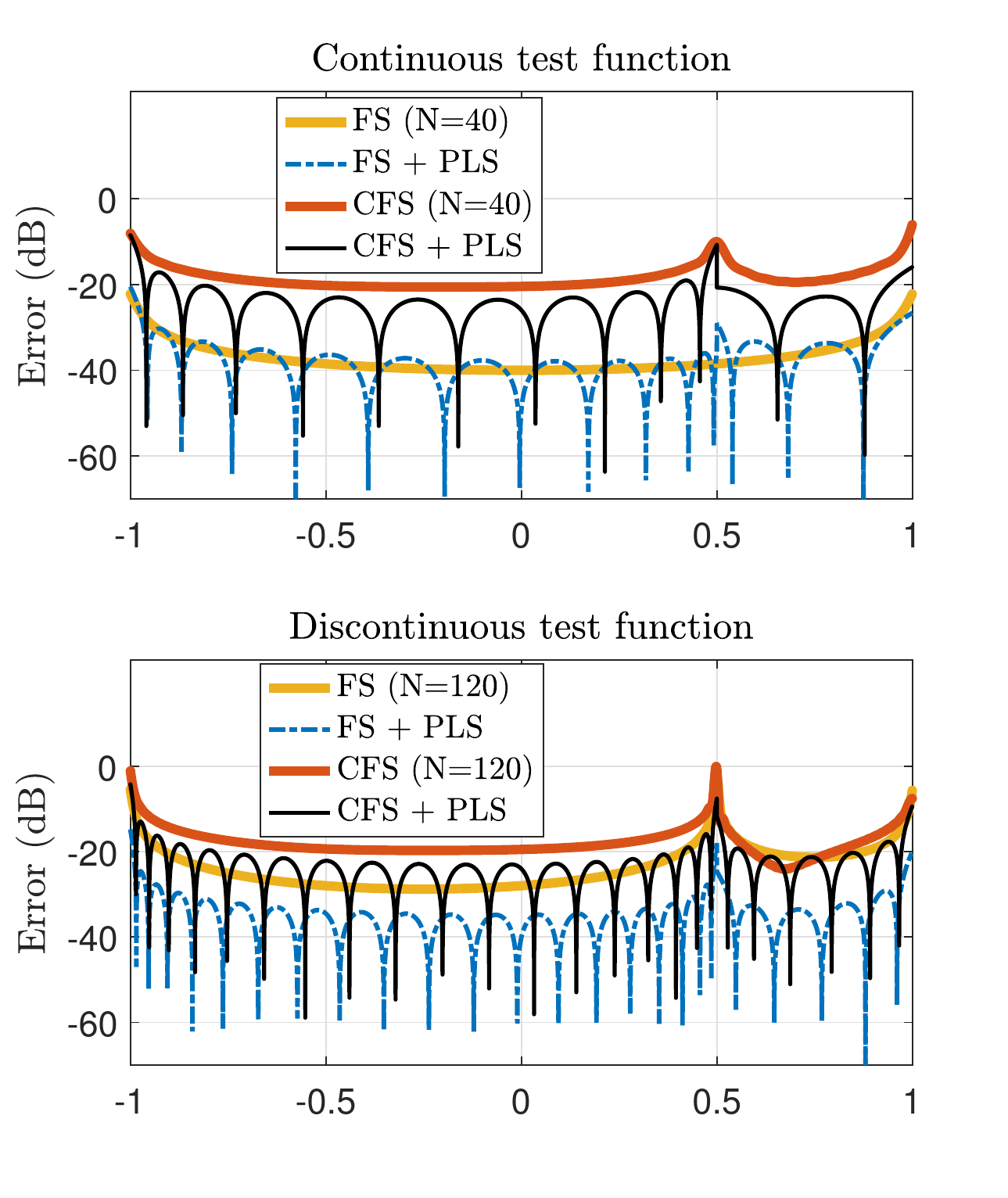}
		\caption{}\label{error_lpls}
	\end{subfigure}	
~
	\begin{subfigure}{0.31\textwidth}
		\centering
		\includegraphics[width=\linewidth]{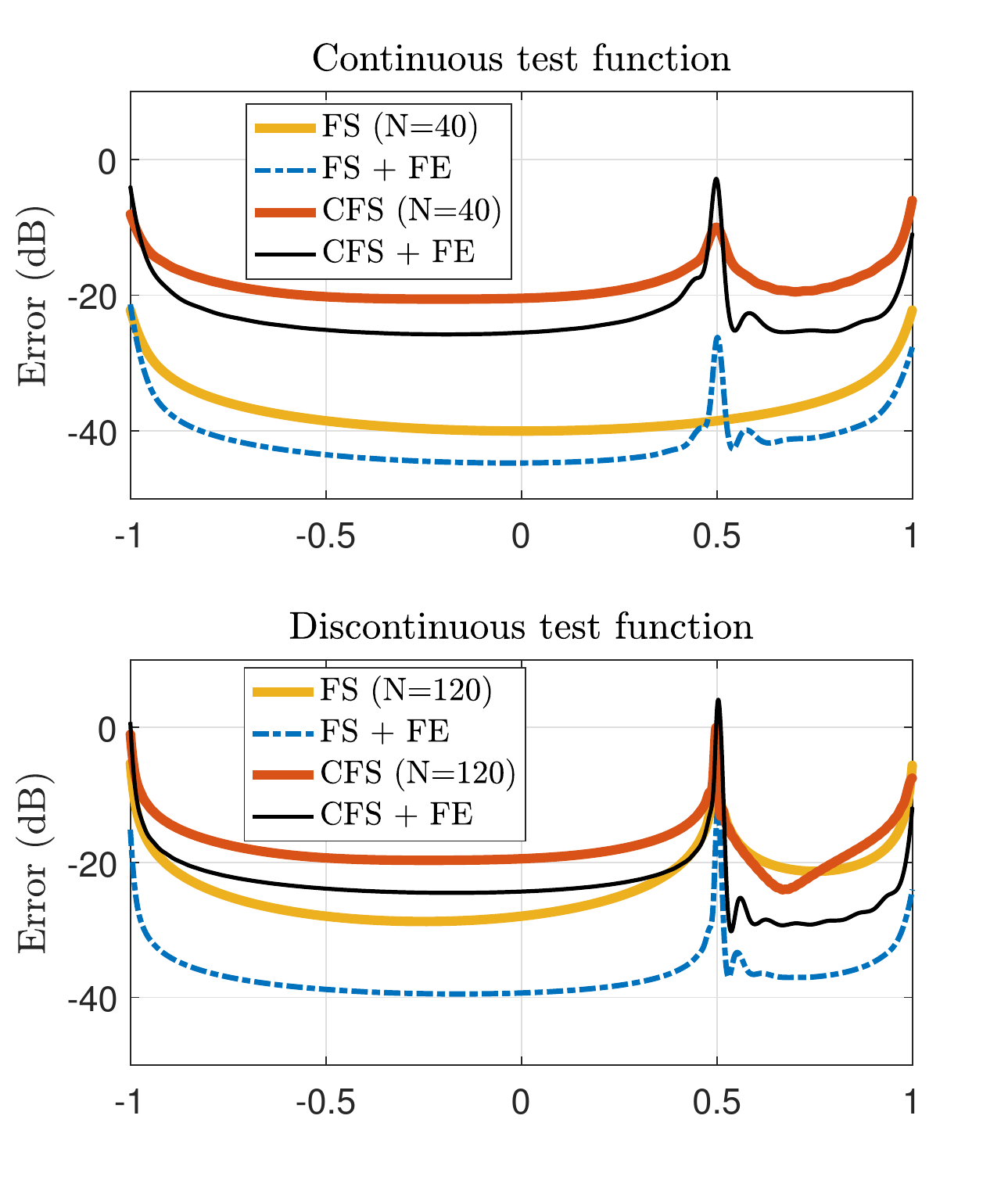}
		\caption{}\label{error_fe}
	\end{subfigure}
\quad
	\begin{subfigure}{0.31\textwidth}
		\centering
		\includegraphics[width=\linewidth]{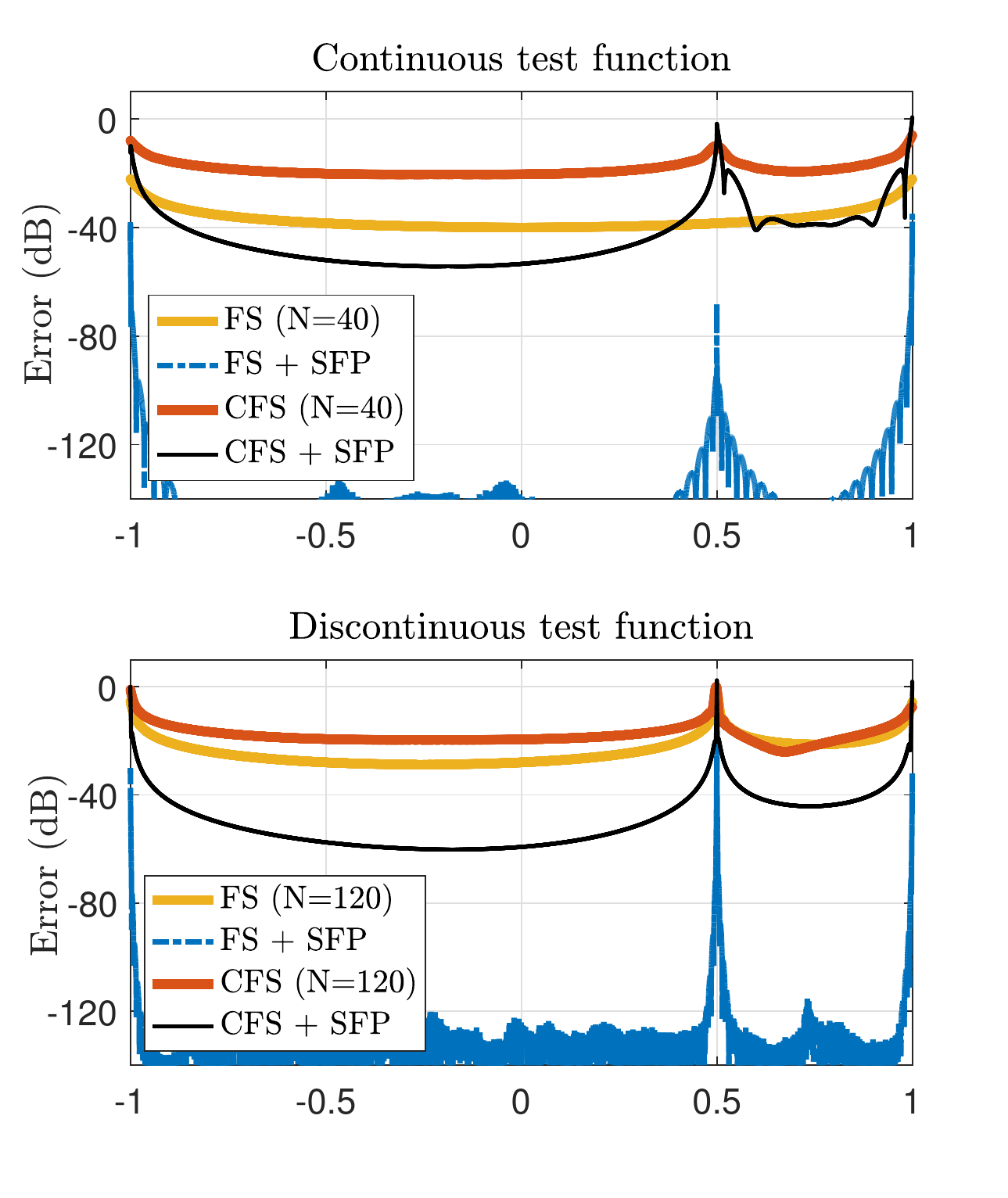}
		\caption{}\label{error_sfp}		
	\end{subfigure}
~
	\begin{subfigure}{0.31\textwidth}
		\centering
		\includegraphics[width=\linewidth]{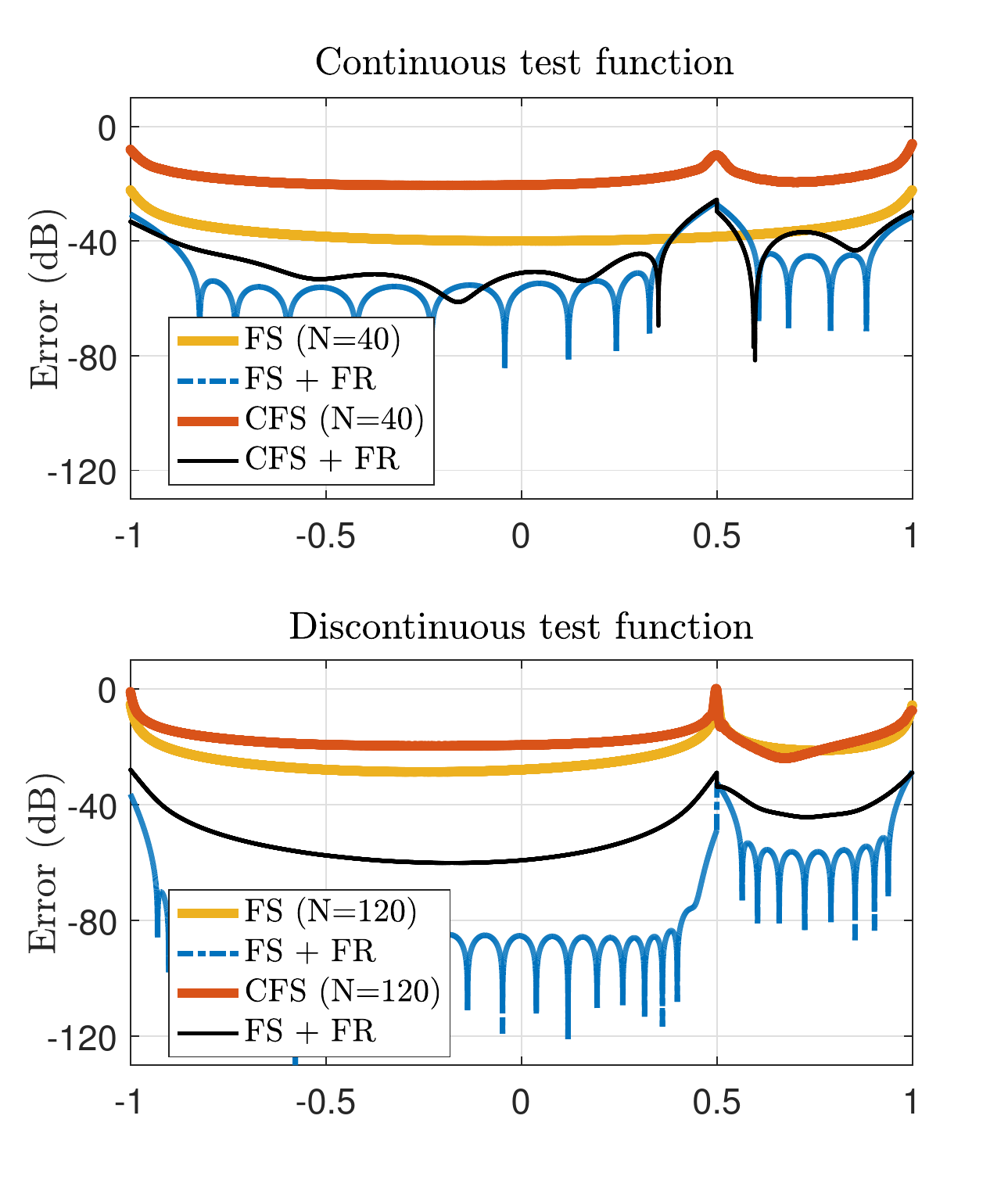}
		\caption{}\label{error_fr}
	\end{subfigure}
~
	\begin{subfigure}{0.31\textwidth}
		\centering
		\includegraphics[width=\linewidth]{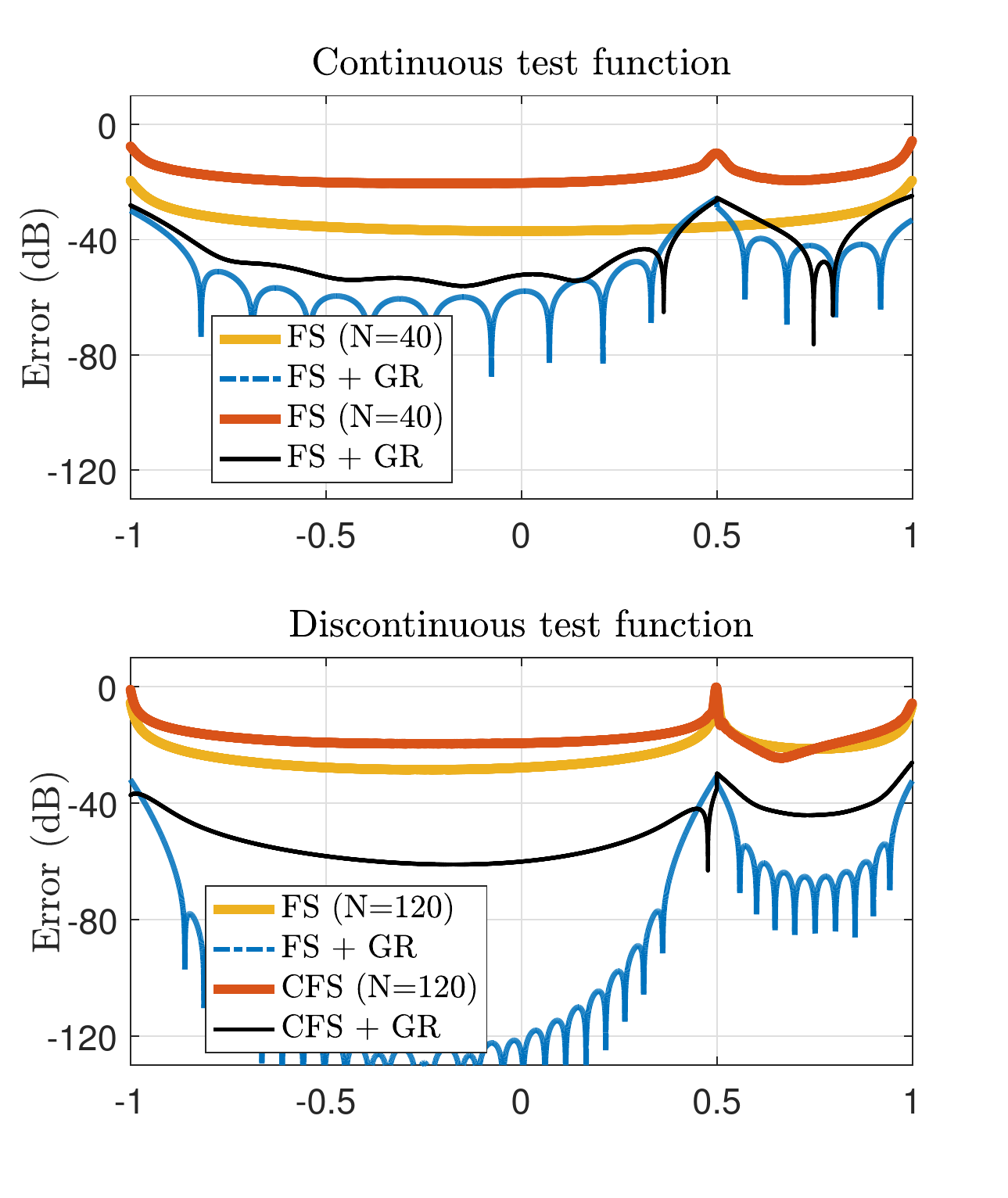}
		\caption{}\label{error_gr}
	\end{subfigure}

	\caption{Point-wise Fourier series and reconstruction error of the test functions \ref{z14} and \ref{z15} vs. the Fourier order (\subref{error_vf}) Vandeven filter, (\subref{error_lpls}) Legendre polynomial least-squares, (\subref{error_fe}) Fourier extension, (\subref{error_sfp}) singular Fourier-Pad\'e, (\subref{error_fr}) Freud reconstruction, and (\subref{error_gr}) Gegenbauer reconstruction methods.}
	\label{fig_error_comp}
\end{figure*}

In figure \ref{fig_error_comp}, we have plotted point-wise error graphs regrading the methods listed above, for both the continuous (with the Fourier order $\Scale[.9]{N=40}$) and discontinuous (with $\Scale[.9]{N=120}$) test functions of equations \ref{z14} and \ref{z15}.
Each plot in this figure includes four logarithmic error graphs (in terms of dB) as follows: the Fourier series error $\Scale[.9]{e(x)=10\log|f-f_N|}$ in yellow, the convolutional Fourier series error $\Scale[.9]{\tilde{e}(x)=10\log|f-\tilde{f}_N|}$ in red, the reconstruction error from the Fourier series $\Scale[.9]{e_R(x)=10\log|f-\mathcal{R}{f_N}|}$ in blue, and the reconstruction error with convolutional Fourier series $\Scale[.9]{\tilde{e}_R(x)=10\log|f-\mathcal{R}{\tilde{f}_N}|}$ in black, wherein $\Scale[.9]{\mathcal{R}}$ denotes the reconstruction operator acting on a Fourier series.

\begin{figure*}[!t]	
	\centering

	\begin{subfigure}{0.31\textwidth}
		\centering
		\includegraphics[width=\linewidth]{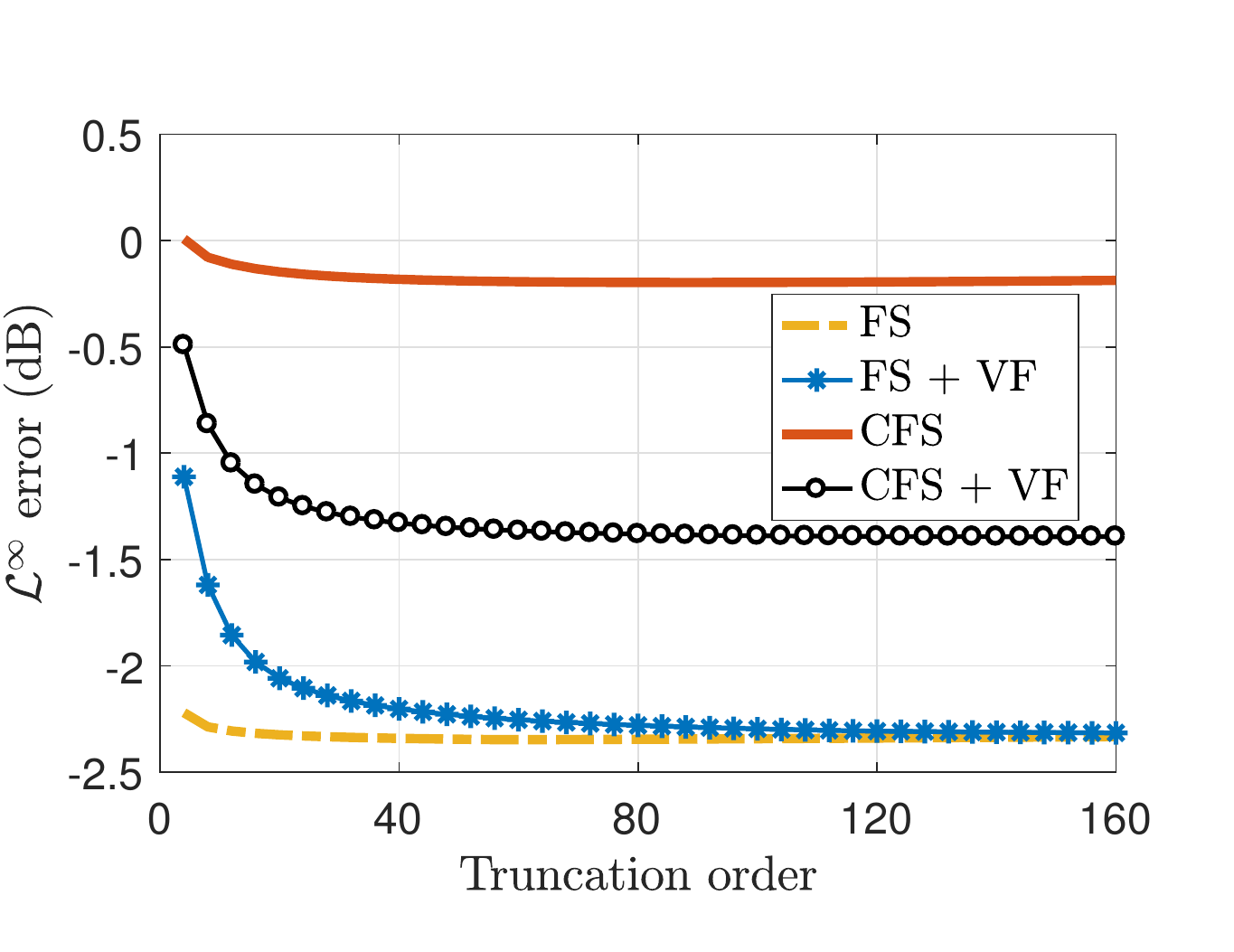}
		\caption{}\label{N_vf}
	\end{subfigure}	
~
	\begin{subfigure}{0.31\textwidth}
		\centering
		\includegraphics[width=\linewidth]{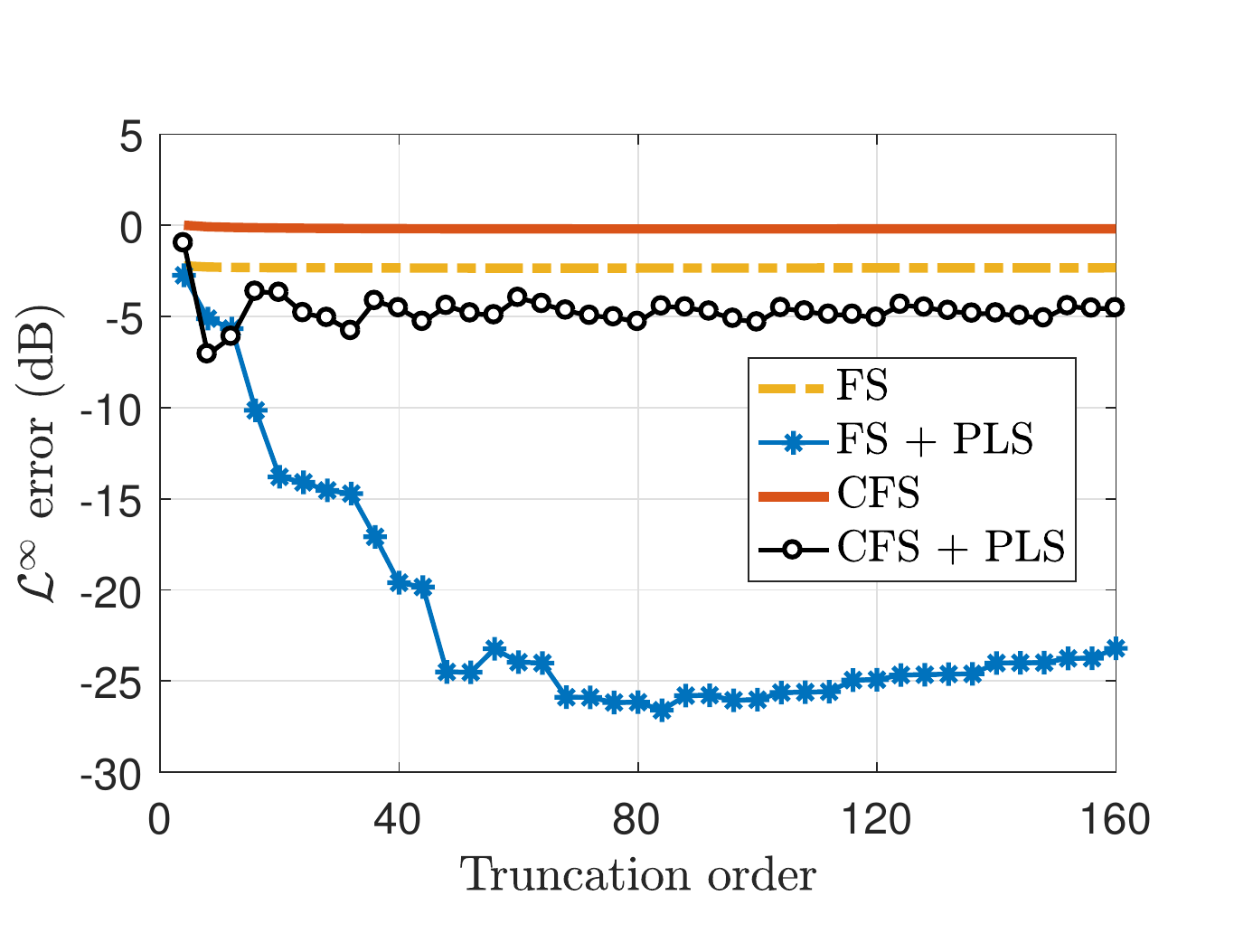}
		\caption{}\label{N_lpls}
	\end{subfigure}	
~
	\begin{subfigure}{0.31\textwidth}
		\centering
		\includegraphics[width=\linewidth]{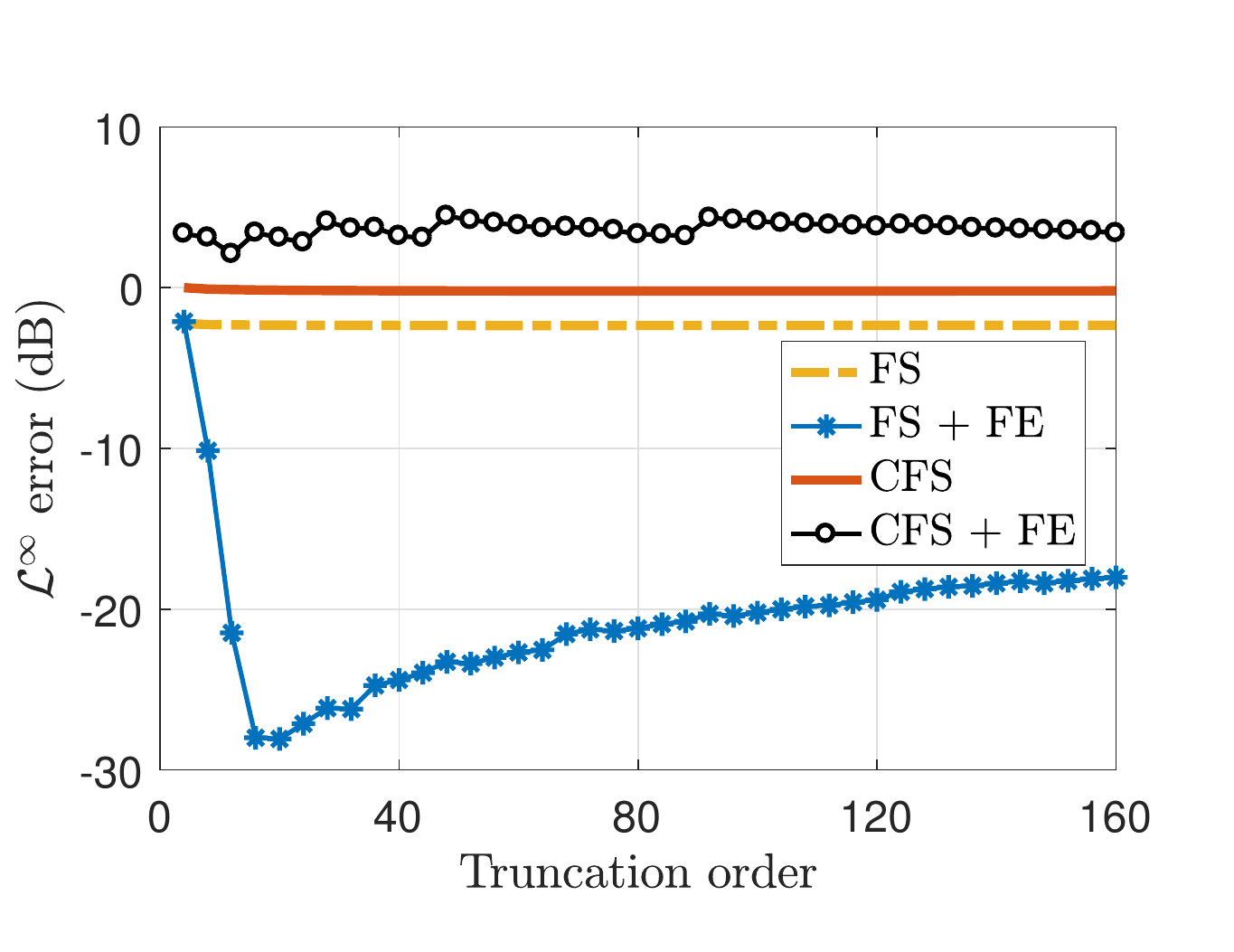}
		\caption{}\label{N_fe}
	\end{subfigure}
\quad
	\begin{subfigure}{0.31\textwidth}
		\centering
		\includegraphics[width=\linewidth]{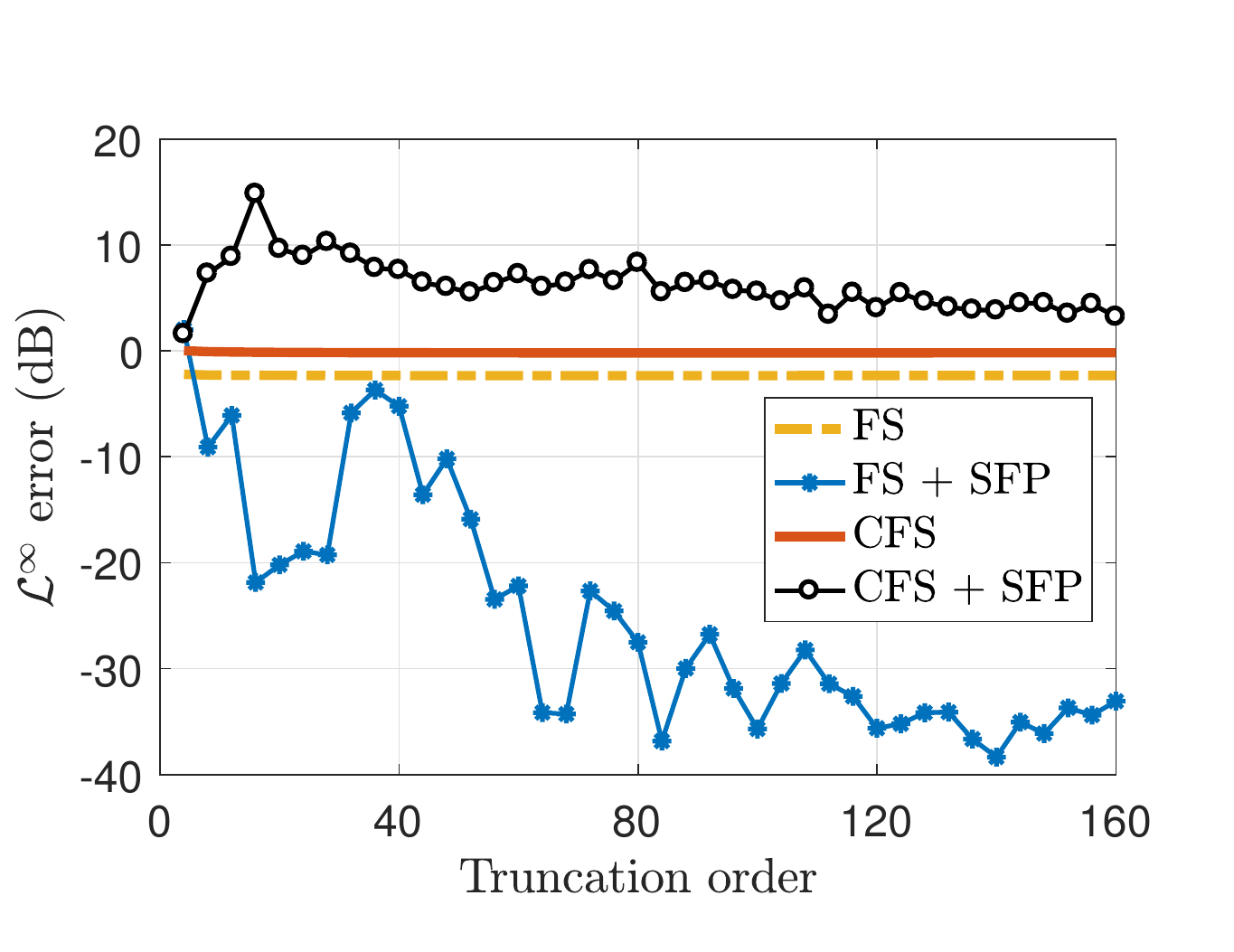}
		\caption{}\label{N_sfp}		
	\end{subfigure}
~
	\begin{subfigure}{0.31\textwidth}
		\centering
		\includegraphics[width=\linewidth]{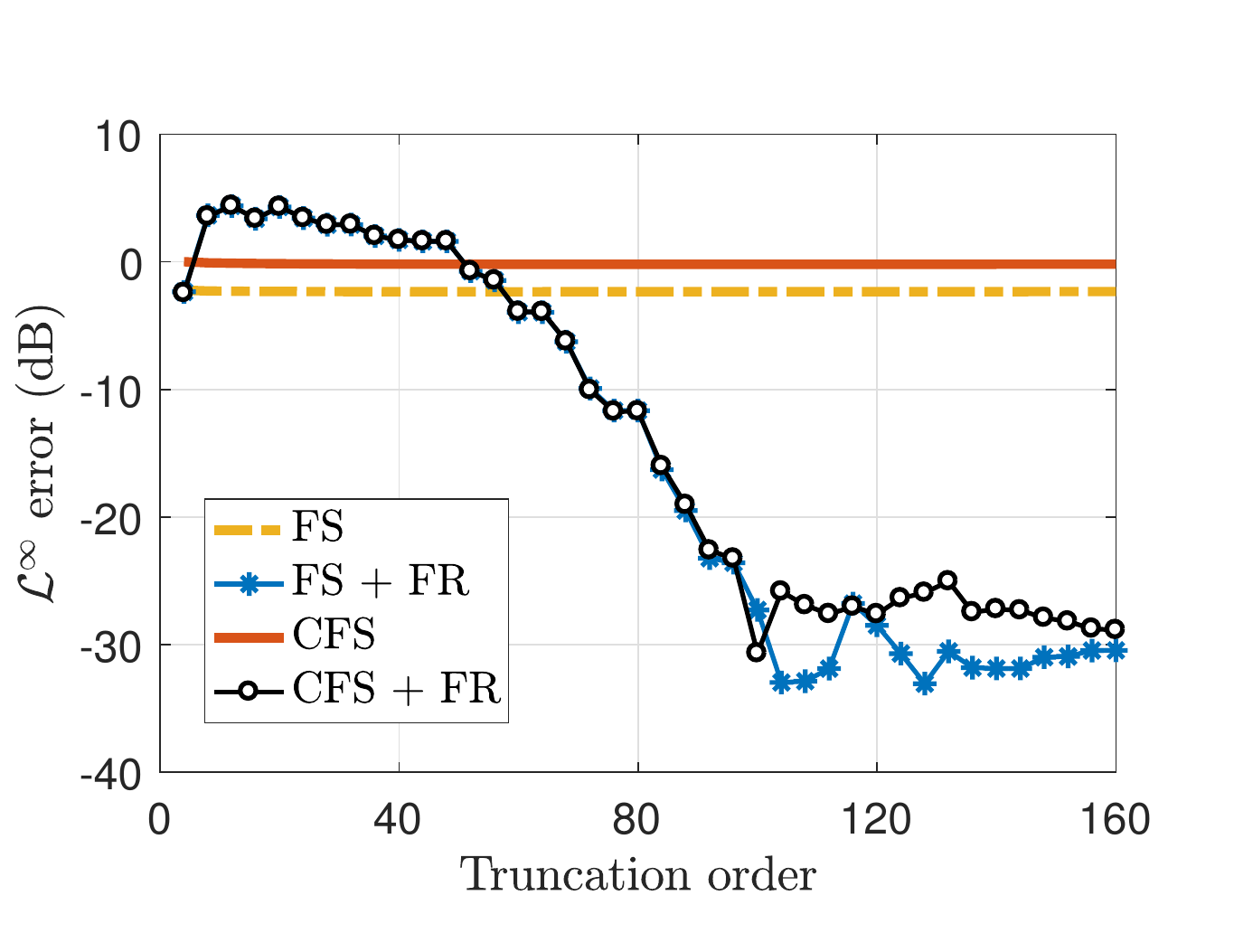}
		\caption{}\label{N_fr}
	\end{subfigure}
~
	\begin{subfigure}{0.31\textwidth}
		\centering
		\includegraphics[width=\linewidth]{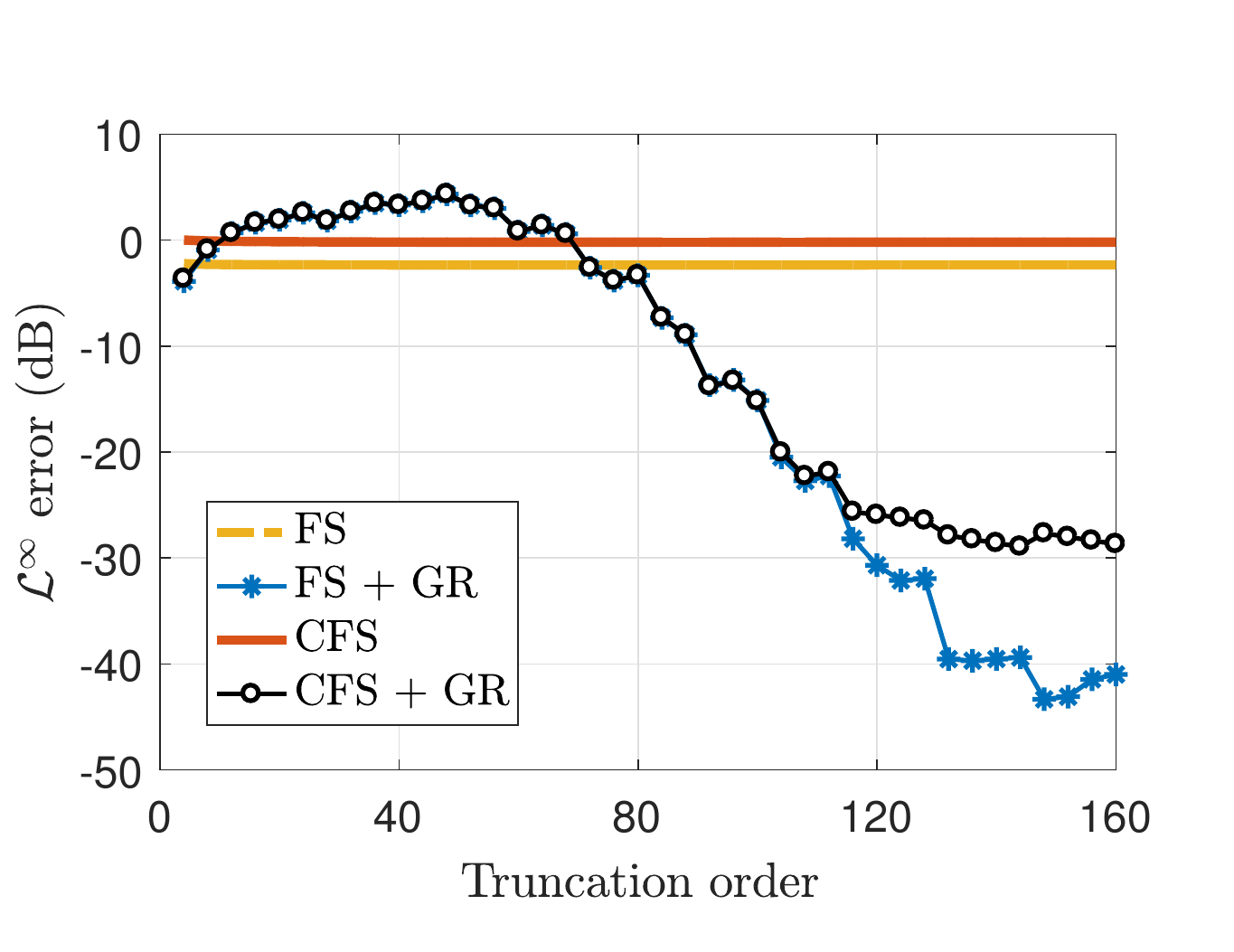}
		\caption{}\label{N_gr}
	\end{subfigure}	

	\caption{Fourier series and reconstruction error of the test function \ref{z15} vs. the Fourier order, for(\subref{N_vf}) Vandeven filter, (\subref{N_lpls}) Legendre polynomial least-squares, (\subref{N_fe}) Fourier extension, (\subref{N_sfp}) singular Fourier-Pad\'e, (\subref{N_fr}) Freud reconstruction, and (\subref{N_gr}) Gegenbauer reconstruction methods.}
	\label{fig_N_comp}
\end{figure*}

In figure \ref{error_vf}, the error graphs are plotted for the Vandeven filter, as one of the most efficient filters. In figure \ref{error_lpls},same graphs are plotted for the polynomial least-squares method. We have chosen the Legendre polynomials as the reconstruction basis due to their stability and simplicity, with the parameters $\Scale[.9]{(\eta_1,\eta_2)=(3\sqrt{3}/4,\sqrt{3}/4)}$ for the two smooth sub-domains. In figure \ref{error_fe}, the corresponding graphs are plotted for the Fourier extension method, with the parameters $\Scale[.9]{(T_1,\eta_1,T_2,\eta_2)=(2,6,6,2)}$ for the two sub-domains.
In figure \ref{error_sfp}, the error graphs are plotted for the singular Fourier-Pad\'e method. Eventually, figures \ref{error_fr} and \ref{error_gr}, pertain to the spectral reprojection via Freud and Gegenbauer polynomials respectively.
Besides the point-wise error plots, we have also plotted the maximum error ($\Scale[.9]{\mathcal{L}^\infty}$ error) of each method versus the Fourier order $\Scale[.9]{N}$ in figure \ref{fig_N_comp}. Similarly, each graph includes four error plots, in the respective order mentioned above. In this figure, we have only plotted the error graphs for the discontinuous test function, as the corresponding graphs of the other function demonstrate quite similar behavior.

Observing the results, let us make some key points on the performance of the above-mentioned methods.
We begin with the point-wise error graphs of figure \ref{fig_error_comp}. First of all, either with the original or convolutional Fourier coefficients, retrieving a function becomes less accurate at the boundaries of the reconstruction interval, regardless of the Gibbs phenomenon or TCE.
Moreover, post-processing methods predominantly do not make a distinction between a continuous or discontinuous function.
Besides, comparing the red and black plots reveals that no method, except for the last two, is capable of retrieving the function around discontinuities from the convolutional Fourier series. Whereas, using the original Fourier series, only the filtering method (as expected) cannot improve the retrieval around these points.

Let us also consider the convergence behavior of the methods in figure \ref{fig_N_comp}. As expected, the Vandeven filter error decreases very slowly with both the original and the convolutional Fourier series. On the other side, excluding the spectral reprojections, the emergence of TCE totally disrupts the convergence of other techniques.
On the downside, given the ideal Fourier coefficients (blue graphs), the beginning points of the convergence in the last two methods appear at comparably higher-orders. This phenomenon is attributed to the poor resolution power of these methods, i.e., the ability to recover the oscillatory behavior of a function \cite{Gottlieb1994}, while for instance, the Fourier extension method demonstrates an excellent resolution power \cite{Adcock2014_2}. Moreover, there are serious stability concerns with the Gegenbauer method stemming from the round-off error \cite{Gelb2004, Gelb2006} or other complications such as the generalized Rung\'e phenomenon \cite{Boyd2005}. In fact, the instability is an intrinsic property to any exponentially fast method \cite{Adcock2014-4}.

We will conclude this part by comparing the two spectral reprojection techniques, i.e. Freud and Gegenbauer methods. The former lacks a rigorous mathematical theory of convergence, and its applicability is only numerically substantiated. Besides, since the Freud polynomials do not have a closed-form, they should be estimated on a discrete stencil using the Stieltjes algorithm \cite{Gelb2006}, making it a much more computationally expensive approach, and leaving the Gegenbauer method as a more plausible approach.

\section{Robustness against TCE}
\label{Sec: Gegen}

Here, we present the basic formulation of the Gegenbauer method and then use it to prove its robustness against TCE. An analytic explanation of the inefficacy of other methods follows.

\subsection{Spectral Reprojection}
\label{Sub: Gegen}

In the spectral reprojection method, the finite expansion of a function in terms of the basis functions (complex exponentials here) is projected onto another space with different bases, e.g. polynomials. Let $\Scale[.9]{\{\psi_k=e^{ik\pi x}\}}_{k=-N}^{N}$ and $\Scale[.9]{\{\varphi_\ell\}_{\ell=0}^{M}}$ characterize the expansion and projection spaces respectively, and $\Scale[.9]{f}$ be an analytic function over $\Scale[.9]{\left[{- 1,1} \right]}$, with truncated finite Fourier series as defined in equation \ref{z1}.
The spectral reprojection $\Scale[.9]{f_{N,M}}$ of the truncated expansion $\Scale[.9]{f_N}$, would be defined as the follows:

\begin{eqnarray}\label{z16}
\Scale[.9]{
f_{N,M}(x) = \displaystyle\sum\nolimits_{\ell=0}^{M}\{{\langle f_N,\varphi_\ell \rangle}_w/{\langle \varphi_\ell,\varphi_\ell \rangle}_w\} \varphi_\ell(x).
}
\end{eqnarray}

The idea is that by selecting the appropriate bases $\Scale[.9]{\varphi_l}$, the new spectral expansion can converge to the main function at a faster rate.
A set of bases that fulfills such a property is called a Gibbs' complement.
Gottlieb \emph{et al.} showed that Gegenbauer polynomials $\Scale[.9]{\varphi_\ell = C_{\ell}^\lambda(x)}$, under certain circumstances, constitute a Gibbs' complement. Later Gelb \emph{et al.} speculated that a properly tailored set of Freud polynomials can be another Gibbs' complement. To date, however, no other bases with such properties have been discovered.

The principles of the spectral reprojection using Gegenbauer polynomials are briefly noted here. Given the weight parameter $\Scale[.9]{\lambda>-\nicefrac{1}{2}}$ (not to be mixed up with the wavelength) and for $\Scale[.9]{\ell\in \mathbb{N}\cup \{0\}}$, Gegenbauer polynomials $\Scale[.9]{C_\ell ^ \lambda(x)}$ are a group of orthogonal polynomials defined over the interval $\Scale[.9]{\left [{- 1,1} \right]}$ with the weight function $\Scale[.9]{w_\lambda(x) = {({1 - {x ^ 2}})^{\lambda - \nicefrac{1}{2}}}}$, satisfying a weighted orthogonality relationship as follows, in which $\Scale[.9]{\delta _{\ell,n}}$ is the Kronecker delta and $\Scale[.9]{\Gamma(\cdot)}$ denotes the famous gamma function:

\begin{eqnarray}\label{z17}
\Scale[.9]{
\left\langle C_\ell,C_n \right\rangle_w = \displaystyle\int_{ - 1}^1 { C_\ell^\lambda C_n^\lambda w_\lambda dx = {\delta _{l,n}}h_n^\lambda},\hspace{50pt}
h_n^\lambda  = {\pi ^{ \nicefrac{1}{2}}}\frac{{\Gamma \left( {n + 2\lambda } \right)}{\Gamma \left( {\lambda  +  \nicefrac{1}{2}} \right)}}{{\Gamma \left( {2\lambda } \right)n!}{\Gamma \left( \lambda  \right)\left( {n + \lambda } \right)}}.
}
\end{eqnarray}

Hopefully, there is a closed-form for the Fourier coefficients of the Gegenbauer polynomials, as follows, which plays a pivotal role in deriving the convergence rate of the method:

\begin{eqnarray}\label{z20}
\Scale[.9]{
\frac{1}{{h_\ell^\lambda }}\left\langle e^{in\pi x},C_\ell^\lambda \right\rangle_w =
\Gamma \left( \lambda  \right){\left( {{\pi n}/{2}} \right)^{-\lambda} }{i^\ell}\left( {\ell + \lambda } \right){{\rm J}_{\ell + \lambda }}\left( {\pi n} \right).
}
\end{eqnarray}

In which $\Scale[.9]{{\rm J}_\alpha(\cdot)}$ denotes the Bessel function of the first kind. Now, we denote the ordinary and special projection of a typical function $\Scale[.9]{u(x)}$ on the space of Gegenbauer polynomials of the maximum order $\Scale[.9]{M+1}$ and the parameter $\Scale[.9]{\lambda}$ with $\Scale[.9]{\mathcal{G}_{M,\lambda}\left\{u \right\}}$ and $\Scale[.9]{{\mathcal{G}}^{\star}_{M,\lambda}\left\{u \right\}}$ respectively, defined as below:

\begin{eqnarray}\label{z21}
\Scale[.9]{
\mathcal{G}_{M,\lambda}\left\{u \right\} = \displaystyle\sum\nolimits_{\ell = 0}^{M} \frac{1}{{h_\ell^\lambda }}{\left\langle {u ,C_\ell^\lambda } \right\rangle_w C_\ell^\lambda(x)},\hspace{50pt}
{\mathcal{G}}^{\star}_{M,\lambda}\left\{u \right\} = {\mathcal{G}}_{M,\lambda}\left\{u \right\}-{\langle {u ,C_0^\lambda}\rangle_w C_0^\lambda /{h_0^\lambda }}.
}
\end{eqnarray}

The special projection clearly differs from the ordinary one in a constant.
Since Gegenbauer polynomials constitute a complete set of bases, the smooth function $\Scale[.9]{f}$ over $\Scale[.9]{[-1,1]}$, can be expressed as $\Scale[.9]{f = \mathcal{G}_{\infty,\lambda}\left\{f \right\}}$. In the spectral reprojection, a truncated Gegenbauer series is approximated, using the accessible Fourier approximation instead of the function itself, i.e. $\Scale[.9]{\mathcal{G}_{M,\lambda}\left\{f_N \right\}}$. This approach can also be generalized for piecewise smooth functions, wherein the function is reconstructed over a smooth subinterval $\Scale[.9]{x \in \left[{a,b}\right]}$, using the mapping parameters $\Scale[.9]{\delta = \left({a + b} \right) / 2} $ and $\Scale[.9]{\varepsilon = \left({b - a} \right) / 2}$ such that $\Scale[.9]{x={\varepsilon \rho + \delta}}$ for $\Scale[.9]{\rho \in \left[{-1, 1} \right]}$. Accordingly, a Gegenbauer expansion for the function can be expressed in terms of the new variable $\Scale[.9]{\rho}$, as follows:

\begin{eqnarray}\label{z25}
\Scale[.9]{
f\left( x = {\varepsilon \rho  + \delta } \right) = \mathcal{G}_\infty^\lambda\left\{f(\varepsilon \rho  + \delta) \right\} = \displaystyle\sum\nolimits_{l = 0}^\infty  {\hat f_\varepsilon ^\lambda \left( l \right)C_l^\lambda \left( \rho  \right)}, \hspace{25pt}
\hat f_\varepsilon ^\lambda \left( l \right) = \frac{1}{{h_l^\lambda }}\left\langle f(\varepsilon\rho+\delta),C_l^\lambda(\rho) \right\rangle_w.
}
\end{eqnarray}

In which $\Scale[.9]{{\hat f_\varepsilon^\lambda }\left( l \right)}$ denote the Gegenbauer coefficients, and are to be approximated using the truncated Fourier series instead of the function, denoted by $\Scale[.9]{{\hat g_\varepsilon^\lambda }\left( l \right)}$. Thanks to the identity \ref{z20}, $\Scale[.9]{{\hat g_\varepsilon^\lambda }\left( l \right)}$ will have a closed form:

\begin{eqnarray}\label{z26}
\Scale[.9]{
\hat g_\varepsilon ^\lambda \left( l \right) = \Gamma \left( \lambda  \right){i^l}\left( {l + \lambda } \right)\displaystyle\sum\nolimits_{\left| k \right| \leqslant  N} {{{\hat f}(k)}{e^{i\pi k\delta }}} {\left( {{\pi k\varepsilon }/{2}} \right)^{-\lambda} }{{\rm J}_{l + \lambda }}\left( {\pi k\varepsilon } \right).
}
\end{eqnarray}

The coefficients $\Scale[.9]{\{\hat g_\varepsilon ^\lambda \left( \ell \right)\}_{\ell=0}^{M}}$ can then be used with corresponding polynomials $\Scale[.9]{C_\ell^\lambda \left\{ \rho=(x-\varepsilon)/\delta \right\}}$ terms to reconstruct the function over $\Scale[.9]{\left[{a,b}\right]}$. Note that the summand with $\Scale[.9]{k=0}$  in equation \ref{z26} should simply be computed as $\Scale[.9]{k\to 0}$. Apart from that, the summation is not difficult to evaluate, though, fast Fourier transform-based methods are also available for computation of Gegenbauer coefficients \cite{Micheli2013}.
What Gottlieb \emph{et al.} proved was that constant parameters $\Scale[.9]{(\alpha,\beta)}$ can be found such that given $\Scale[.9]{M=\beta N}$ and $\Scale[.9]{\lambda=\alpha N}$, the reconstruction error satisfies $\Scale[.9]{\|f-\mathcal{G}_{M,\lambda}\left\{f_N \right\}\|_\infty=\mathcal{O}(q^{\varepsilon N})}$ as $\Scale[.9]{N\to\infty}$, for some $\Scale[.9]{q<1}$.
In particular, they showed that for $\Scale[.9]{\alpha=\beta=0.25}$, the exponential convergence is guaranteed, though, the choice might not be optimum. The reliance of the Gegenbauer convergence to the choice of parameters is known to be significant, and despite scarce reports on the optimization of Gegenbauer parameters  \cite{Gelb2004}, the underlying theory still lacks a systematic study.

\subsection{Robustness of Gegenbauer Method}

As noted, the overall efficacy of spectral reconstructions in a finite computational framework, largely depends on their robustness against TCE, which can be more effectively studied using an appropriately shifted form of its first-order approximation, i.e. Li's function. A robust method should be able to resolve this function as an input, as the Fourier order tends to infinity. For the Gegenbauer method, reconstruction of $\Scale[.9]{{\Phi_N}(x-1) = {\Phi_N}(x + 1)}$ should accordingly be considered under the $\Scale[.9]{\mathcal{L}^\infty}$-norm. Before we begin, let us denote the ordinary and special Gegenbauer reconstruction of the shifted Li's function as follows, according to the definition \ref{z4} and \ref{z21}:

\begin{eqnarray}\label{z28}
\Scale[0.9]{
\Psi_N(x)= \mathcal{G}_{M,\lambda}\left\{{\Phi_N}\left( {x - 1} \right) \right\},\hspace{50pt}
\Psi_N^\star(x)= \mathcal{G}_{M,\lambda}^\star\left\{{\Phi_N}\left( {x - 1} \right) \right\}. 	
}
\end{eqnarray}

\begin{figure}[!t]
\centering
\includegraphics[width=.73\linewidth]{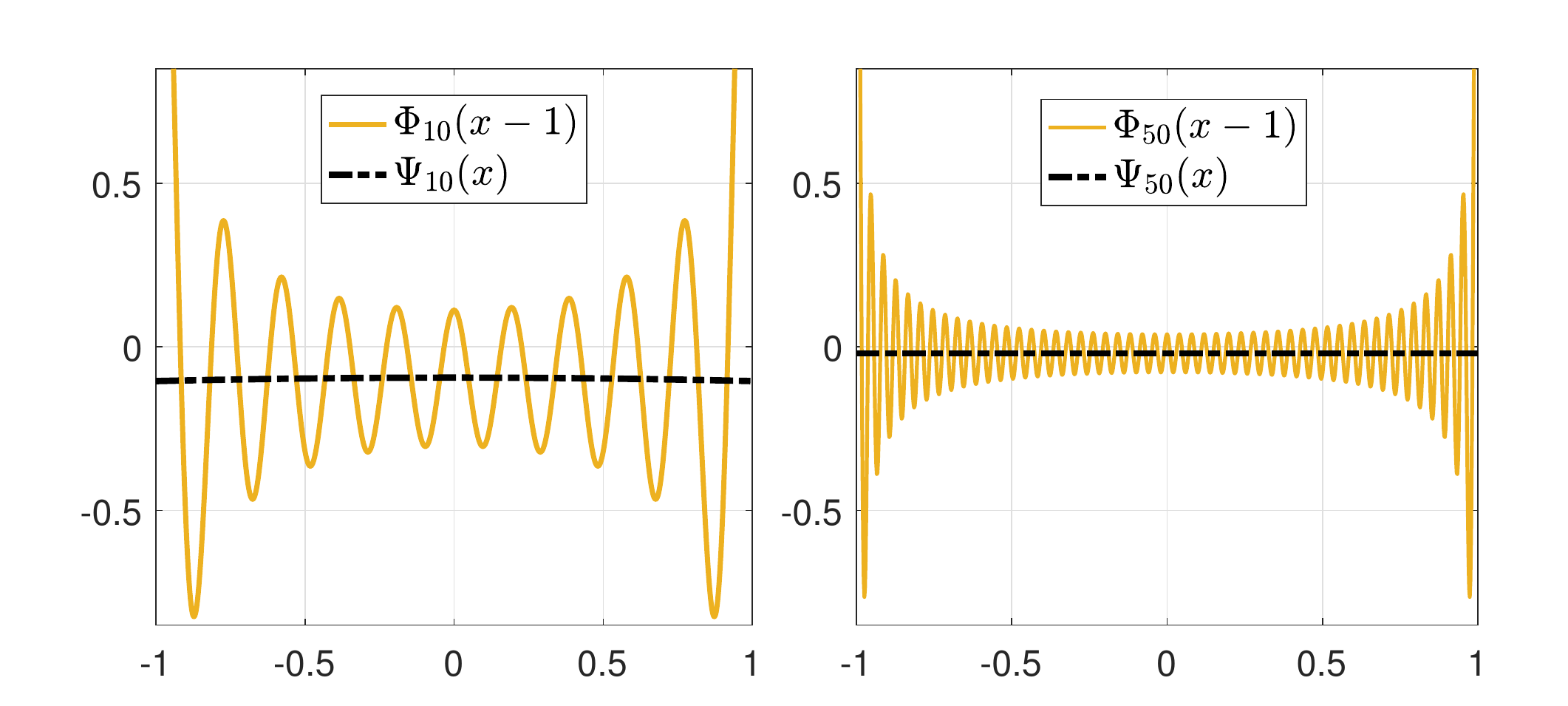}
\caption{Shifted Li's function $\Phi_N(x-1)$ and its Gegenbauer image $\Psi_N(x)$ for $N=10$ and $N=50$.}
\label{Li_func}
\end{figure}

In figure \ref{Li_func}, the shifted Li's function $\Scale[.9]{{\Phi_N}(x-1)}$ and its projection on the Gegenbauer space for $\Scale[.9]{N = 10}$ and $\Scale[.9]{N = 50}$ are plotted, which seems to become smaller overall by increasing the Fourier order $\Scale[.9]{N}$. Here, we will quantify and prove this observation as the following theorem:

\textbf{Theorem 3.1.} \textit{The ordinary and special Gegenbauer projection of the Li's shifted function diminish as $\Scale[.9]{N\to\infty}$, satisfying the following asymptotic relationships:}

\begin{eqnarray}\label{z29}
\Scale[.9]{
\left\|\Psi_N(x)\right\|_\infty = \mathcal{O}\left(N^{-1}\right),\hspace{50pt}
\left\|\Psi^\star_N(x)\right\|_\infty = \mathcal{O}\left(\ln(N) N^{-\nicefrac{1}{2}} e^{-0.232N} \right).
}
\end{eqnarray}

\textbf{Proof.} Henceforth, we will use the notations $\Scale[.9]{\mathcal{E}=\left\|\Psi_N(x)\right\|_\infty}$ and $\Scale[.9]{\mathcal{E}^\star=\left\|\Psi^\star_N(x)\right\|_\infty}$ for simplicity.
We start by defining  a weighted summation of the Bessel functions for nonnegative real $\Scale[.9]{\lambda}$ and nonnegative integer $\Scale[.9]{m}$, as follows:

\begin{eqnarray}\label{z30}
\Scale[0.9]{
\mathcal{S}_{m,\lambda}(n)= \displaystyle\sum\nolimits_{\ell=1}^{n}{(-1)^{\ell} {\ell}^{-\lambda}{\rm J}_{m+\lambda}\left(\ell\pi\right)}.
}
\end{eqnarray}

Our proof depends on a property of the above summation, expressed as the following lemma:

\textbf{Lemma 3.1.} \textit{Given the aforementioned prerequisites, the following relationship holds:}

\begin{eqnarray}\label{z31}
\Scale[0.9]{
\begin{array}{l}
\lim \limits_{n\to \infty} \mathcal{S}_{m,\lambda}(n)= -\frac{\left({\pi}/{2}\right)^\lambda}{2\Gamma\left(\lambda+1\right)}\delta_m + C_{m,\lambda}e_m,\hspace{50pt}C_{m,\lambda}\neq 0.
\end{array}
}
\end{eqnarray}

\textit{In which $\Scale[.9]{C_{m,\lambda}}$ is a constant with respect to $\Scale[.9]{n}$, $\Scale[.9]{e_m=\sin^2\left({m\pi}/{2}\right)}$ returning $\Scale[.9]{0}$ for even subscripts and $\Scale[.9]{1}$ for odd ones, and $\Scale[.9]{\delta_{m}=\delta_{m,0}}$ denotes the Kronecker delta.}

As the rest of the arguments do not depend on the proof of this lemma, we will bring it at the end of this section. A difference form of equation \ref{z30}, can be expressed as follows:

\begin{eqnarray}\label{z32}
\Scale[0.9]{
\Delta\mathcal{S}_{m,\lambda} = \mathcal{S}_{m,\lambda}(n)-\mathcal{S}_{m,\lambda}(n-1)= (-1)^n n^{-\lambda}{\rm J}_{m+\lambda}\left(n\pi\right).
}
\end{eqnarray}

Bessel functions exhibit the asymptotic behavior $\Scale[.9]{{\rm J}_{\alpha}\left(x\right)\approx (\pi x/2)^{-\nicefrac{1}{2}}\cos\left(x-\alpha\pi/2-\pi/4\right)}$ for $\Scale[.9]{x/\alpha \gg 1}$. Hence, given that $\Scale[.9]{n}$ is sufficiently larger than $\Scale[.9]{m+\lambda}$, the difference term in equation \ref{z32} can be approximated with $\Scale[.9]{(-1)^n n^{-\lambda}{\rm J}_{m+\lambda}\left(n\pi\right) \sim c n^{-\lambda-\nicefrac{1}{2}}}$ for some constant $\Scale[.9]{c}$.
As a result, after a certain point, say $\Scale[.9]{\ell=n_0}$, the summands do not change sign (without loss of generality assume they are positive), and the summation of equation \ref{z30} behaves as a p-series, converging as $\Scale[.9]{n\to\infty}$, if and only if $\Scale[.9]{\lambda>\nicefrac{1}{2}}$. On the other hand, irrespective of the limit to the series, the rate of convergence/divergence for $\Scale[.9]{\mathcal{S}_{m,\lambda}(n)}$ may be evaluated through the summation (approximated by integration) over the difference-terms of equation \ref{z32}. In other words, for sufficiently large $\Scale[.9]{n\pi/(m+\lambda)}$:

\begin{eqnarray}\label{z33}
\Scale[0.9]{
\begin{array}{l}
\mathcal{S}_{m,\lambda}(n) = A_0 + \displaystyle\sum\nolimits_{\ell=n_0}^{n}{\Delta\mathcal{S}_{m,\lambda}}(\ell) \approx A_0 + A_1 \displaystyle\sum\nolimits_{\ell=n_0}^{n}{\ell^{-\lambda-\nicefrac{1}{2}}}\\
\hspace{37pt}\approx  A_0 + A_1 \displaystyle\int\nolimits_{n_0}^{n} {\ell^{-\lambda-\nicefrac{1}{2}}d\ell} \sim A + \mathcal{O}\left(|\lambda|^{-1}n^{-\lambda+\nicefrac{1}{2}}\right).
\end{array}
}
\end{eqnarray}

In which $\Scale[.9]{A_0}$, $\Scale[.9]{A_1}$, and $\Scale[.9]{A}$ are constants. Note that unlike the aforementioned asymptotic relationship for Bessel functions which holds for sufficiently large arguments, we could have used Landau's inequality that holds everywhere, i.e. $\Scale[.9]{\left|{\rm J}_{\alpha}\left(x\right)\right|\leqslant b{|x|}^{-\nicefrac{1}{3}}}$ for all positive $\Scale[.9]{\alpha}$ and real $\Scale[.9]{x}$ with $\Scale[.9]{b\approx 0.7875}$ \cite{Landau2000}, though slightly weakening the bound. In other words, by replacing Landau's bound, equation \ref{z33} would have held for relatively smaller values of $\Scale[.9]{n}$ (it still needs to be large enough for terms of equation \ref{z32} not to change sign).
In practice, by taking the square root relationship, one should have at least $\Scale[.9]{x>3(m+\lambda)}$, yet by taking the cubic root bound, even $\Scale[.9]{x>m+\lambda}$ seems to suffice. In our final application, the argument will be large enough to utilize the first bound, however, for the sake of generality, we may rewrite equation \ref{z33} using $\Scale[.9]{\left|{\rm J}_{\alpha}\left(x\right)\right|\leqslant b|x|^{-\epsilon}}$ for $\Scale[.9]{\epsilon \in [\nicefrac{1}{3},\nicefrac{1}{2}]}$.

According to Lemma 4.1. for nonzero even values of $\Scale[.9]{m}$ and real $\Scale[.9]{\lambda>\nicefrac{1}{2}}$, the constant $\Scale[.9]{A}$ in equation \ref{z33} should be zero to satisfy the zero limit. On the other hand, for $\Scale[.9]{\lambda<\nicefrac{1}{2}}$, the sum tends to infinity and the amount of the constant does not matter anyway. Consequently, for large enough $\Scale[.9]{n\pi/(m+\lambda)}$, equation \ref{z33} can be rewritten as follows:

\begin{eqnarray}\label{z34}
\Scale[0.9]{
\mathcal{S}_{m,\lambda}(n) = (e_m + \delta_m) A_{m,\lambda} + \mathcal{O}\left(|\lambda|^{-1}n^{-\lambda-\epsilon+1}\right).	
}
\end{eqnarray}

Where $\Scale[.9]{A_{m,\lambda}}$ is a nonzero constant. Now, let $\Scale[.9]{{\mathfrak{p}_\nu}(\cdot)}$ denote a polynomial of the order $\Scale[.9]{\nu}$, which depending on the parity of $\Scale[.9]{\nu}$, comprises only even or odd order terms. Moreover, assume $\Scale[.9]{{\mathfrak{p}_\nu}(n)}$ is a nonnegative increasing function for $\Scale[.9]{n\in \mathbb{N}}$. We will denote the set all such polynomials with $\Scale[.9]{\mathbb{P}^\ast}$. Note that any $\Scale[.9]{{\mathfrak{p}_\nu}\in\mathbb{P}^\ast}$ can be expressed as below:

\begin{eqnarray}\label{z35}
\Scale[0.9]{
{\mathfrak{p}_\nu}(x)= \displaystyle\sum\limits_{0 \leqslant r \leqslant \nu \hfil\atop
2\mid \nu-r\hfil}{\rho_{r}x^{r}}.	
}
\end{eqnarray}

Now, we would like to generalize the Bessel sums of equation \ref{z30}, based on a polynomial $\Scale[.9]{{\mathfrak{p}_\nu}\in\mathbb{P}^\ast}$. We define:

\begin{eqnarray}\label{z36}
\Scale[0.9]{
\mathcal{S}_{m,\lambda}\left(n;{\mathfrak{p}_\nu}\right)= \displaystyle\sum\nolimits_{\ell=1}^{n}{(-1)^{\ell}{\mathfrak{p}_\nu}(\ell) {\ell}^{-\lambda}{\rm{J}}_{m+\lambda}\left(\ell\pi\right)}.
}
\end{eqnarray}

Notice that the generalized form of equation \ref{z36} can be expressed in terms of the simple forms of equation \ref{z30}, as follows:

\begin{eqnarray}\label{z37}
\Scale[0.9]{
\mathcal{S}_{m,\lambda}\left(n;{\mathfrak{p}_\nu}\right) = \displaystyle\sum\limits_{0 \leqslant r \leqslant \nu \hfil\atop
2\mid \nu-r\hfil}{{\rho_r}\mathcal{S}_{m+r,\lambda-r}(n)}.
}
\end{eqnarray}

If we assume $\Scale[.9]{\nu}$ is an even number and $\Scale[.9]{n\pi/(m+\lambda)}$ is large enough, it is straightforward using equations \ref{z34} and \ref{z37} to conclude $\Scale[.9]{\mathcal{S}_{m,\lambda}\left(\cdot;{\mathfrak{p}_\nu}\right)}$ should have the following asymptotic form, wherein $\Scale[.9]{A_{m,\lambda}}$ and $\Scale[.9]{c}$ are some constants:

\begin{eqnarray}\label{z38}
\Scale[0.9]{
\mathcal{S}_{m,\lambda}\left(n;{\mathfrak{p}_\nu}\right) = \left(e_{m+\nu} + e_{\nu+1}\delta_{m}\right)A_{m,\lambda} + \mathcal{O}(n^c).
}
\end{eqnarray}

The above equation specifies the form of the constant term accurately enough. However, to further characterize the $\Scale[.9]{n}$-dependant term $\Scale[.9]{\mathcal{O}(n^c)}$, we need to use a similar approach to what we practiced earlier for the derivation of equation \ref{z34}. Writing down the difference-form of equation \ref{z36} accordingly, we have:

\begin{eqnarray}\label{z39}
\Scale[0.9]{
\Delta\mathcal{S}_{m,\lambda}\left(n;{\mathfrak{p}_\nu}\right) = \mathcal{S}_{m,\lambda}\left(n;{\mathfrak{p}_\nu}\right) - \mathcal{S}_{m,\lambda}\left(n-1;{\mathfrak{p}_\nu}\right)
 =(-1)^n{\mathfrak{p}_\nu}(n) n^{-\lambda}{\rm J}_{m+\lambda}\left(n\pi\right)\sim {\mathfrak{p}_\nu}(n) n^{-\lambda-\epsilon}.
}
\end{eqnarray}

Wherein we have assumed for the sake of simplicity (though without loss of generality) that the difference-terms are positive. Now, similar to equation \ref{z33}, we may conclude:

\begin{eqnarray}\label{z40}
\Scale[0.9]{
\begin{array}{l}
\mathcal{S}_{m,\lambda}\left(n;{\mathfrak{p}_\nu}\right) = A_0 + \displaystyle\sum\nolimits_{\ell=n_0}^{n}{\Delta\mathcal{S}_{m,\lambda}}\left(\ell;{\mathfrak{p}_\nu}\right) \approx A + B \displaystyle\sum\nolimits_{\ell=n_0}^{n}{{\mathfrak{p}_\nu}(\ell)\ell^{-\lambda-\epsilon}}\\
\hspace{55pt} \leqslant A + B \mathfrak{p}_\nu (n) \displaystyle\sum\nolimits_{\ell=n_0}^{n}{\ell^{-\lambda-\epsilon}}\sim A + B \mathfrak{p}_\nu(n)|\lambda|^{-1}{n^{-\lambda-\epsilon+1}}.
\end{array}
}
\end{eqnarray}

In which $\Scale[.9]{A_0}$, $\Scale[.9]{A}$, and $\Scale[.9]{B}$ are constants, and we have used the fact that $\Scale[.9]{\mathfrak{p}_\nu}$ is positive and increasing over positive integers.
If we let $\Scale[.9]{\lambda =\mathcal{O}(N)}$ for some $\Scale[.9]{N}$ such that $\Scale[.9]{n \leqslant N}$, then:

\begin{eqnarray}\label{z41}
\Scale[0.9]{
\mathcal{S}_{m,\lambda}\left(n;{\mathfrak{p}_\nu}\right) \sim \left(e_{m+\nu} + e_{\nu+1}\delta_{m}\right)A_{m,\lambda} + \mathcal{O}\left({\mathfrak{p}_\nu}(n) n^{-\lambda-\epsilon}\right).
}
\end{eqnarray}

Provided that $\Scale[.9]{\nu}$ and $\Scale[.9]{m}$ are both even, the following result will consequently hold, using Lemma 4.1:

\begin{eqnarray}\label{z42}
\Scale[0.9]{
\mathcal{S}_{m,\lambda}\left(n;{\mathfrak{p}_\nu}\right) = A_{\lambda} \delta_m \mathfrak{p}_\nu(0) + \mathcal{O}\left(  \mathfrak{p}_\nu(n) n^{-\lambda-\epsilon} \right),\hspace{40pt}
A_{\lambda}\sim \lim \limits_{n\to \infty} \mathcal{S}_{0,\lambda}(n) \sim {\left(\nicefrac{\pi}{2}\right)^\lambda}/{\Gamma\left(\lambda+1\right)}.	
}
\end{eqnarray}

Considering a polynomial $\Scale[.9]{{\mathfrak{p}_\nu}\in\mathbb{P}^\ast}$ with an even $\Scale[.9]{\nu}$, now we define a perturbative test function as below, an try to evaluate how the Gegenbauer method reprojects it:

\begin{eqnarray}\label{z43}
\Scale[0.9]{
\Delta _N \left( x;{\mathfrak{p}_\nu} \right) =  {\displaystyle\sum\limits_{0 < |n| \leqslant N} (-1)^n{\mathfrak{p}_\nu}\left(|n|\right){e^{in\pi x}}}.
}
\end{eqnarray}

We denote the corresponding reconstruction of the above function as follows:

\begin{eqnarray}\label{z44}
\Scale[0.9]{
\Psi\left( x;{\mathfrak{p}_\nu} \right) = \mathcal{G}_{M,\lambda} \left\{\Delta_N \left( {x};{\mathfrak{p}_\nu} \right) \right\}.
}
\end{eqnarray}

Considering the above function, let us define $\Scale[.9]{\mathcal{E}^{{\mathfrak{p}_\nu}} =\Psi(1;{\mathfrak{p}_\nu})}$.
Notice that since both the perturbative functions $\Scale[.9]{\Delta_N(x;{\mathfrak{p}_\nu})}$ and the Gegenbauer polynomials take their maximum absolute value at the boundaries, we may conclude:

\begin{eqnarray}\label{z45}
\Scale[0.9]{
\|\Psi(\cdot;{\mathfrak{p}_\nu})\|_\infty = \left|\Psi(1;{\mathfrak{p}_\nu})\right| = \left|\mathcal{E}^{\mathfrak{p}_\nu}\right|.
}
\end{eqnarray}

Our strategy henceforth would be to estimate $\Scale[.9]{\mathcal{E}^{\mathfrak{p}_\nu}}$, expand $\Scale[.9]{\Phi(x-1)}$ in terms of $\Scale[.9]{\Delta _N \left( x;{\mathfrak{p}_\nu} \right)}$ functions, and try to estimate $\Scale[.9]{\mathcal{E}}$ in terms of $\Scale[.9]{\mathcal{E}^{\mathfrak{p}_\nu}}$.
To estimate the former, we begin by noting that $\Scale[.9]{\Delta _N \left( x;{\mathfrak{p}_\nu} \right)}$ and $\Scale[.9]{w_\lambda(x)=(1-x^2)^{\lambda-\nicefrac{1}{2}}}$ are both even functions. As a result, the following orthogonality relationship holds for all the odd-ordered Gegenbauer polynomials (being odd functions):

\begin{eqnarray}\label{z46}
\Scale[0.9]{
\left\langle {\Delta _N(\cdot;{\mathfrak{p}_\nu}) ,C_{m = 2k + 1}^\lambda } \right\rangle_w  = 0.
}
\end{eqnarray}

Now, using the above property, equation \ref{z44} could be expanded as follows:

\begin{eqnarray}\label{z47}
\Scale[0.9]{
\Psi\left( x;{\mathfrak{p}_\nu} \right) = \displaystyle\sum\nolimits_{k = 0}^{{M}/{2}} \frac{1}{{h_{2k}^\lambda }} {\left\langle {\Delta_N(\cdot;{\mathfrak{p}_\nu}) ,C_{2k}^\lambda } \right\rangle_w C_{2k}^\lambda \left( x \right)}.
}
\end{eqnarray}

As a result and by definition:

\begin{eqnarray}\label{z48}
\Scale[0.9]{
\mathcal{E}^{\mathfrak{p}_\nu} = \displaystyle\sum\limits_{k = 0}^{{M}/{2}} \frac{1}{{h_{2k}^\lambda }}\left\langle {\Delta_N(\cdot;{\mathfrak{p}_\nu}) ,C_{2k}^\lambda } \right\rangle_w C_{2k}^\lambda(1)  =
2\displaystyle\sum\limits_{k = 0}^{{M}/{2}} {\sum\limits_{n = 1}^N \frac{\left(-1 \right)^n \mathfrak{p}_\nu(n)}{{h_{2k}^\lambda }} \left\langle {{e^{in\pi x}},C_{2k}^\lambda } \right\rangle_w \frac{{\Gamma \left( {2k + 2\lambda } \right)}}{{\left( {2k} \right)!\Gamma \left( {2\lambda } \right)}}}.
}
\end{eqnarray}

In which we have made use of the fact that $\Scale[.9]{C_m^\lambda \left( 1 \right) = \frac{{\Gamma \left( {m + 2\lambda } \right)}}{m!{\Gamma \left( {2\lambda } \right)}}}$, and that $\Scale[.9]{{\left( {\frac{x}{2}} \right)^{ -\lambda }}{{\rm J}_{2k + \lambda }}\left( {x} \right)}$ is an even function.
Now, using equation \ref{z20} to evaluate $\Scale[.9]{\langle {{e^{i n\pi x}},C_{2k}^\lambda } \rangle_w}$, equation \ref{z48} can be expressed in terms of $\Scale[.9]{{\mathcal{S}_{2k,\lambda}(\cdot;\mathfrak{p}_\nu)}}$ terms, as follows:

\begin{eqnarray}\label{z50}
\Scale[0.9]{
{\mathcal{E}^{\mathfrak{p}_\nu}} = 2\left(\nicefrac{\pi}{2}\right)^{-\lambda}\displaystyle\sum\nolimits_{k = 0}^{{M}/{2}} {(-1)^{k}{\sigma_{k,\lambda}}{\mathcal{S}_{2k,\lambda}(N;\mathfrak{p}_\nu)}}, \hspace{40pt}\sigma_{k,\lambda} = \frac{{\Gamma \left( {2k + 2\lambda } \right)\Gamma \left( \lambda  \right)\left( {2k + \lambda } \right)}}{{\left( {2k} \right)!\Gamma \left( {2\lambda } \right)}}.
}
\end{eqnarray}

Measuring the absolute value of both sides of the above equation, applying the triangle inequality, and using the estimation given by equation \ref{z42}, we get:

\begin{eqnarray}\label{z52}
\Scale[0.9]{
\left|\mathcal{E}^{\mathfrak{p}_\nu}\right|\leqslant {A_0} \left({\pi}/{2}\right)^{-\lambda}\displaystyle\sum\nolimits_{k = 0}^{{M}/{2}} {\sigma_{k,\lambda} \left\{\mathfrak{p}_\nu(N) N^{-\lambda -\epsilon} + A_{\lambda} \mathfrak{p}_\nu(0)\delta_k \right\}}. 	
}
\end{eqnarray}

Using the Stirling's approximation as $\Scale[.9]{\Gamma \left( x \right) = \left(x-1\right)! \sim {x^{x - \nicefrac{1}{2}}}{e^{-x}}}$, and separating the $\Scale[.9]{\sigma_{0,\lambda}}$ summand from the rest, equation \ref{z52} can be simplified as follows:

\begin{eqnarray}\label{z54}
\Scale[0.9]{
\begin{array}{l}
\left| \mathcal{E}^{\mathfrak{p}_\nu} \right| \leqslant A_0 {\left( {{{\pi}}/{2}} \right)^{-\lambda}} \left\{ A_{\lambda} \mathfrak{p}_\nu(0) \sigma_{0,\lambda} + \mathfrak{p}_\nu(N){N^{-\lambda -\epsilon}} e^{-\lambda}\lambda^\lambda \displaystyle\sum\nolimits_{k = 1}^{{M}/{2}} {\tilde\sigma_{k,\lambda}}\right\}, \\
{\tilde\sigma_{k,\lambda}} = \left\{k \left( {k + \lambda } \right)\right\}^{ -\nicefrac{1}{2}}\left( {2k + \lambda } \right){\left( {k + \lambda} \right)^{2k + 2\lambda}}{k^{-2k}}{\lambda ^{-2\lambda }}.
\end{array}	
}
\end{eqnarray}

In which $\Scale[.9]{\sigma_{k,\lambda}\sim e^{-\lambda}\lambda^\lambda \tilde\sigma_{k,\lambda}}$ and $\Scale[.9]{A_0}$ is a constant. Now, by defining $\Scale[.9]{\gamma = k/\lambda}$, the terms $\Scale[.9]{\tilde\sigma_{k,\lambda}}$ can be expressed as follows:

\begin{eqnarray}\label{z55}
\Scale[0.9]{
{\tilde\sigma_{k,\lambda}} = \left\{\gamma \left( 1 + \gamma \right)\right\}^{ - \nicefrac{1}{2}}\left( {1 + 2\gamma } \right){\left\{ {{{\left( {1 + \gamma } \right)}^{2 + 2\gamma }}{\gamma ^{ - 2\gamma }}} \right\}^\lambda}.	
}
\end{eqnarray}

Wherein $\Scale[.9]{{\lambda^{-1}} \leqslant \gamma \leqslant \nicefrac{1}{2}}$ .
It is not difficult to see that for sufficiently large $\Scale[.9]{\lambda}$, $\Scale[.9]{\tilde\sigma_{k,\lambda}}$ would be an increasing function with respect to $\Scale[.9]{\gamma}$ within the above boundaries. The reader may simply compare the magnitude of the first and last terms for $\Scale[.9]{{\tilde\sigma_{1,\lambda}} \sim {\lambda ^{\nicefrac{5}{2}}}}$ and $\Scale[.9]{{\tilde\sigma_{\nicefrac{M}{2},\lambda}} \sim {{\left(\nicefrac{27}{4}\right)}^\lambda}}$, for instance.
Now, by noting that $\Scale[.9]{\sigma_{0,\lambda} = \Gamma(\lambda+1)}$, replacing all the $\Scale[.9]{k}$ summands in equation \ref{z54} with $\Scale[.9]{\tilde\sigma_{\nicefrac{M}{2},\lambda}}$, and the asymptotic form of $\Scale[.9]{A_\lambda}$ as in equation \ref{z42}, the following estimate is obtained using the Gegenbauer classical parameters $\Scale[.9]{(M,\lambda)= ({N}/{4},{N}/{4})}$, for some nonzero constants $\Scale[.9]{A}$ and $\Scale[.9]{B}$:

\begin{eqnarray}\label{z57}
\Scale[0.9]{
\left| {{\mathcal{E}^{\mathfrak{p}_\nu}}} \right| \leqslant {A}\mathfrak{p}_\nu(0) + {B}\mathfrak{p}_\nu(N){N^{-\lambda-\epsilon}} {\left( {\frac{\pi e}{2}} \right)^{ -\lambda }}\lambda^\lambda(\frac{M}{2}){\left(\frac{27}{4}\right)^\lambda}
= {A}\mathfrak{p}_\nu(0) + \mathcal{O}\left(\mathfrak{p}_\nu(N) N^{-\epsilon+1} \left(\frac{8\pi e}{27}\right)^{-{N}/{4}}\right).	
}
\end{eqnarray}

On the other side, given the numerical criterion noted earlier for the asymptotic behavior of Bessel functions, with the classical Gegenbauer parameters we obtain $\Scale[.9]{N\pi/(m+\lambda)\geqslant 2\pi}$, that is large enough to utilize $\Scale[.9]{\varepsilon=\nicefrac{1}{2}}$ in our estimations. Accordingly, the above relationship may be simplified as follows:

\begin{eqnarray}\label{z58}
\Scale[0.9]{
\left| \mathcal{E}^{\mathfrak{p}_\nu} \right| = {A}\mathfrak{p}_\nu(0) + \mathcal{O}\left(\mathfrak{p}_\nu(N) N^{\nicefrac{1}{2}} e^{-0.232N}\right), \hspace{50pt}{2\mid \nu}.	
}
\end{eqnarray}

Now, we try to express $\Scale[.9]{\Phi_N(x-1)}$ in terms of $\Scale[.9]{\Delta _N \left( x;{\mathfrak{p}_\nu} \right)}$ functions. Let us begin with expanding $\Scale[.9]{\Phi_N(x-1)}$ and reformulating its Fourier coefficients $\Scale[.9]{\alpha_N}$ form equation \ref{z4}:

\begin{eqnarray}\label{z59}
\Scale[0.9]{
\Phi_N(x-1)=\displaystyle\sum\nolimits_{|n| \leqslant N} {{{\left( { - 1} \right)}^n}{\alpha _n}{e^{in\pi x}}},\hspace{50pt}
{\alpha _n} = \frac{1}{2n(N+\nicefrac{1}{2})}{\displaystyle\sum\limits_{r=1-n}^{n} {\frac{1}{1-\frac{r-\nicefrac{1}{2}}{N+\nicefrac{1}{2}}}}}.
}
\end{eqnarray}

By elaborating each summand of the above relationship as a geometric series and recombining similar terms of each expansion, we may write:

\begin{eqnarray}\label{z60}
\Scale[0.9]{
\begin{array}{l}
{\alpha _n} = \frac{1}{2n(N+\nicefrac{1}{2})}\displaystyle\sum\nolimits_{r=1-n}^{n} \displaystyle\sum\nolimits_{\ell=0}^{\infty}
\left({\frac{r-\nicefrac{1}{2}}{N+\nicefrac{1}{2}}}\right)^\ell = \displaystyle\sum\nolimits_{\ell=0}^{\infty}{\frac{\mathfrak{u}_{2\ell}(n)}{(N+\nicefrac{1}{2})^{2\ell+1}}}.
\end{array}
}
\end{eqnarray}

In which $\Scale[.9]{\mathfrak{u}_{2\ell}(\cdot)}$ is defined as follows, with $\Scale[.9]{B_{2\ell+1}(\cdot)}$ denoting the $\Scale[.9]{2\ell+1}$-th Bernoulli polynomial:

\begin{eqnarray}\label{z61}
\Scale[0.9]{
\mathfrak{u}_{2\ell}(n)=\frac{1}{n}\displaystyle\sum\nolimits_{r=1}^{n} {(r-\nicefrac{1}{2})^{2\ell}} =
\frac{B_{2\ell+1}(n+\nicefrac{1}{2})}{n(2\ell+1)}.	
}
\end{eqnarray}

A generalized form of this identity can be found in \cite{olver2010nist}. The reader should note that $\Scale[.9]{\mathfrak{u}_{0}(n)=1}$. Now, since $\Scale[.9]{x=0}$ is a root of the equation $\Scale[.9]{B_{2k+1}(x+\nicefrac{1}{2})=0}$ for $\Scale[.9]{\ell>0}$, $\Scale[.9]{n^{-1}B_{2\ell+1}(n+\nicefrac{1}{2})}$ and consequently $\Scale[.9]{\mathfrak{u}_{2\ell}(n)}$ are also polynomials of $\Scale[.9]{n}$. Furthermore, from equation \ref{z61}, it follows trivially that $\Scale[.9]{\mathfrak{u}_{2\ell}(n)}$ is an increasing function over positive integers.
On the other side, since Bernoulli polynomials satisfy $\Scale[.9]{B_{m}(1-x)=(-1)^m B_{m}(x)}$ for any real $\Scale[.9]{x}$, it can be easily found that $\Scale[.9]{n^{-1}B_{2\ell+1}(n+\nicefrac{1}{2})}$ is an even polynomial in terms of $\Scale[.9]{n}$, which means that the polynomial representation of $\Scale[.9]{\mathfrak{u}_{2\ell}(n)}$ is also comprised of even-order terms only, and that $\Scale[.9]{\mathfrak{u}_{2\ell}\in \mathbb{P}^\ast}$.
Eventually, we would also need to find $\Scale[.9]{\mathfrak{u}_{2\ell}(0)}$ as the following limit, which is easily calculated using L'Hospital's rule:

\begin{eqnarray}\label{z62}
\Scale[0.9]{
\mathfrak{u}_{2\ell}(0)=\lim\limits_{x \to 0}{\left\{(2\ell+1)x\right\}^{-1}B_{2\ell+1}(x+\nicefrac{1}{2})}=B_{2\ell}(\nicefrac{1}{2})=(2^{1-2\ell}-1)B_{2\ell}.	
}
\end{eqnarray}

In which $\Scale[.9]{B_{2\ell}}$ denotes the $\Scale[.9]{2\ell}$-th Bernoulli number. Now, let us rewrite equation \ref{z60} for the Fourier coefficients as follows, in which, $\Scale[.9]{\mathfrak{q}_{2L}}$ is clearly a polynomial itself:

\begin{eqnarray}\label{z63}
\Scale[0.9]{
{\alpha _n} = \lim\limits_{L\to \infty}{\mathfrak{q}_{2L}(n)},\hspace{50pt}
\mathfrak{q}_{2L}(n)=\displaystyle\sum\nolimits_{\ell=0}^{L} (N+\nicefrac{1}{2})^{-2\ell-1} \mathfrak{u}_{2\ell}(n).
}
\end{eqnarray}

Consequently by definition (from equations \ref{z59} and \ref{z43}):

\begin{eqnarray}\label{z64}
\Scale[0.9]{
{\Phi_N}(x-1) =
\lim\limits_{L\to \infty}\displaystyle\sum\nolimits_{\ell=0}^{L} {(N+\nicefrac{1}{2})^{-2\ell-1}}{\Delta_N(x;\mathfrak{u}_{2\ell})}.
}
\end{eqnarray}

Projecting both sides of the above equation on the Gegenbauer space:

\begin{eqnarray}\label{z65}
\Scale[0.9]{
{\Psi_N}(x) = \lim\limits_{L\to \infty}\displaystyle\sum\nolimits_{\ell=0}^{L} {(N+\nicefrac{1}{2})^{-2\ell-1}}{\Psi(x;\mathfrak{u}_{2\ell})}.
}
\end{eqnarray}

By evaluating the $\Scale[.9]{\mathcal{L}^\infty}$-norm from both sides, applying the triangle inequality, and using equations \ref{z45} and \ref{z58}, we get:

\begin{eqnarray}\label{z66}
\Scale[0.9]{
\mathcal{E} \leqslant \lim\limits_{L\to \infty}\displaystyle\sum\nolimits_{\ell=0}^{L} {(N+\nicefrac{1}{2})^{-2\ell-1}}\left| \mathcal{E}^{\mathfrak{u}_{2\ell}} \right| \leqslant \mathcal{E}_1 + \mathcal{E}_2.
}
\end{eqnarray}

In which, for some constant $\Scale[.9]{A}$:

\begin{eqnarray}\label{z67}
\Scale[0.9]{
\begin{array}{l}
\mathcal{E}_1 = A \lim\limits_{L\to\infty} \displaystyle\sum\nolimits_{\ell=0}^{L} {(N+\nicefrac{1}{2})^{-2\ell-1}\mathfrak{u}_{2\ell}(0)}, \\
\mathcal{E}_2 =
\lim\limits_{L\to\infty}\displaystyle\sum\nolimits_{\ell=0}^{L}{(N+\nicefrac{1}{2})^{-2\ell-1} \mathcal{O}\left( \mathfrak{u}_{2\ell}(N) N^{\nicefrac{1}{2}} e^{-0.232N} \right)}.
\end{array}
}
\end{eqnarray}

Note that for a sequence of functions $\Scale[.9]{f_i:\mathbb{N}\rightarrow \mathbb{R}^{+}}$, the relationship $\Scale[.9]{\sum_{i}^{\infty}{\mathcal{O}(f_i)} = \mathcal{O}(\sum_{i}^{\infty}{f_i})}$ holds provided that $\Scale[.9]{\sup_{i} \|\mathcal{O}(f_i)/ f_i\|_\infty}$ is finite. As a result, since $\Scale[.9]{\lim\nolimits_{L\to \infty}\mathfrak{q}_{2L}(N)=\alpha_N\sim \ln(N)/N}$, we may express $\Scale[.9]{\mathcal{E}_2}$ as follows through collapsing the summation into the parentheses:

\begin{eqnarray}\label{z68}
\Scale[0.9]{
\mathcal{E}_2 =	\lim\limits_{L \to \infty} \mathcal{O}\left(\mathfrak{q}_{2L}(N)N^{\nicefrac{1}{2}} e^{-0.232N} \right) = \mathcal{O}\left(  \alpha_N N^{\nicefrac{1}{2}} e^{-0.232N} \right)
= \mathcal{O}\left(\ln(N) N^{-\nicefrac{1}{2}} e^{-0.232N} \right).
}
\end{eqnarray}

On the other side:

\begin{eqnarray}\label{z69}
\Scale[0.9]{
\mathcal{E}_1 = A\lim\limits_{L \to \infty} {\mathfrak{q}_{2L}(0)} \sim \displaystyle\sum\nolimits_{\ell=0}^{\infty}{\frac{\mathfrak{u}_{2\ell}(0)}{(N+\nicefrac{1}{2})^{2\ell+1}}} = \displaystyle\sum\nolimits_{\ell=0}^{\infty}{\frac{(2^{1-2\ell}-1)B_{2\ell}}{(N+\nicefrac{1}{2})^{2\ell+1}}}.
}
\end{eqnarray}

We need to show the above series is finite (as Bernoulli numbers grow faster than exponentially), and find the rate at which its value decreases with $\Scale[.9]{N}$. To do so, note the following asymptotic expansion of the trigamma function $\Scale[.9]{\psi'(z) = \frac{d^2}{dz^2} \ln \Gamma(z)}$, as a Laurent series:

\begin{eqnarray}\label{z70}
\Scale[0.9]{
\psi'(z)\sim \frac{1}{z} + \frac{1}{2z^2}+ \displaystyle\sum\nolimits_{\ell=1}^{\infty}{\frac{B_{2\ell}}{z^{2\ell+1}}},\hspace{50pt}z \to \infty.
}
\end{eqnarray}

The trigamma function is decreasing for positive arguments such that for large arguments $\Scale[.9]{\psi'(z)\to 0}$, and its descent rate is dominated by the first-order term $\Scale[.9]{z^{-1}}$. Accordingly, the series appearing in equation \ref{z69} can be asymptotically expressed in terms of trigamma functions as follows:

\begin{eqnarray}\label{z71}
\Scale[0.9]{
\mathcal{E}_1 \sim 4\psi'(2N+1) - \psi'(N+\nicefrac{1}{2}),\hspace{50pt}N\to\infty.
}
\end{eqnarray}

On the other side, the following identity holds for all the polygamma functions \cite{olver2010nist}:

\begin{eqnarray}\label{z72}
\Scale[0.9]{
\psi'(2x) =\frac{1}{4}\{ \psi'(x) + \psi'(x+\nicefrac{1}{2})\}.
}
\end{eqnarray}

By replacing $\Scale[.9]{x = N+\nicefrac{1}{2}}$ in the above equation, and using equations \ref{z66}, \ref{z68}, and \ref{z71}, it follows that for large enough $\Scale[.9]{N}$, the following result holds:

\begin{eqnarray}\label{z73}
\Scale[0.9]{
\mathcal{E} \leqslant A \psi'(N+1) + \mathcal{O}\left(\ln(N) N^{-\nicefrac{1}{2}} e^{-0.232N} \right) = \mathcal{O}\left(N^{-1} \right).
}
\end{eqnarray}

This result concludes the proof of the first proposition in Theorem 4.1.
For the second proposition, let us point out an important implication of our derivations. As we observed, the reconstruction error of the shifted Li's function incorporates diminishing terms of the order $\Scale[.9]{\mathcal{O}(N^{-k})}$ and $\Scale[.9]{\mathcal{O}(e^{-cN})}$. Notice that the former terms arise from the appearance of the constant part of $\Scale[.9]{\mathfrak{u}_{2\ell}}$ polynomials in $\Scale[.9]{\mathcal{E}_1}$. On the other side, according to equation \ref{z52}, the constant parts of the polynomials are sustained only if the first Gegenbauer polynomial $\Scale[.9]{C_0^\lambda(x)}$ exists among the reprojection bases. In other words, by omitting the first Gegenbauer polynomial from the reconstruction space, only the $\Scale[.9]{\mathcal{E}_2}$ term will appear in $\Scale[.9]{\mathcal{E}}$, and hence:

\begin{eqnarray}\label{z74}
\Scale[.9]{
\mathcal{E}^\star = \left\|\Psi_N^\star(x)\right\|_\infty = \mathcal{O}\left(\ln(N) N^{-\nicefrac{1}{2}} e^{-0.232N} \right).
}
\end{eqnarray}

Which proves the second part of Theorem 4.1. and concludes it.
$\blacksquare$

\textbf{Proof of Lemma 3.1.} The relationship \ref{z31} will be tackled with for $\Scale[.9]{m=0}$, $\Scale[.9]{m=2p>0}$, and $\Scale[.9]{m=2p+1}$ separately, and the results will eventually be combined together.
According to Poisson's integral \cite{olver2010nist}, the following property holds for $\Scale[.9]{w_\alpha(t) = {({1 - {t ^ 2}})^{\alpha - \nicefrac{1}{2}}}}$ with $\Scale[.9]{{\Re}\{\alpha\}>-\nicefrac{1}{2}}$:

\begin{eqnarray}\label{z75}
\Scale[0.9]{
{\rm J}_{\alpha}(z)= 2{K_\alpha}\left(\frac{z}{\pi}\right)^\alpha \displaystyle\int\nolimits_{0}^{1}{\cos(zt)w_\alpha(t) dt},\hspace{40pt}K_\alpha = {\pi^{-\nicefrac{1}{2}} {\left({\pi}/{2}\right)^\alpha}\Gamma\left(\alpha+\nicefrac{1}{2}\right)^{-1}}.
}
\end{eqnarray}

Now, if we define:

\begin{eqnarray}\label{z77}
\Scale[0.9]{
\mathcal{Q}_{m,n}(t)=\displaystyle\sum\nolimits_{\ell=l}^{n}{(-1)^\ell \ell^m}{\cos\left(\ell \pi t\right)}.
}
\end{eqnarray}

Then, equation \ref{z75} may be used to express the summation of equation \ref{z30} as an integral, with $\Scale[.9]{K_{m+\lambda}}$ as defined in equation \ref{z75}:

\begin{eqnarray}\label{z78}
\Scale[0.9]{
\mathcal{S}_{m,\lambda}(n)= \displaystyle\sum\nolimits_{\ell=1}^{n}{(-1)^{\ell} {\ell}^{-\lambda}{\rm J}_{m+\lambda}\left(\ell\pi\right)}
= 2 K_{m+\lambda} \displaystyle\int\nolimits_{0}^{1}{\mathcal{Q}_{m,n}(t)w_{m+\lambda}(t) dt}.
}
\end{eqnarray}

First of all, note that due to Lagrange's trigonometric identity, $\Scale[.9]{\mathcal{Q}_{0,n}(t)}$ constitutes of two terms, as follows:

\begin{eqnarray}\label{z79}
\Scale[0.9]{
\mathcal{Q}_{0,n}(t) = -\frac{1}{2}+\tilde{\mathcal{Q}}_{0,n}(t),\hspace{50pt}
\tilde{\mathcal{Q}}_{0,n}(t) = \frac{(-1)^n\cos\left((n+\nicefrac{1}{2})\pi t\right)}{2\cos(\pi t/2)}.
}
\end{eqnarray}

By replacing the above relationship in equation \ref{z78} and simplifying the integral for each constituent, $\Scale[.9]{\mathcal{S}_{0,\lambda}(n)}$ can be expressed as follows:

\begin{eqnarray}\label{z80}
\Scale[0.9]{
\mathcal{S}_{0,\lambda}(n) = -\frac{\left({\pi}/{2}\right)^\lambda}{2\Gamma\left(\lambda+1\right)} + {(-1)^n K_\lambda} \displaystyle\int\nolimits_{0}^{1}{f(t)\cos\left(\kappa t\right) dt}.
}
\end{eqnarray}

Wherein $\Scale[.9]{\kappa=(n+\nicefrac{1}{2})\pi}$ and $\Scale[.9]{f(t)= \sec(\pi t/2)w_\lambda(t)}$ is a nonnegative function. Note that $\Scale[.9]{\sec(\pi t/2)}$ is singular at $\Scale[.9]{t=1}$ and behaves as $\Scale[.9]{\sim(1-t)^{-1}}$ at the vicinity of the singularity. Besides, the weight function $\Scale[.9]{w_\lambda(\cdot)}$ behaves as $\Scale[.9]{\sim(1-t)^{\lambda-\nicefrac{1}{2}}}$. Now, since we had assumed $\Scale[.9]{\lambda>\nicefrac{1}{2}}$, the limiting behavior of $\Scale[.9]{f(t)}$ would be as $\Scale[.9]{\sim(1-t)^{\epsilon-1}}$ for a positive $\Scale[.9]{\epsilon}$, and hence the function would be $\Scale[.9]{\mathcal{L}_1}$-integrable over the interval. On the other hand, the latter statement would be a sufficient condition for the Riemann-Lebesgue lemma to hold for the integral term of the above equation \cite{olver2010nist}. As a result, the integral tends to zero as $\Scale[.9]{\kappa}$ approaches infinity and:

\begin{eqnarray}\label{z81}
\Scale[0.9]{
\lim \limits_{n\to \infty} \mathcal{S}_{0,\lambda}(n) = -\frac{\left({\pi}/{2}\right)^\lambda}{2\Gamma\left(\lambda+1\right)} \neq 0.
}
\end{eqnarray}

Having concluded the case $\Scale[.9]{m=0}$, we continue by considering the case of positive even $\Scale[.9]{m}$. Notice that for $\Scale[.9]{m>0}$, the function $\Scale[.9]{\mathcal{Q}_{m,n}(t)}$ of equation \ref{z77}, may be obtained as a coefficient of the $\Scale[.9]{m}$-th derivative of $\Scale[.9]{\mathcal{Q}_{0,n}(t)}$ or $\Scale[.9]{\tilde{\mathcal{Q}}_{0,n}(t)}$:

\begin{eqnarray}\label{z82}
\Scale[0.9]{
\begin{array}{l}
(-1)^{\nicefrac{m}{2}}{\pi^{m}}\mathcal{Q}_{m,n}(t)= \mathcal{Q}_{0,n}^{(m)}(t)= \tilde{\mathcal{Q}}_{0,n}^{(m)}(t).
\end{array}
}
\end{eqnarray}

By replacing this equivalent form in equation \ref{z78}, we get:

\begin{eqnarray}\label{z83}
\Scale[0.9]{
\mathcal{S}_{m,\lambda}(n)= 2(-1)^{\nicefrac{m}{2}} {\pi^{-m}} K_{m+\lambda} \displaystyle\int\nolimits_{0}^{1}{\mathcal{Q}_{0,n}^{(m)}(t) w_{m+\lambda}(t) dt}.
}
\end{eqnarray}

Here, we use induction to prove that for any positive even number $\Scale[.9]{m}$, the following relationship holds, in which $\Scale[.9]{\mu>m+\nicefrac{1}{2}}$ is a constant:

\begin{eqnarray}\label{z84}
\Scale[0.9]{
\lim\limits_{n \to \infty}\displaystyle\int\nolimits_{0}^{1}{\mathcal{Q}_{0,n}^{(m)}(t)w_{\mu}(t) dt}=0,\hspace{50pt}\mu>m+\nicefrac{1}{2}.
}
\end{eqnarray}

Using integration by parts two consecutive times, we get:

\begin{eqnarray}\label{z85}
\Scale[0.9]{
\displaystyle\int\nolimits_{0}^{1}{\mathcal{Q}_{0,n}^{(m)}w_{\mu} dt}=
\left\{ \mathcal{Q}_{0,n}^{(m-1)} w_{\mu} - \mathcal{Q}_{0,n}^{(m-2)} w'_{\mu} \right\}_{0}^{1}
+\displaystyle\int\nolimits_{0}^{1}{\mathcal{Q}_{0,n}^{(m-2)}w''_{\mu} dt}.
}
\end{eqnarray}

Since $\Scale[.9]{m \geqslant 2}$, it follows that $\Scale[.9]{\mu > \nicefrac{5}{2}}$ and hence, $\Scale[.9]{w_\mu(1)=w'_\mu(1)=w'_\mu(0)=0}$. On the other side, since $\Scale[.9]{\mathcal{Q}_{0,n}^{(m-1)}}$ is an odd order derivative of a sum of $\Scale[.9]{\cos(\ell \pi t)}$-terms, it will only comprise $\Scale[.9]{\sin(\ell \pi t)}$-terms, meaning that $\Scale[.9]{\mathcal{Q}_{0,n}^{(m-1)}(0)=0}$. As a result of this argument, the content of the braces in the above equation simply vanishes. Now, applying two times integration by parts on the remaining right-hand side integral in equation \ref{z85}, we obtain:

\begin{eqnarray}\label{z86}
\Scale[0.9]{
\displaystyle\int\nolimits_{0}^{1}{\mathcal{Q}_{0,n}^{(m)}w_{\mu} dt}=
4\mu(\mu-1)\displaystyle\int\nolimits_{0}^{1}{\mathcal{Q}_{0,n}^{(m-2)}w_{\mu-2} dt}
-2\mu(2\mu-1)\displaystyle\int\nolimits_{0}^{1}{\mathcal{Q}_{0,n}^{(m-2)}w_{\mu-1} dt}.
}
\end{eqnarray}

Hypothesizing equation \ref{z84} holds with any pair $\Scale[.9]{(m',\mu')}$ that $\Scale[.9]{m'<m}$ is a positive even number and $\Scale[.9]{\mu'>m'+\nicefrac{1}{2}}$, we need to prove that a similar relationship holds for the pair $\Scale[.9]{(m,\mu)}$ with the specified conditions. However, this argument is quite clear according to the above equation, as both the right-hand side integrals of equation \ref{z86} satisfy the induction hypothesis, and hence the induction step holds.
It only remains to verify the induction base case for $\Scale[.9]{m=2}$. Let us go back to equation \ref{z85} and expand the right-hand side integral using equation \ref{z79}:

\begin{eqnarray}\label{z87}
\Scale[0.9]{
\displaystyle\int\nolimits_{0}^{1}{\mathcal{Q}_{0,n}^{(2)}w_{\mu} dt}=
\displaystyle\int\nolimits_{0}^{1}{\mathcal{Q}_{0,n}w''_{\mu} dt}
= -\frac{1}{2}\left\{w'_{\mu}(t)\right\}_{0}^{1} + \frac{(-1)^n}{2}\displaystyle\int\nolimits_{0}^{1}{f(t)\cos\left(\kappa t\right) dt}.
}
\end{eqnarray}

Wherein $\Scale[.9]{\kappa = (n+\nicefrac{1}{2})\pi}$ and $\Scale[.9]{f(t)= \sec(\pi t/2)w''_{\mu}(t)}$. The content of the braces clearly equals to zero. Now, note that $\Scale[.9]{\mu>\nicefrac{5}{2}}$ and $\Scale[.9]{w''_{\mu} = A w_{\mu-2}- B w_{\mu-1}}$ for some constant $\Scale[.9]{A}$ and $\Scale[.9]{B}$. Accordingly, it would be straightforward to tailor the argument we presented under equation \ref{z80}, and show the right-hand side integral also satisfies the Riemann-Lebesgue lemma prerequisites and consequently tends to zero. The latter step completes the induction and concludes the following result:

\begin{eqnarray}\label{z88}
\Scale[0.9]{
\lim \limits_{n\to \infty} \mathcal{S}_{m,\lambda}(n) = 0,\hspace{50pt}{m>0}\hspace{5pt}{\rm and}\hspace{5pt} 2\mid m.
}
\end{eqnarray}

Eventually, let us consider a last case where $\Scale[.9]{m}$ is and odd number. We start by defining the following series:

\begin{eqnarray}\label{z89}
\Scale[0.9]{
\mathcal{P}_{m,n}(t)=\displaystyle\sum\nolimits_{\ell=l}^{n}{(-1)^\ell \ell^m}{\sin\left(\ell \pi t\right)}.
}
\end{eqnarray}

Another Lagrange's trigonometric identity provides an explicit relationship for $\Scale[.9]{\mathcal{P}_{0,n}}$, as follows:

\begin{eqnarray}\label{z90}
\Scale[0.9]{
\mathcal{P}_{0,n}(t) = -\frac{1}{2}\tan({\pi t}/{2}) +  \mathcal{R}_{n}(t),\hspace{50pt}
\mathcal{R}_{n}(t) = \frac{(-1)^n\sin\left((n+\nicefrac{1}{2})\pi t\right)}{2\cos(\pi t/2)}.
}
\end{eqnarray}

Here, a similar relationship to equation \ref{z82}, may be expressed as below for an odd $\Scale[.9]{m}$:

\begin{eqnarray}\label{z91}
\Scale[0.9]{
(-1)^{\frac{m-1}{2}}{\pi^{m}}\mathcal{Q}_{m,n} = \mathcal{P}_{0,n}^{(m)}=
-\frac{1}{2}\left\{{\tan({\pi t}/{2})}\right\}^{(m)} + \mathcal{R}_{n}^{(m)}.
}
\end{eqnarray}

Again, we use induction to prove the following feature, provided that $\Scale[.9]{m}$ is an odd number and $\Scale[.9]{\mu>m+\nicefrac{1}{2}}$ a constant:

\begin{eqnarray}\label{z92}
\Scale[0.9]{
\lim\limits_{n \to \infty}\displaystyle\int\nolimits_{0}^{1}{\mathcal{R}_{n}^{(m)}(t)w_{\mu}(t) dt}=0,\hspace{50pt}\mu>m+\nicefrac{1}{2}.
}
\end{eqnarray}

Let us begin by considering the base case $\Scale[.9]{m=1}$. Using integration by parts, we get:

\begin{eqnarray}\label{z93}
\Scale[0.9]{
\displaystyle\int\nolimits_{0}^{1}{\mathcal{R}_{n}^{(1)}(t)w_{\mu}(t) dt}=
\left\{\mathcal{R}_{n}(t)w_{\mu}(t)\right\}_{0}^{1}
-(-1)^{n}(\mu-\nicefrac{1}{2})\displaystyle\int\nolimits_{0}^{1}{f(t)\sin(\kappa t) dt}.
}
\end{eqnarray}

In which $\Scale[.9]{\kappa = (n+\nicefrac{1}{2})\pi}$ and $\Scale[.9]{f(t)= t\sec(\pi t/2)w_{\mu-1}(t)}$. We know $\Scale[.9]{\mathcal{R}_{n}(0)=0}$ and $\Scale[.9]{w_{\mu}(1)=0}$, hence, the content of the braces vanishes.
As for the right-hand side integral of above equation, since we assumed $\Scale[.9]{\mu-1>\nicefrac{1}{2}}$, a similar argument to the previous one shows that $\Scale[.9]{f(t)}$ is absolutely integrable over the interval (a sufficient condition for the Riemann-Lebesgue lemma) and consequently its value tends to zero as $\Scale[.9]{n\to\infty}$. This result affirms the induction base case.
Now, assuming $\Scale[.9]{m\geqslant 3}$, we perform two times integration by parts on the integral of equation \ref{z92} to reach:

\begin{eqnarray}\label{z94}
\Scale[0.9]{
\displaystyle\int\nolimits_{0}^{1}{\mathcal{R}_{n}^{(m)}w_{\mu} dt}=
\left\{ \mathcal{R}_{n}^{(m-1)} w_{\mu} - \mathcal{R}_{n}^{(m-2)} w'_{\mu} \right\}_{0}^{1}
+\displaystyle\int\nolimits_{0}^{1}{\mathcal{R}_{n}^{(m-2)}w''_{\mu} dt}.
}
\end{eqnarray}

One easily finds that $\Scale[.9]{w_\mu(1)=w'_\mu(1)=w'_\mu(0)=0}$. Moreover, since $\Scale[.9]{m-1}$ is a positive even number, it is easy to check from equation \ref{z90} that $\Scale[.9]{\mathcal{R}_{n}^{(m-1)}(t)}$ can be expressed as a finite sum of $\Scale[.9]{\sin(\ell\pi t/2)}$-terms and the $\Scale[.9]{(m-1)}$-th derivative of a $\Scale[.9]{\tan(\pi t/2)}$-term. The former terms clearly equal to zero at $\Scale[.9]{t=0}$. As for the latter, note that in the Taylor series $\Scale[.9]{\tan(x)=x+x^3/3+2x^5/12+\cdots}$ with $\Scale[.9]{|x|<\pi/2}$, all the coefficients are positive numbers.
Since only odd orders appear in this series, all the even order derivatives at zero must equal to zero. This point concludes that $\Scale[.9]{\mathcal{R}_{n}^{(m-1)}(0)=0}$ and hence, the content of the braces in equation \ref{z94} also equals to zero. Now, we perform another two times of integration by part on the right-hand side integral of equation \ref{z94} and simplify it:

\begin{eqnarray}\label{z96}
\Scale[0.9]{
\displaystyle\int\nolimits_{0}^{1}{\mathcal{R}_{n}^{(m)}w_{\mu} dt}=
4\mu(\mu-1)\displaystyle\int\nolimits_{0}^{1}{\mathcal{R}_{n}^{(m-2)}w_{\mu-2} dt}
-2\mu(2\mu-1)\displaystyle\int\nolimits_{0}^{1}{\mathcal{R}_{n}^{(m-2)}w_{\mu-1} dt}.
}
\end{eqnarray}

Having obtained a similar relationship to equation \ref{z86}, a quite similar argument easily proves the induction step, concluding the relationship \ref{z92}. Accordingly, replacing the representation \ref{z91} in the integral form of \ref{z78} leads to the following:

\begin{eqnarray}\label{z97}
\Scale[0.9]{
(-1)^{\frac{m+1}{2}}{\pi^{m}}\mathcal{S}_{m,\lambda}(n)= {K_{m+\lambda}}\displaystyle\int\nolimits_{0}^{1}{\tan\left({\pi t}/{2}\right)^{(m)} w_{m+\lambda}(t) dt} + o(1).
}
\end{eqnarray}

Wherein $\Scale[.9]{\lim_{n\to\infty} o(1)=0}$. It is clear that the integral appearing in the right-hand side of the above equation is independent from $\Scale[.9]{n}$. Hence it only remains to show the value of this integral is nonzero, which is quite clear according to the Taylor series of the tangent function. In fact, the expansion of $\Scale[.9]{\tan\left({\pi t}/{2}\right)}$ comprises only positive coefficients for $\Scale[.9]{t\in (0,1)}$, which means that any derivative of the function is positive over the integration interval $\Scale[.9]{(0,1)}$. Furthermore, the weight function $\Scale[.9]{w_{m+\lambda}(t)}$ is also positive over $\Scale[.9]{(0,1)}$. Consequently, the product of two positive functions has a positive integral, proving the following result:

\begin{eqnarray}\label{z98}
\Scale[0.9]{
\lim \limits_{n\to \infty} \mathcal{S}_{m,\lambda}(n) \neq 0,\hspace{50pt}{m>0}\hspace{5pt}{\rm and}\hspace{5pt} 2\nmid m.
}
\end{eqnarray}

Now, combining the results of equations \ref{z81}, \ref{z88}, and \ref{z98}, concludes the relationship \ref{z31} of Lemma 4.1. $\blacksquare$

\subsection{Incompetence of Other Methods}

Let us at the end of this section, concisely discuss the ineptitude of other spectral reconstructions against TCE. These methods predominantly make use of filtering, discontinuous functions, or most notably miscellaneous inverse techniques.
Having interpreted TCE as the emergence of the Li's function or the appearance of a point discontinuity, it would be to a substantial extent straightforward to explain analytically why the rest of methods cannot resolve this phenomenon, at least as effectively as the spectral reprojection.

The inefficacy of filters, including the Vandeven filter, is quite easy to show. It suffices to notice in the following equation, the value of $\Scale[.9]{\lim_{N\to\infty}\mathcal{V}\Phi_{N}(1)}$ does not tend to zero at all, which means $\Scale[.9]{\|\mathcal{V}\Phi_{N}\|_{\infty}=\mathcal{O}(1)}$:

\begin{eqnarray}\label{n1}
\Scale[0.9]{
\mathcal{V}\Phi_N (1) = 2\displaystyle\sum\nolimits_{0 < n \leqslant N} {\sigma \left({n}/{N}\right){\hat \alpha_n}} \geqslant 2\displaystyle\sum\nolimits_{0 < n \leqslant N} {\frac{\sigma \left({n}/{N}\right)}{N}} \approx 2\displaystyle\int\nolimits_{0}^{1} \sigma(t)dt > 0.
}
\end{eqnarray}

As for the singular Fourier-Pad\'e method, application of $\Scale[.9]{\Phi_{N}(x)}$ as the input function results in $\Scale[.9]{f^{+}=f^{-}}$ and $\Scale[.9]{\rho = 1}$. Accordingly, the rational functions $\Scale[.9]{g_{0,1}}$ are to be found satisfying $\Scale[0.9]{
{f^+ }\left( z \right) = g_0 \left( z \right) + g_1 \left( z \right)\ln \left( {1 - z} \right)}$ as best as possible. On the other side however, computing both sides of this equation at $\Scale[.9]{x=0^+}$ and $\Scale[.9]{x=0^-}$ and subtracting, one obtains:

\begin{eqnarray}\label{n2}
\Scale[0.9]{
{f^+ }\left( 1+ 0^{+}i \right) - {f^+ }\left( 1+ 0^{-}i \right) = g_1 \left( 1 \right)\left\{\ln \left( 0^+ i \right)- \ln \left( 0^- i \right)\right\} = g_1(1)\pi i.
}
\end{eqnarray}

Note that $\Scale[.9]{g_{0,1}}$ should be analytic at $\Scale[.9]{z=1}$, not to mention $\Scale[.9]{g_{1}(1)\neq 0}$, otherwise the singularity is removed from the method. Hence, one may conclude that $\Scale[.9]{f^{+}(1+0^+ i)}$ and $\Scale[.9]{f^{+}(1+0^- i)}$ should be unequal. However, that does not happen for the input function $\Scale[.9]{\Phi_N}$, as it produces a point discontinuity while complex logarithms describe a jump discontinuity.
For the sake of clarity, one may use the approximation $\Scale[.9]{\alpha_n \approx 1/N}$ to find that the function $\Scale[.9]{f^{+}=z(1-z^N)/N(1-z)}$ gives the same value for both $\Scale[.9]{z = 1 + 0^\pm i}$.

As for the inverse techniques which characterize one of the most significant categories of post-processing methods, one should initially note that they are primarily data-fitting methods that, unlike the Gegenbauer method, do not exploit the structure of Fourier data. In essence, all these methods end in a linear equation of the type $\Scale[.9]{\bm{A}\bm{v} = \bm{U}'\hat{\bm{f}}}$ in which, $'$ denotes the conjugate transpose, $\Scale[.9]{\hat{\bm{f}}}$ is the vector of Fourier coefficients, $\Scale[.9]{\bm{v}}$ is the vector of coefficients in the reconstruction space, $\Scale[.9]{\bm{U}}$ is a matrix where $\Scale[.9]{[\bm{U}]_{j,k}=\langle \phi_k, \psi_j \rangle}$, and $\Scale[.9]{\bm{A} = \bm{U}\bm{U}'}$. In such a problem, an error of the magnitude $\Scale[.9]{{\varepsilon}}$ in the inputs, i.e. Fourier coefficients, yields an error of magnitude roughly $\Scale[.9]{\kappa(\bm{A}){\varepsilon}}$ in the output entries of $\Scale[.9]{\bm{v}}$, in which $\Scale[.9]{\kappa(\cdot)}$ denotes the condition number of a matrix. It is proved that $\Scale[.9]{\kappa(\bm{A})}$ is of the same order as $\Scale[.9]{\kappa(\tilde{\bm{A}})}$ with $\Scale[.9]{[\tilde{\bm{A}}]_{j,k} = \langle \phi_j, \phi_k \rangle}$ denoting the Gram matrix of the reconstruction bases \cite{Adcock2012}. Moreover, it is shown that for a general Gegenbauer least squares reconstruction of the parameter $\Scale[.9]{\lambda}$ (not to be mixed up with spectral reprojection via Gegenbauer polynomials) $\Scale[.9]{\kappa(\tilde{\bm{A}}) = \mathcal{O}(M^{|2\lambda-1|})}$ in which $\Scale[.9]{M}$ is reconstruction space dimension \cite{Adcock2012}. For the Fourier extension approach the latter condition number could be exponentially large, due to the natural redundancy of the frames \cite{Huybrechs2010}. As a result, in the most stable case of Legendre polynomial least squares reconstruction, the condition number satisfies $\Scale[.9]{\kappa(\bm{A})=\mathcal{O}(1)}$, which means at best, the error magnitude at the output is of the same order as the input error.
Now, note that TCE can be approximated as an additive error of the form $\Scale[.9]{\Phi_N(x)}$. Hence, using an inverse spectral reconstruction scheme, a vector comprising the coefficients $\Scale[.9]{\alpha_n \sim 1/N}$ can be considered as the input error of the method with the magnitude
$\Scale[.9]{\varepsilon = (\sum_{|n|\leqslant N} \alpha_n^2)^{\nicefrac{1}{2}} = \mathcal{O}(N^{-\nicefrac{1}{2}})}$.
Accordingly, if we denote the error terms in the output vector $\Scale[.9]{\bm{v}}$ with $\Scale[.9]{\beta_k}$, then the magnitude of the error in the output vector coefficients will be of similar order, i.e. $\Scale[.9]{(\sum_{k=1}^{M}{\beta_k^2})^{\nicefrac{1}{2}} = \mathcal{O}(\varepsilon)}$. Given the fact that $\Scale[.9]{\|\phi_k\|_2\|w^{\nicefrac{1}{2}}\|_2 \geqslant \|\phi_k\|_w = 1}$ and $\Scale[.9]{\|w^{\nicefrac{1}{2}}\|_2= \mathcal{O}(1)}$, one may deduce the following result, in which $\Scale[.9]{\mathcal{R}}$ denotes the polynomial least squares operator:

\begin{eqnarray}\label{n3}
\Scale[0.9]{
\left\|\mathcal{R}{\Phi_N} \right\|_\infty \geqslant \left\|\mathcal{R}{\Phi_N} \right\|_2 = \left(\displaystyle\sum\nolimits_{k=1}^{M}{\beta_k^2}\|\phi_k\|_2^2 \right)^{\nicefrac{1}{2}}
\geqslant \|w^{1/2}\|_2^{-1}\left(\displaystyle\sum\nolimits_{k=1}^{M}{\beta_k^2}\right)^{\nicefrac{1}{2}} = \mathcal{O}\left(N^{-\nicefrac{1}{2}}\right).
}
\end{eqnarray}

The obtained result, though does not prove the total inefficacy of inverse methods against TCE, still indicates they cannot surpass the performance of the Gegenbauer method.

\section{Case Study: TCE in Grating Modes}
\label{Sec: Grat}

In this section, we show how TCE manifests and how it might be eliminated in the eigenmodes of a grating, which constitute the building blocks for the Fourier modal method and derivation of far-field parameters.

\subsection{Emergence of TCE in Eigenproblems}
\label{Sub: Wave Eq}

Before explaining how TCE emerges in the eigenmodes of a grating medium, let us consider it in a simple scalar periodic eigenvalue equation of the form $\Scale[.9]{(\mathcal{D}+\epsilon)f=\gamma f}$, where $\Scale[.9]{\mathcal{D}}$ is a differential operator, $\Scale[.9]{\epsilon}$ is a periodic partially smooth function, $\Scale[.9]{f}$ is an eigenfunction, and $\Scale[.9]{\gamma}$ denotes the corresponding eigenvalue. We assume the only discontinuity of $\Scale[.9]{\epsilon(x)}$ over the whole interval to be at $\Scale[.9]{x=0}$. The spectral form of the equation can be expressed as $\Scale[.9]{\left(\mathcal{D}f\right)_N + \left(\epsilon f \right)_N = \gamma f_N}$, however, it is not manageable in this form. Using the convolutional Fourier series and provided that $\Scale[.9]{f(x)}$ is also discontinuous at $\Scale[.9]{x=0}$, the latter equation can be restated as follows:

\begin{eqnarray}\label{z5}
\Scale[0.9]{\mathcal{D}_N f_N + \epsilon_N \star f_N  + {\mu_0}\Phi_N(x) = \gamma f_N.}
\end{eqnarray}

Wherein $\Scale[.9]{\mu_0}$ is a constant and the truncated operator $\Scale[.9]{\mathcal{D}_N}$ can typically be modeled using a diagonal matrix.
This would be the ideal form of the equation, however, what in practice one solves is as $\Scale[0.9]{\mathcal{D}_N \tilde{f}_{N} + \epsilon_N \star \tilde{f}_{N} = \gamma \tilde{f}_{N}}$, without the $\Scale[.9]{\Phi_N}$ terms.
Comparing the latter with equation \ref{z5}, it seems a relationship of the form $\Scale[.9]{f_N = \tilde{f}_{N} + a\Phi_N}$ approximately holds. To assert this claim and obtain $\Scale[.9]{a}$, we will perform a first-order approximation in our calculations, which will be explained subsequently.
Recall that for large values of $\Scale[.9]{N}$, the function $\Scale[.9]{\Phi_N}$ acts like a delta function. Hence, performing any differential operation on it results in higher-order distributions which we neglect in our first-order approximation. On the other side, $\Scale[.9]{\epsilon_N\star\Phi_N}$ can be approximated by $\Scale[.9]{c_0\Phi_N}$ where $\Scale[.9]{c_0 = {\epsilon(0^-)}/{2}+{\epsilon(0^+)}/{2}}$.
Accordingly, by replacing $\Scale[.9]{f_N = \tilde{f}_{N} + a\Phi_N}$ in equation \ref{z5} and trying to balance both sides of the above equation with respect to $\Scale[.9]{\Phi_N}$, we obtain $\Scale[.9]{a = \mu_0(\gamma - c_0)^{-1}}$.
Now, since no first-order TCE should be in the original truncated solution $\Scale[.9]{f_N}$, it follows that a perturbative term $\Scale[.9]{-\mu_0(\gamma - c_0)^{-1}{\Phi_N}}$ should appear in $\Scale[.9]{\tilde{f}_{N}}$.

Now, let us consider the wave equation in a grating problem. If we denote the free-space wavenumber with $\Scale[.9]{k= 2\pi/{\lambda}}$, in which $\Scale[.9]{\lambda}$ is the free-space wavelength, Fourier modal methods commonly lead to an eigenproblem as $\Scale[.9]{\left( {\bm P - {\xi^2}\bm I} \right){\bm{e_t}} = 0}$, wherein
$\Scale[.9]{\bm P}$ is the modal matrix, $\Scale[.9]{\bm{e_t}}$ the lateral electric field eigenvector, and $\Scale[.9]{\xi= {k_z}/{k}}$ the normalized longitudinal propagation constant and the square root of an eigenvalue. The detailed expansion of the modal matrix $\Scale[.9]{\bm P}$ is expressed in \ref{Apx_A}. For a lamellar grating, the modal matrix can be elaborated as follows, in which $\Scale[.9]{\bm{F}=\mathcal{T}(\epsilon)}$ is a Toeplitz matrix of the Fourier coefficients of the periodic dielectric function $\Scale[.9]{\epsilon(x)}$, $\Scale[.9]{\bm{N_x}}$ is diagonal matrix equivalent to the operator $\Scale[.9]{\partial = k^{-1}\partial/{\partial x}}$, $\Scale[.9]{N_{y0} = {k_{y0}}/{k}}$ is a constant, and $\Scale[.9]{\gamma = N_{y0}^2+\xi^2}$ is the translated eigenvalue:

\begin{eqnarray}\label{z10}
\Scale[0.9]{
\left[ {\begin{array}{*{20}{c}}
{{\bm F} - \bm{N_x}{{\bm F}^{ - 1}}\bm{N_x}{\bm F} - {\gamma}{\bm I}}&{\bm 0}\\
{N_{y0}\left( {\bm{N_x} - {{\bm F}^{ - 1}}\bm{N_x}{\bm F}} \right)}&{{\bm F} - \bm{N_x}^2 - {\gamma}{\bm I}}
\end{array}} \right]
\left[ {\begin{array}{*{20}{c}}
{\bm{e_x}}\\
{\bm{e_y}}
\end{array}} \right]=\bm{0}.
}
\end{eqnarray}

Here, we will consider the equation for the $\Scale[.9]{x}$-component of the electric field, denoted by $\Scale[.9]{E_x= E_x(x)}$. The equivalent differential form of the equation would be $\Scale[.9]{(\epsilon - \partial \epsilon^{-1} \partial \epsilon - \gamma)E_x=0}$. Depending on the methodology in Fourier modal methods, the inverse matrix $\Scale[.9]{\bm{F}^{-1}}$ may be replaced with $\Scale[.9]{\mathcal{T}(\epsilon^{-1})}$, asymptotically producing the same matrix \cite{Gray2006}. For the sake of simplicity, let us use the latter case and write the unperturbed truncated equation for $\Scale[.9]{E_{x,N}}$:

\begin{eqnarray}\label{z11}
\Scale[0.9]{
{\epsilon_N}\star E_{x,N} + \mu_0 \Phi_N -\gamma E_{x,N} =
\partial_N\left\{\epsilon^{-1}_N \star \{\partial_N\{\epsilon_N\star E_{x,N} + \mu_0 \Phi_N\}\} + \mu_1 \Phi_N \right\}.
}
\end{eqnarray}

Where $\Scale[.9]{\mu_0}$ and $\Scale[.9]{\mu_1}$ are constants with respect to $\Scale[.9]{x}$. On the other side, what we practically solve in an algebraic form with $\Scale[.9]{\tilde{E}_{x,N}}$ as the eigenfunction, is the same as the above equation without the $\Scale[.9]{\Phi_N}$ terms.
Hence, assuming the relationship $\Scale[.9]{E_{x,N}=\tilde{E}_{x,N} +a \Phi_N}$ for the original and perturbed solutions, the constant $\Scale[.9]{a}$ should be found such that $\Scale[.9]{\Phi_N}$ terms are omitted from equation \ref{z11}.
Given the estimating rules of the previous example, it would be straightforward to show $\Scale[.9]{a = \mu_0(\gamma-c_0)^{-1}}$, where $\Scale[.9]{c_0 = {\epsilon(0^-)}/{2}+{\epsilon(0^+)}/{2}}$. Now, since $\Scale[.9]{\Phi_N}$ does not manifest in $\Scale[.9]{E_{x,N}}$, it should inevitably appear in $\Scale[.9]{\tilde{E}_{x,N}}$. Same process could be repeated to analyze $\Scale[.9]{E_y = E_y(x)}$, using the following equation:

\begin{eqnarray}\label{z13}
\Scale[0.9]{
N_{y0}\left\{ \partial_N E_{x,N} - \epsilon^{-1}_N\star \left\{ \partial_x \left\{ \epsilon_x \star E_{x,N} + \mu_0 \Phi_N \right\}\right\} + \mu_1\Phi_N \right\}
+\epsilon_N \star E_{y,N} - \partial^2_N \{E_{y,N}\} - \gamma E_{y,N} = 0.
}
\end{eqnarray}

Note that since $\Scale[.9]{E_{y}}$ is continuous, no additional $\Scale[.9]{\Phi_N}$ term needs to be placed in the right-hand side convolutional series of this equation.
Now, repeating the above process with $\Scale[.9]{E_{x,N}=\tilde{E}_{x,N} +a\Phi_N}$ and $\Scale[.9]{E_{y,N}=\tilde{E}_{y,N}+b \Phi_N}$ leads to the relationship $\Scale[.9]{b = N_{y0}\mu_1(\gamma-c_0)^{-1}}$.
Note that $\Scale[.9]{E_{x}}$ is discontinuous and besides the Gibbs phenomenon, it incorporates TCE as well. On the other hand, $\Scale[.9]{E_{y}}$ is a continuous function and hence does not include the (first-order) Gibbs phenomenon, however, depending on the formulation, it can still incorporate TCE.

\subsection{Reconstruction of Grating Modes}
\label{Sub: Modes}

\begin{figure}[!t]	
	\centering
	\begin{subfigure}{0.37\textwidth}
		\centering
		\includegraphics[width=\linewidth]{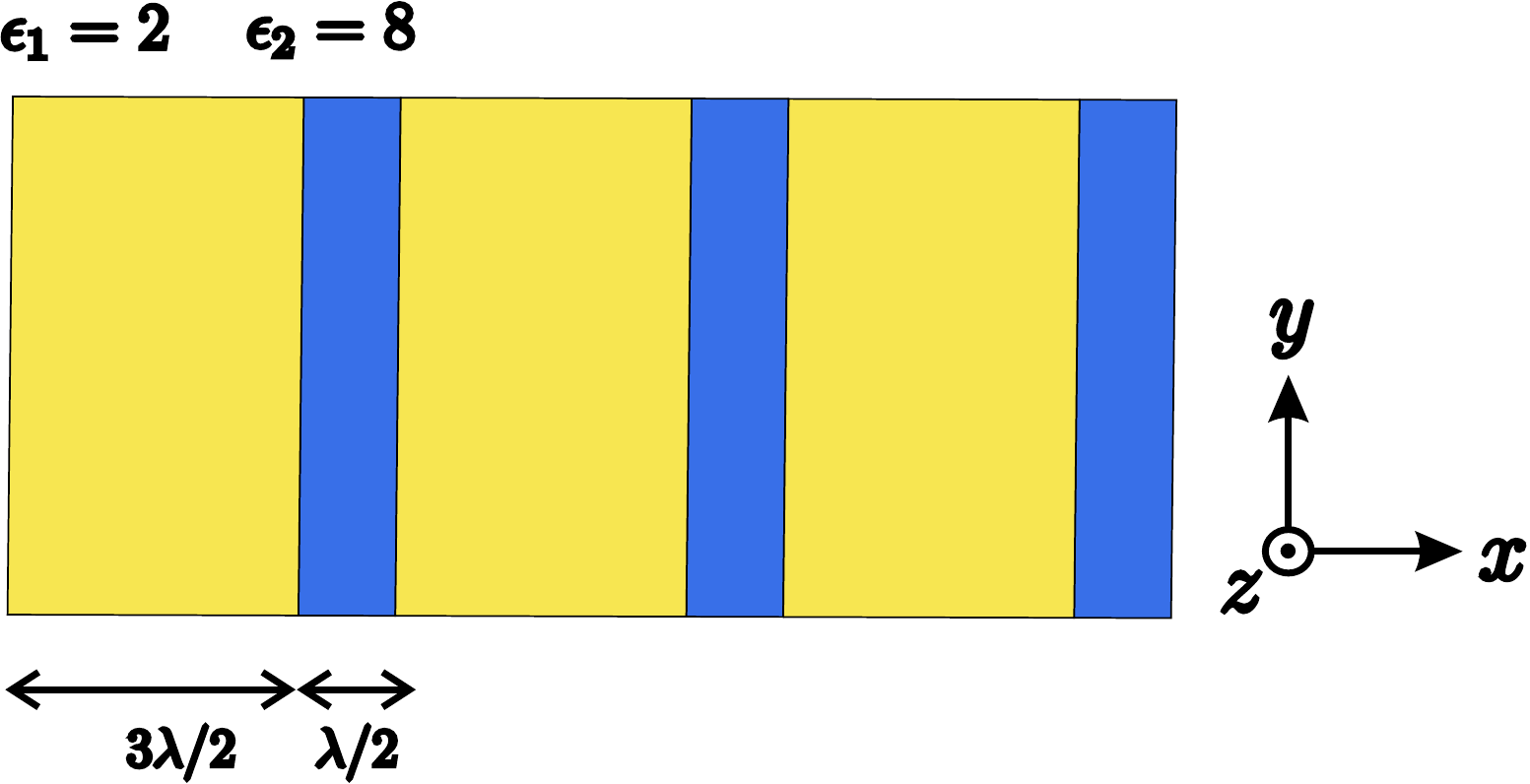}
		\caption{}\label{lam_grat}		
	\end{subfigure}
~
	\begin{subfigure}{0.30\textwidth}
		\centering
		\includegraphics[width=\linewidth]{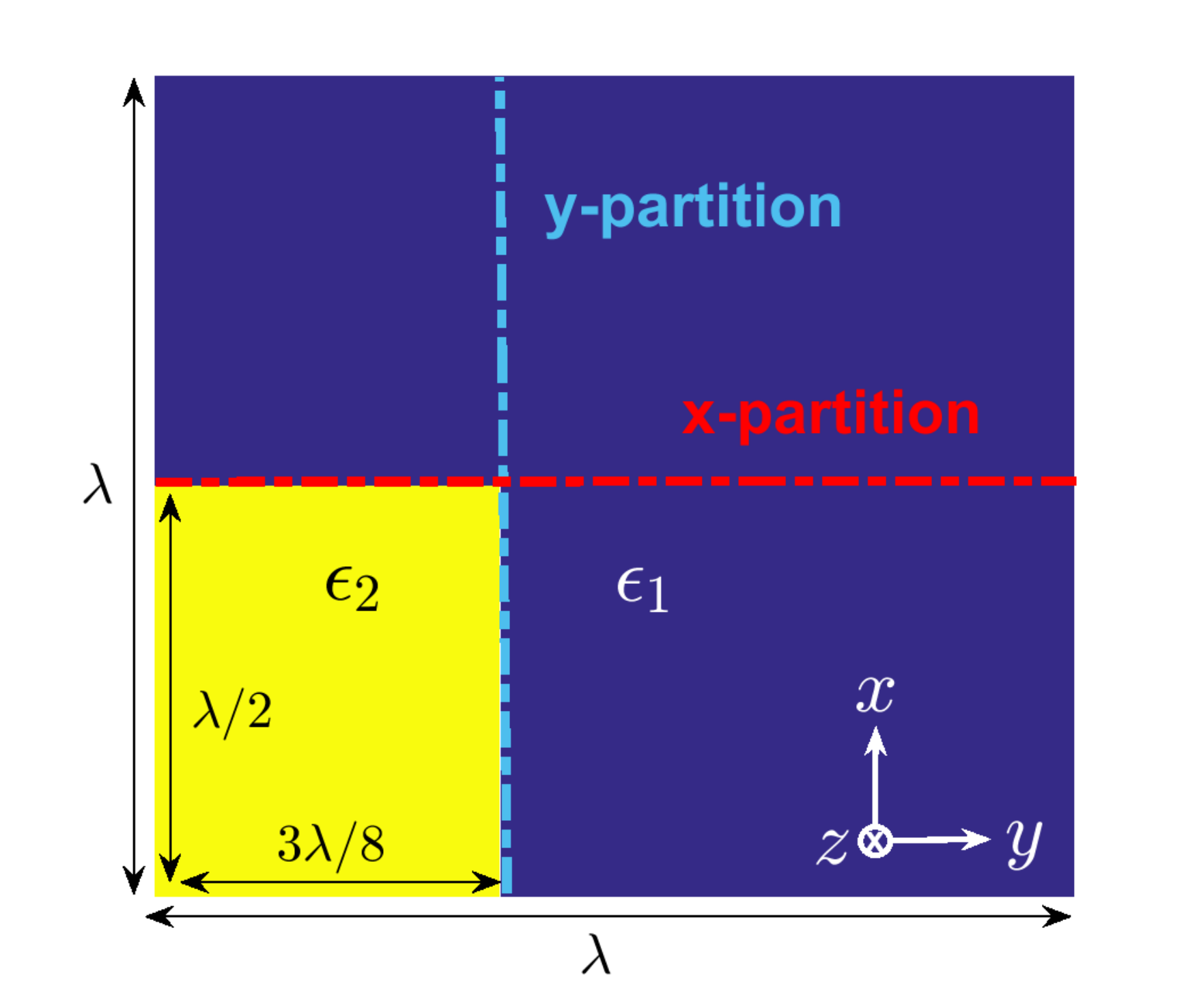}
		\caption{}\label{cros_grat}
	\end{subfigure}

	\caption{The unit cell of a (\subref{lam_grat}) lamellar grating, and a (\subref{cros_grat}) crossed grating.}
	\label{fig_g}
\end{figure}

\begin{figure}[!t]	
	\centering
	\begin{subfigure}{0.41\textwidth}
		\centering
		\includegraphics[width=\linewidth]{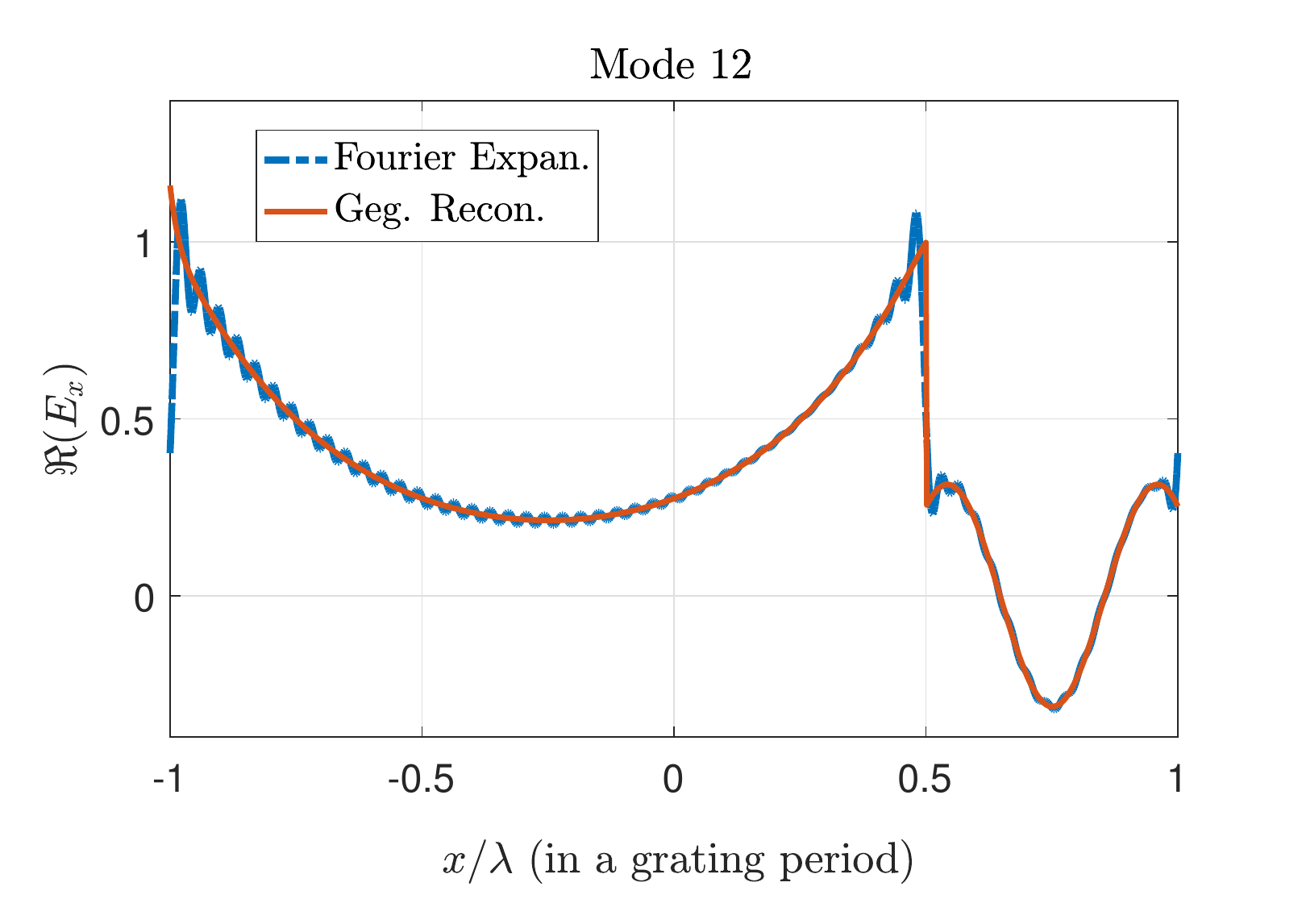}
		\caption{}\label{lg_mode1}		
	\end{subfigure}
~
	\begin{subfigure}{0.41\textwidth}
		\centering
		\includegraphics[width=\linewidth]{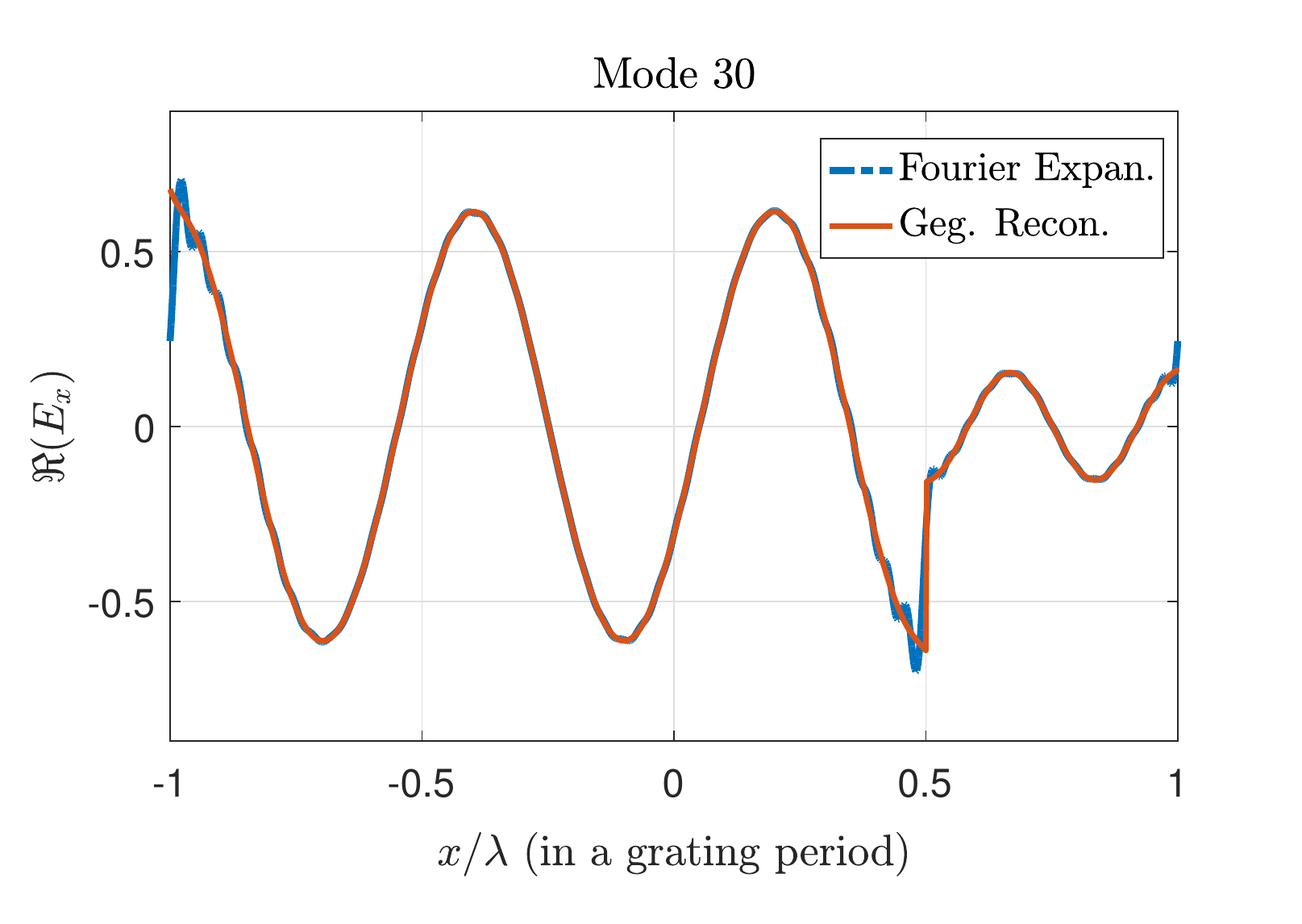}
		\caption{}\label{lg_mode2}
	\end{subfigure}

	\caption{The Fourier expansion ($\Scale[.9]{N=48}$) and reconstruction of $\Scale[.9]{{\Re}(E_x)}$ for the (\subref{lg_mode1}) mode 12, and (\subref{lg_mode2}) mode 30 of the lamellar grating figure \ref{lam_grat}.}
	\label{fig_lg}
\end{figure}

The modes of a grating are the constituents of any field expression, hence, tackling the Gibbs phenomenon in individual modes could be a key step towards application of spectral reconstruction in a far-field grating problems as well. Consider a grating structure in the $xy$-plane, uniformly stretched along the $z$-axis, as depicted in figure \ref{fig_3d}, with the relative permittivity $\Scale[0.9]{\epsilon(x)}$ or $\Scale[0.9]{\epsilon(x,y)}$ for lamellar or crossed gratings. We define $\Scale[.9]{T_x}$ and $\Scale[.9]{T_y}$ to be the normalized periods along $x$ and $y$-axes, so that $\Scale[0.85]{\epsilon (x+ \lambda T_x) = \epsilon(x)}$ and $\Scale[0.9]{\epsilon (x+ \lambda T_x,y) =\epsilon( x,y+ \lambda T_y) = \epsilon(x,y)}$, for the lamellar and crossed gratings respectively.
The resultant eigenvalue equation of the wave propagation in the grating is expanded in \ref{Apx_A}.
The electric field expansion of each component in a grating mode can be expressed as $\Scale[.9]{E_u(\bm r) = {e^{-i{k_{z}}z}}{e^{-i({k_{x}}x + {k_{y}}y)}}\tilde{E}_u\left({\bm r}/\lambda\right)}$, in which $\Scale[.9]{(k_{x},k_{y})}$ is the lateral component of the incident plane wave, $\Scale[.9]{k_{z}}$ the longitudinal propagation constant and the corresponding eigenvalue, and $\Scale[0.9]{\tilde{E}_u}$ a periodic function over the normalized unit cell, with $\Scale[0.9]{u \in \{x,y,z\}}$.
Note that the component $\Scale[0.9]{E_z}$ is continuous over the unit cell and converges practically fast enough. Hence, the problem of retrieving the eigenmodes predominantly attributes to the lateral components $\Scale[0.9]{E_{x/y}}$.

\begin{figure*}[!t]	
	\centering

	\begin{subfigure}{0.31\textwidth}
		\centering
		\includegraphics[width=\linewidth]{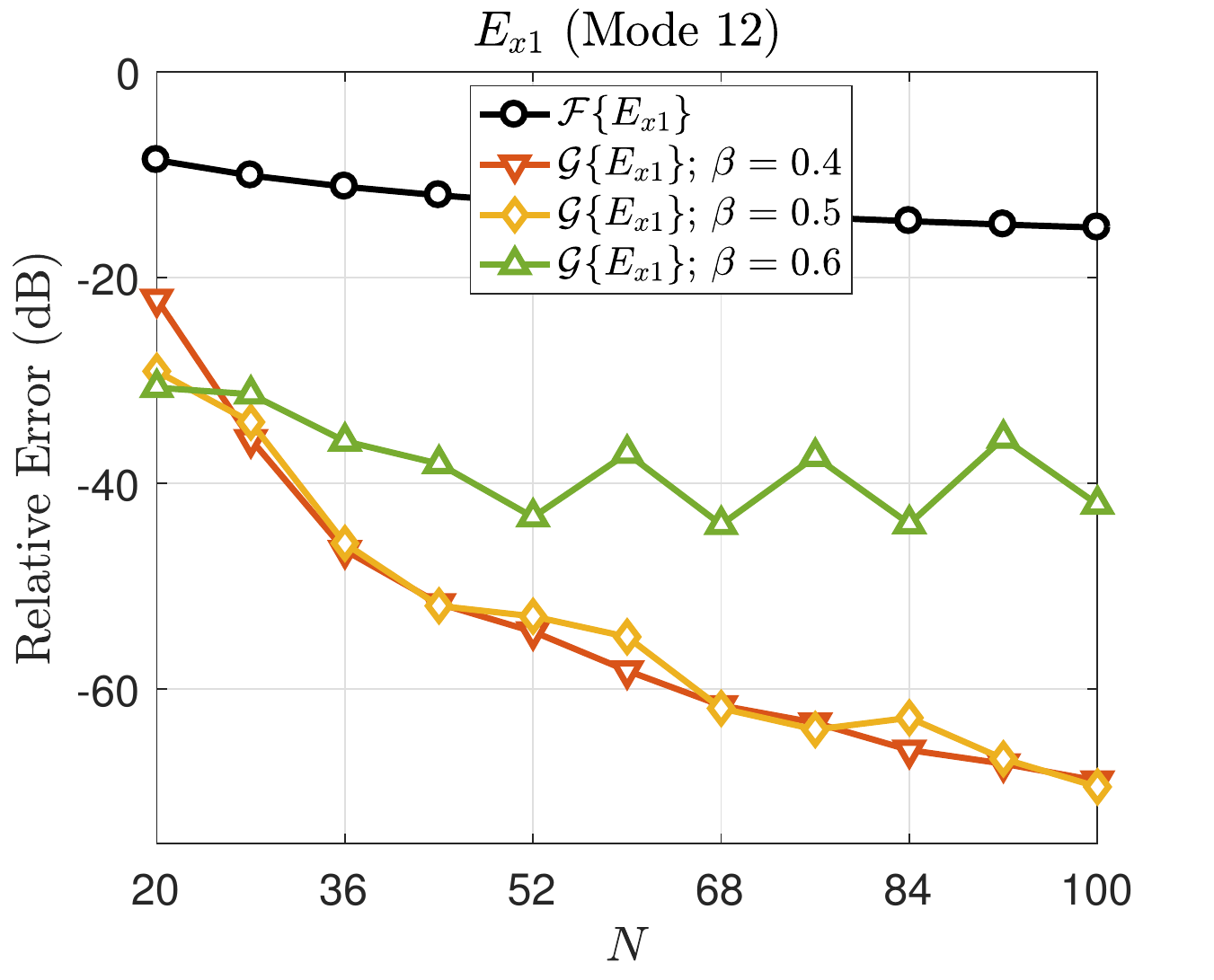}
		\caption{}\label{m12_x1}
	\end{subfigure}	
~
	\begin{subfigure}{0.31\textwidth}
		\centering
		\includegraphics[width=\linewidth]{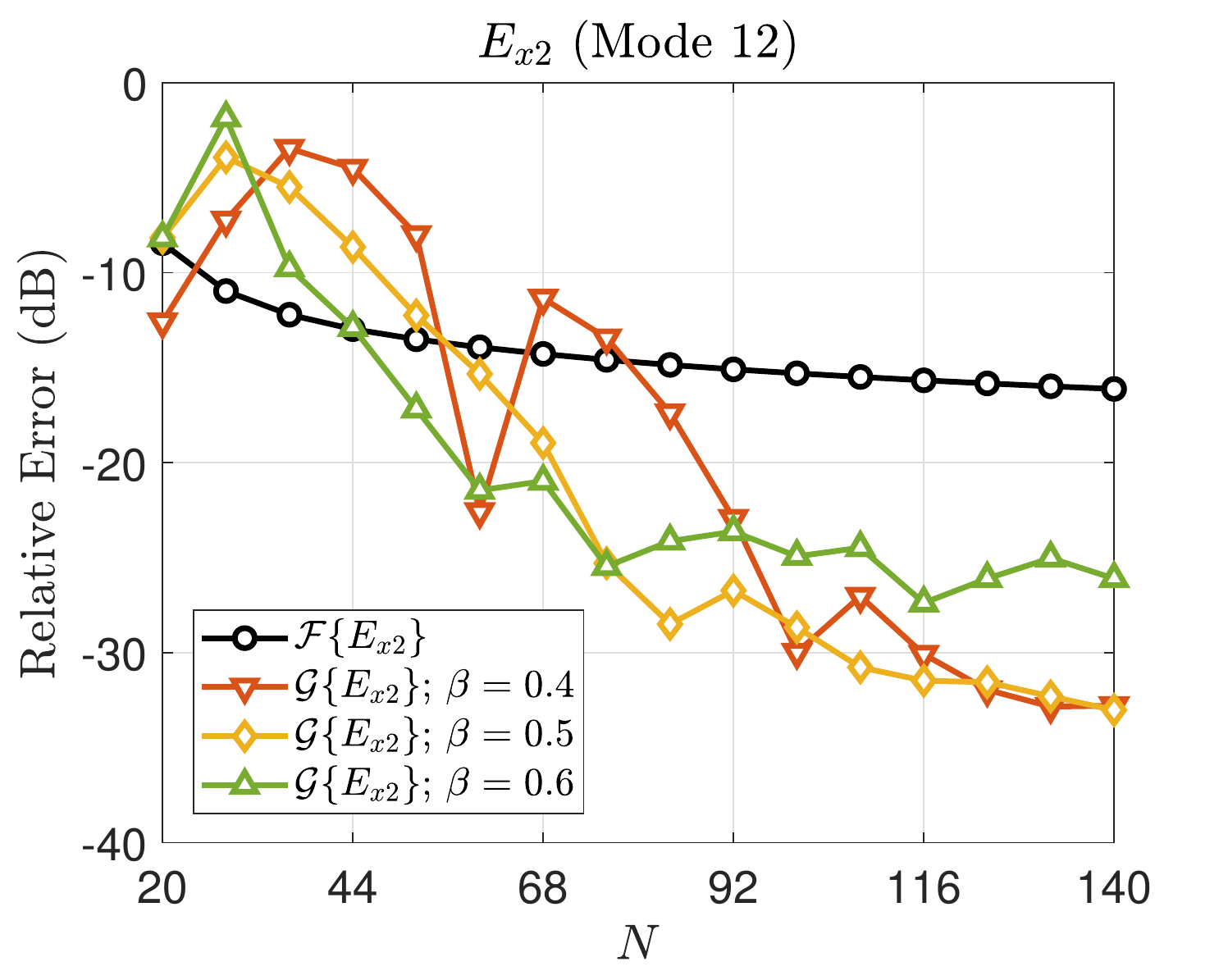}
		\caption{}\label{m12_x2}
	\end{subfigure}
~
	\begin{subfigure}{0.31\textwidth}
		\centering
		\includegraphics[width=\linewidth]{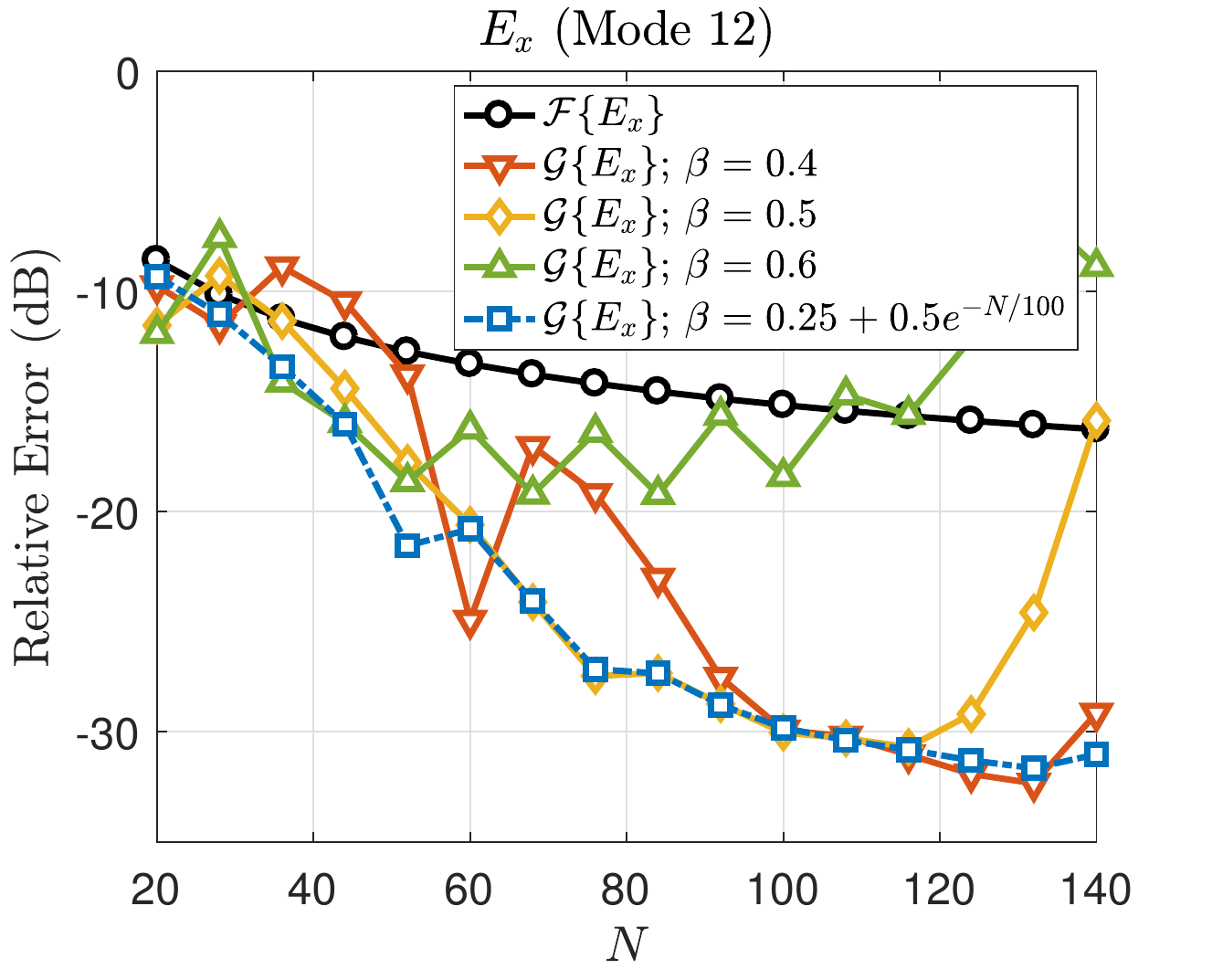}
		\caption{}\label{m12_x}
	\end{subfigure}	
\quad
	\begin{subfigure}{0.31\textwidth}
		\centering
		\includegraphics[width=\linewidth]{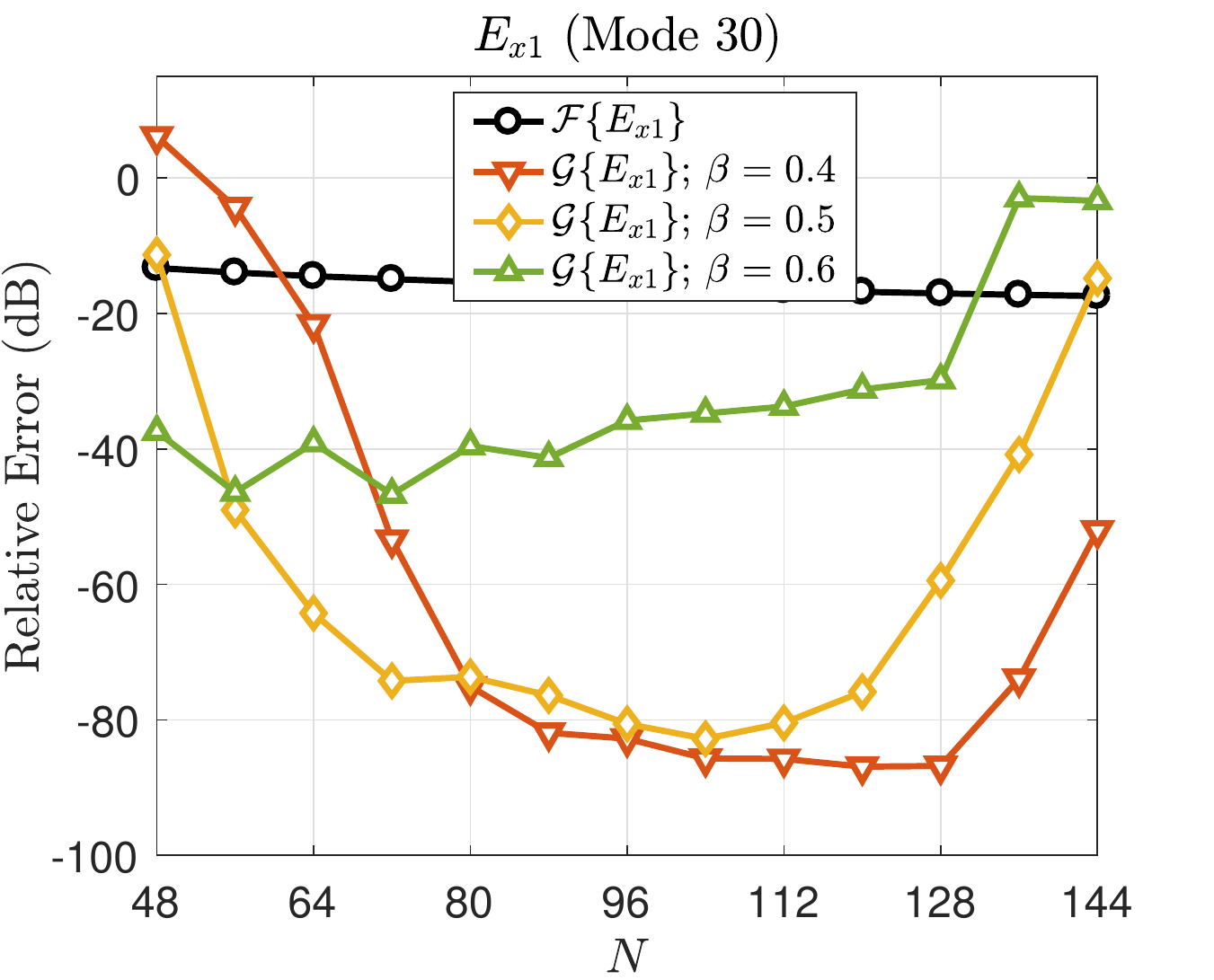}
		\caption{}\label{m30_x1}
	\end{subfigure}
~
	\begin{subfigure}{0.31\textwidth}
		\centering
		\includegraphics[width=\linewidth]{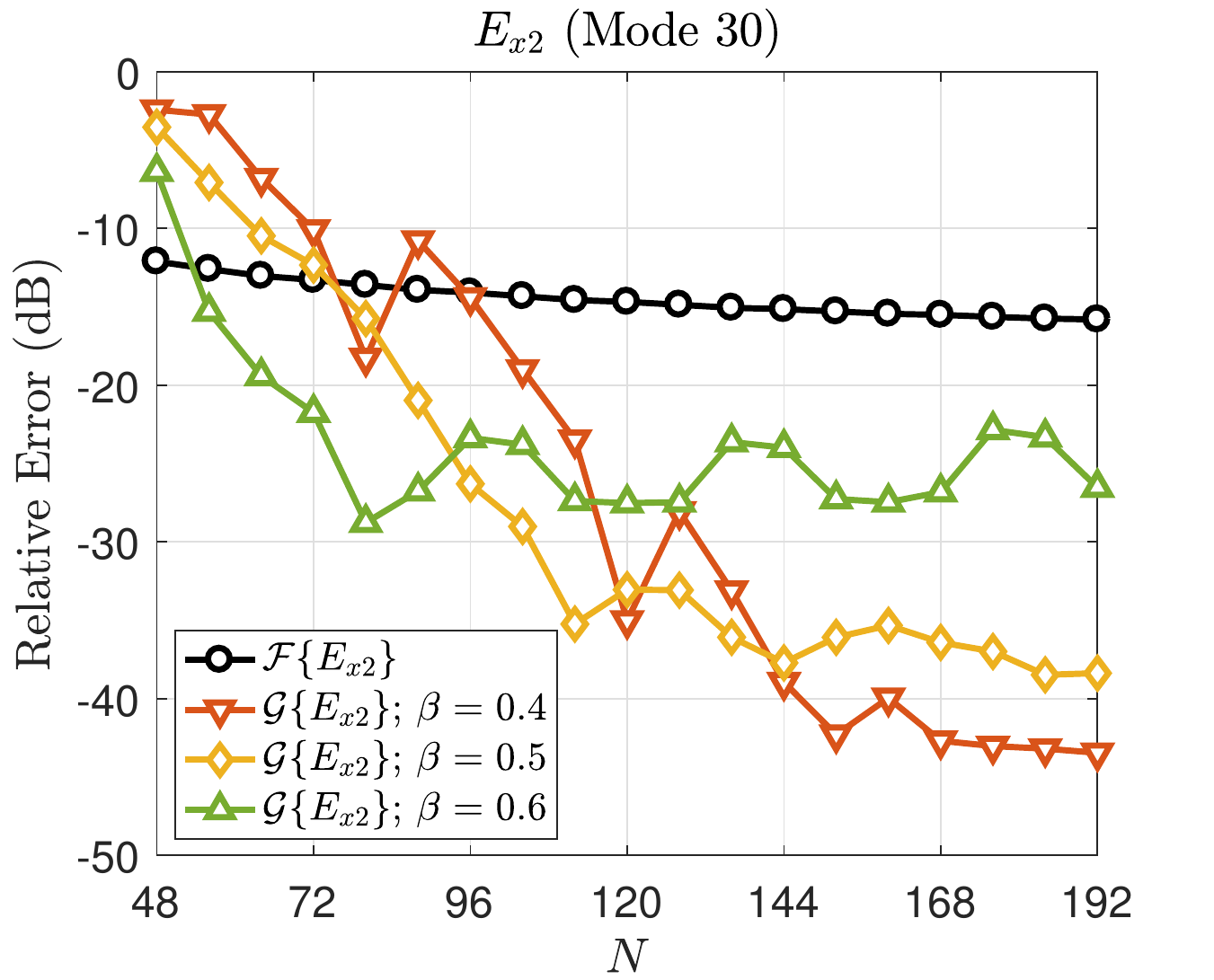}
		\caption{}\label{m30_x2}		
	\end{subfigure}
~
	\begin{subfigure}{0.31\textwidth}
		\centering
		\includegraphics[width=\linewidth]{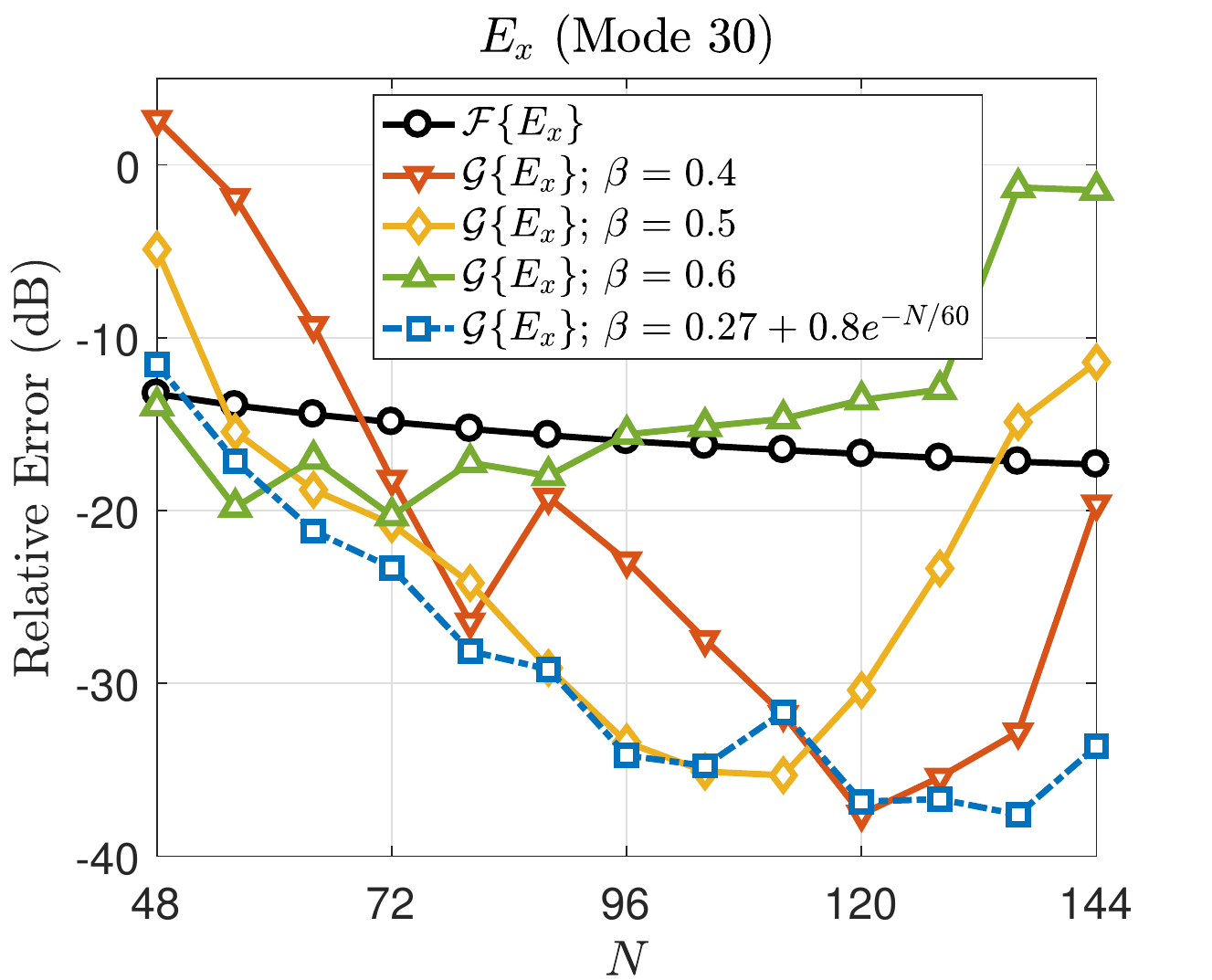}
		\caption{}\label{m30_x}		
	\end{subfigure}

	\caption{Relative convergence error of the Fourier expansion ($\Scale[.9]{\mathcal{F}}$) and the Gegenbauer reconstruction ($\Scale[.9]{\mathcal{G}}$) with various $\Scale[.9]{\beta}$ values for (\subref{m12_x1}) the first subinterval, (\subref{m12_x2}) the second subinterval, and (\subref{m12_x}) the whole interval of the mode 12. Similarly for (\subref{m30_x1}) the first subinterval, (\subref{m30_x2}) the second subinterval, and (\subref{m30_x}) the whole interval of the mode 30 in lamellar grating of figure \ref{lam_grat}.}
	\label{fig_lg_N}
\end{figure*}

Let us indicate the use of Gegenbauer method for the resolution of the Gibbs phenomenon as well as TCE in grating eigenmodes with two numerical examples. Note in a grating problem, if we sort the eigenmodes according to the real part of the associated squared eigenvalues, i.e. $\Scale[.9]{{\Re}(k_z^2)}$, an asymptotically well-defined ordinal relationship is obtained to characterize the modes independent from truncation orders \cite{Faghihifar2020}.
Consider the lamellar grating of figure \ref{lam_grat} under a plane wave with the incidence and azimuthal angles $\Scale[.9]{\theta=\phi=45^\circ}$.
We'd like to reconstruct the discontinuous electric field component $\Scale[.9]{E_x}$ for the 12th and 30th modes of this structure with corresponding $\Scale[.9]{k_z/k = 1.405}$ and $\Scale[.9]{k_z/k = 1.023i}$. Considering a truncation order $\Scale[.9]{N=48}$ and discretizing the integrations using $\Scale[.9]{L=1001}$ points, we obtain the expansion of the modes and reconstruct them using the Gegenbauer method.
The second parameter is selected higher than the popular choice $\Scale[.9]{\beta=0.25}$ to accelerate the convergence, while changing the first parameter could cause instabilities and hence avoided. The Fourier expansion and Gegenbauer reconstruction of $\Scale[.9]{{\Re}(E_x)}$ are plotted for both modes in figures \ref{lg_mode1} and \ref{lg_mode2}.

\begin{figure}[!t]
\centering
\includegraphics[width=0.93\linewidth]{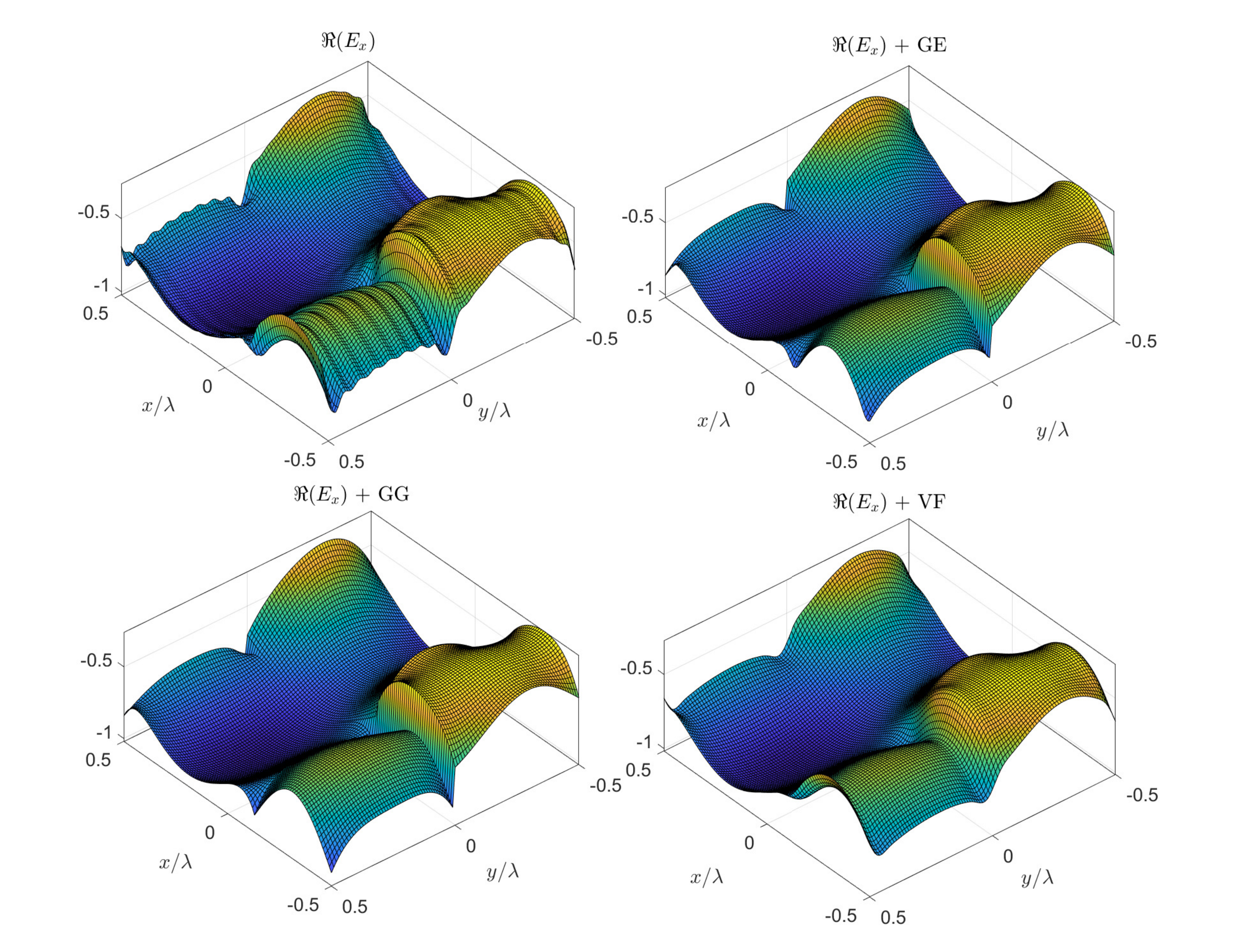}
\caption{3D depiction of the Fourier representation of $\Scale[.9]{\Re\{E_x\}}$ in mode 10 of the crossed grating of figure \ref{cros_grat}, along with its Gegenbauer-exponential reconstruction (GE), double-Gegenbauer reconstruction (GG), and Vandeven-filtered expansion (VF), with $\Scale[.9]{N_1=N_2 = 16}$ and $\Scale[.9]{\beta = 0.55}$.}
\label{cg_mode_x}
\end{figure}

In order to illustrate how the eigenmodes reconstruction error behaves with the truncation order $\Scale[.9]{N}$ for different values of $\Scale[.9]{\beta}$, one should first note that unlike what we did in subsection \ref{Sub: Perf}, the eigenmodes here do not have a closed relationship to compare the results with. Hence, the convergence rate here is assessed in the Cauchy sense, via a relative error relationship of $\Scale[.9]{\Delta_k = \left\||E_{k+1}|-|E_{k}|\right\|_2/\|E_{k+1}\|_2}$, with $\Scale[.9]{k}$ subscripts denoting $\Scale[.9]{k}$-th stage solutions. Note that the eigenmodes should be normalized at each stage, e.g. to the retrieved value at a subinterval midpoint which is prone to the least computational deviations, so that the effect of random complex coefficients at each stage is canceled out. On the other side, since the subinterval reconstructions are independent yet corresponding errors are aggregated, it would be elucidative to plot the error graphs in different subintervals as well.

Three error plots for each mode have been depicted in figure \ref{fig_lg_N}. For each mode, the relative reconstruction error has been plotted in each subinterval separately and in total, for $\Scale[.9]{\beta=0.4,0.5}$ and $\Scale[.9]{0.6}$. All the reconstruction error plots indicate a faster convergence than the Fourier series itself, yet with a different behavior. As a first observation, note that larger values for $\Scale[.9]{\beta}$ lead to a faster convergence, though with a smaller maximum achievable accuracy, and a faster point of divergence afterwards due to the domination of the round-off error. As noted before, the Gegenbauer method has a poor resolution power, and since electromagnetic modes usually show oscillatory behavior, cautious selection of larger values of $\Scale[.9]{\beta}$ (practically $\Scale[.9]{0.25 \sim 0.75}$) could have a significant effect on the recovery of oscillatory modes. Another observation that should be noted is the different convergence behavior in the larger and smaller subintervals. The polynomial order and the convergence rate of the Gegenbauer reconstruction in different subintervals is related to the scale parameter $\Scale[.9]{\varepsilon}$, according to the formulation presented in subsection \ref{Sub: Gegen}. Hence, generally speaking, with the same choice of $\Scale[.9]{(\alpha,\beta)}$, the convergence in smaller subintervals is achieved more slowly, and the divergence happening for comparably larger values of $\Scale[.9]{N}$. On the other side, note in figures \ref{m12_x} and \ref{m30_x}, how different error graphs minima occur at different values of $\Scale[.9]{N}$, meaning that probably using a properly tailored parameter sequence $\Scale[.9]{\beta_N}$ that remains $\Scale[.9]{\sim 1}$ for $\Scale[.9]{N\to 0}$ and tends to $\Scale[.9]{\sim 0.25}$ as $\Scale[.9]{N\to \infty}$, could be a more efficient approach.
In that regard, we have successfully employed the parameter sequences of $\Scale[.9]{\beta_N=0.25+0.5\exp(-N/100)}$ and $\Scale[.9]{\beta_N=0.27+0.8\exp(-N/60)}$ for the mode 12 and for the mode 30, respectively, and plotted the corresponding error graphs.

\begin{figure}[!t]
\centering
\includegraphics[width=0.84\linewidth]{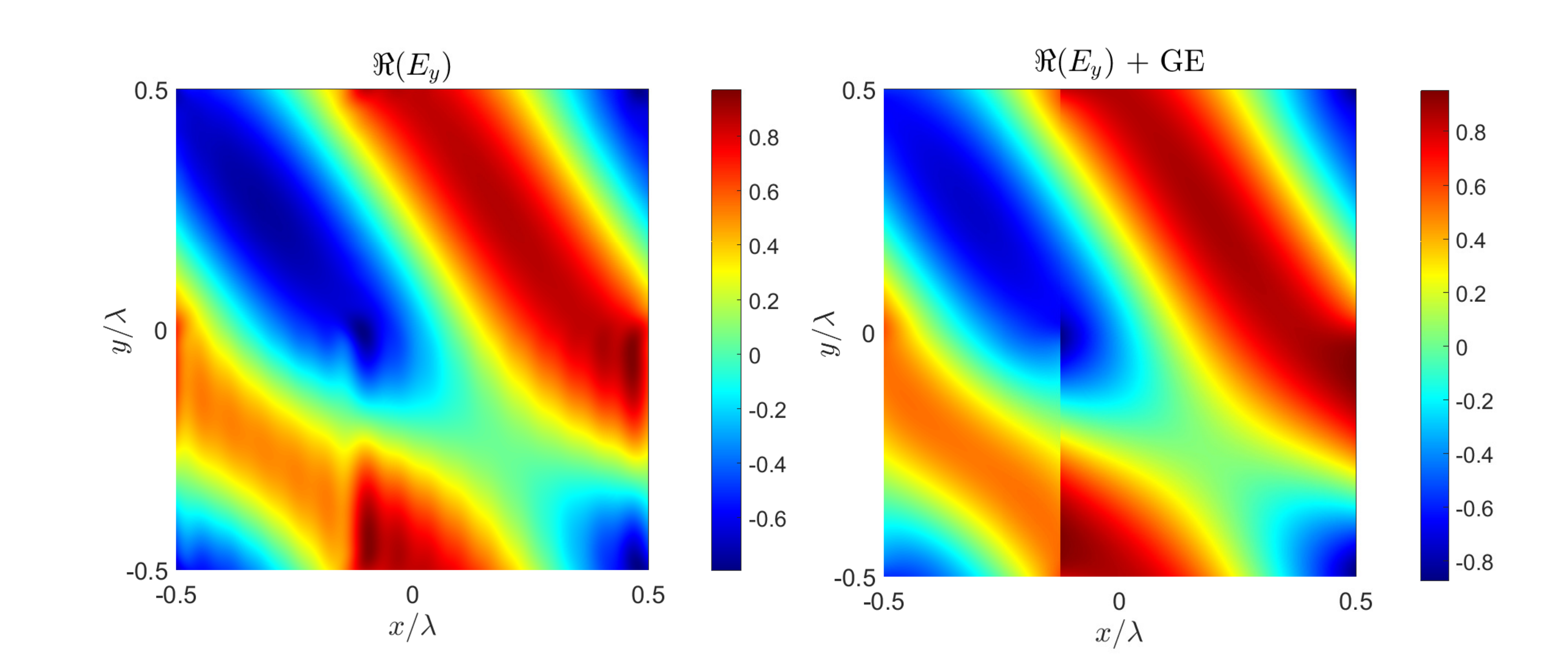}
\caption{2D depiction of the Fourier representation of $\Scale[.9]{\Re\{E_y\}}$ in mode 10 of the crossed grating of figure \ref{cros_grat}, along with its Gegenbauer-exponential reconstruction (GE), for $\Scale[.9]{N_1=N_2 = 16}$ and $\Scale[.9]{\beta = 0.55}$.}
\label{cg_mode_y}
\end{figure}

For the second example, let us consider the crossed grating of figure \ref{cros_grat} comprising anisotropic dielectrics $\Scale[.9]{\epsilon_1 = [1.15\hspace{5pt}0.05\hspace{5pt}0\hspace{2pt}, \hspace{2pt}0.05\hspace{5pt}1.15\hspace{5pt}0\hspace{2pt}, \hspace{2pt}0\hspace{5pt}0\hspace{5pt}1]}$ and $\Scale[.9]{\epsilon_2 = [2.2\hspace{5pt}0.2\hspace{5pt}0\hspace{2pt}, \hspace{2pt}0.2\hspace{5pt}2.2\hspace{5pt}0\hspace{2pt}, \hspace{2pt}0\hspace{5pt}0\hspace{5pt}2.2]}$ wherein ``$,$'' denotes vertical concatenation. Assuming an incident plane wave with $\Scale[.9]{(\theta,\phi)=(45^\circ,60^\circ)}$, the eigenproblem is solved with the truncations orders $\Scale[.9]{N_1=N_2=16}$ and $\Scale[.9]{L_1=L_2=1001}$ points in both directions. We will consider the recovery of both the $\Scale[.9]{E_x}$ and $\Scale[.9]{E_y}$ components of the mode 10, with the associated eigenvalue of $\Scale[.9]{k_z/k=-0.85}$.

Hopefully, a two-dimensional Gegenbauer reconstruction is feasible on rectangular smooth regions \cite{Gelb1997}, however, due to the heterogeneous convergence rates in different subintervals, it might be inefficient.
To increase the efficiency, we will use four subintervals in this scheme (characterized in figure \ref{cros_grat} using both partitions), otherwise three would have sufficed. The reconstruction terms would accordingly be like $\Scale[.9]{C_{m}^{\lambda}\left\{(x-\delta)/\epsilon T_x \right\} C_{n}^{\lambda'}\left\{(y-\delta')/\epsilon' T_y \right\}}$.
A more elegant approach on the other hand, would be to employ the continuity property of the tangential field components, and reconstruct each component using one-dimensional Gegenbauer terms along the axis of discontinuity, while keeping the complex exponential functionality along the other axis. Accordingly, an x-partition would be used for x-component and a y-partition for the y-components (as characterized in figure \ref{cros_grat}). Using this scheme, $\Scale[.9]{E_x}$ components will comprise $\Scale[.9]{C_{m}^{\lambda}\left\{(x-\delta)/\epsilon T_x \right\}\exp\{in\pi y/T_y\}}$ terms in each subinterval, and similarly $\Scale[.9]{E_y}$ will be expressed in terms of $\Scale[.9]{\exp\{im\pi x/T_x\} C_{n}^{\lambda'}\left\{(y-\delta')/\epsilon' T_y \right\}}$ terms.

\begin{figure}[!t]	
	\centering
	\begin{subfigure}{0.41\textwidth}
		\centering
		\includegraphics[width=\linewidth]{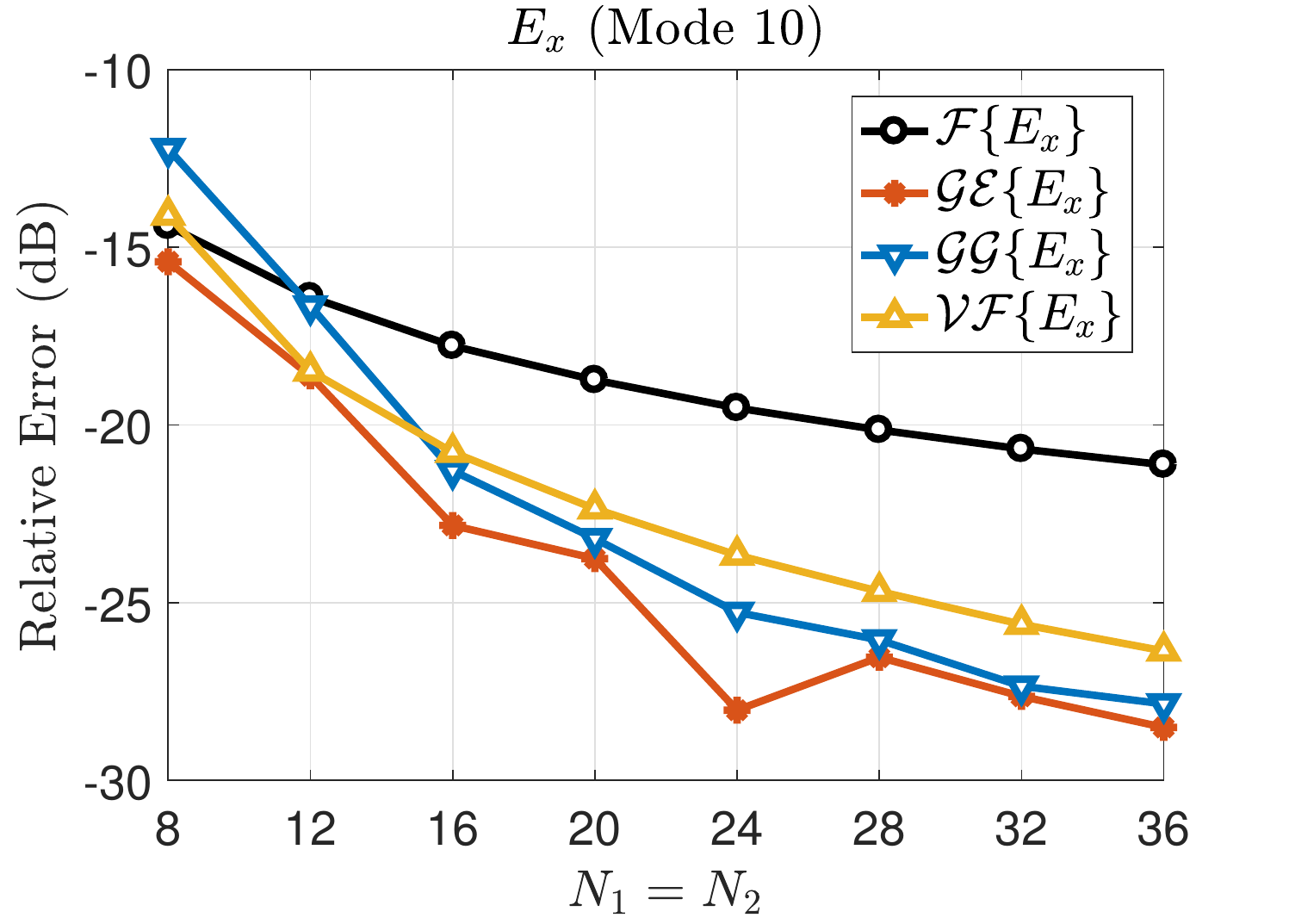}
		\caption{}\label{2g_Nx}		
	\end{subfigure}
~
	\begin{subfigure}{0.41\textwidth}
		\centering
		\includegraphics[width=\linewidth]{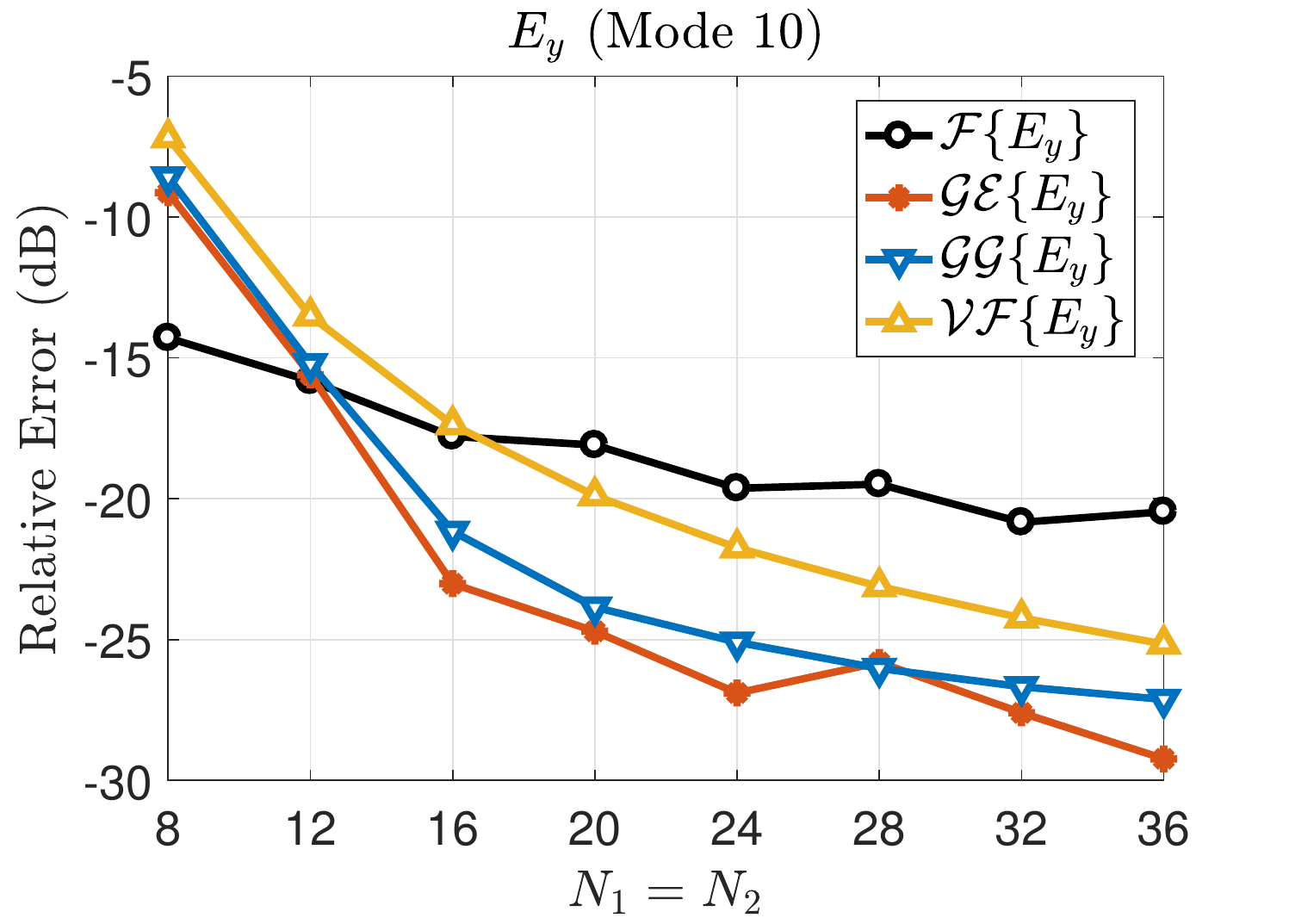}
		\caption{}\label{2g_Ny}
	\end{subfigure}

	\caption{Relative convergence error for the Fourier representation (\Scale[.9]{\mathcal{F}}), Gegenbauer-exponential reconstruction (\Scale[.9]{\mathcal{GE}}), double-Gegenbauer reconstruction (\Scale[.9]{\mathcal{GG}}), and the Vandeven-filtered expansion (\Scale[.9]{\mathcal{VF}}) of (\subref{2g_Nx}) $\Scale[.9]{E_x}$, and (\subref{2g_Ny}) $\Scale[.9]{E_y}$ of mode 10 in crossed grating of figure \ref{cros_grat}.}
	\label{fig_cg_N}
\end{figure}

For the x-component, a 3D plot of $\Scale[.9]{\Re(E_x)}$ in the Fourier form, along with its reconstruction using the two prescribed techniques and the Vandeven filter for comparison, are illustrated in figure \ref{cg_mode_x}. We have used $\Scale[.9]{(\alpha,\beta)=(0.25,0.55)}$ for the Gegenbauer reconstructions. Moreover, a 2D plot of the Fourier expansion and the Gegenbauer-exponential reconstruction for the $\Scale[.9]{\Re(E_y)}$ component, is also depicted in figure \ref{cg_mode_y}. Comparing the graphs, the resolution or amelioration of the Gibbs/TCE phenomenon is quite clear from the graphs, however, in order to qualify the convergence of the utilized techniques, we have also assessed their convergence behavior as well. In figure \ref{fig_cg_N}, the error graphs of both components are plotted for the Fourier series and the reconstruction schemes versus truncation orders $\Scale[.9]{N_1=N_2}$. While all methods improve the convergence, the proposed Gegenbauer-exponential scheme seems superior.

In the end, let us comment on the feasibility of post-processing techniques for the far-field grating problems. Solving the far-field problems relies on both the eigenvalues and eigenmodes, however, reconstruction techniques only apply to the latter. Nonetheless, knowing that the collective distribution of eigenvalues stabilizes fast \cite{Faghihifar2020}, accessing very precise values as provided by subsectional methods \cite{Morf1995, Edee2011}, could not be comparably important. Not to mention reconstruction of continuous field components, though they still incorporate higher-order discontinuities.
More importantly, the computational cost of employing reconstruction techniques in far-field problems, despite what it seems, would not be significantly higher than that of single a mode. In other words, one does not need to reconstruct all the modes individually, but instead, it suffices to solve the unmodified Fourier modal method, obtain the field distributions at the top/bottom boundaries, reconstruct them using post-processing techniques, and recalculate the scattering parameters. The practical implementation, however, would be the subject of another study.

\section{Conclusion}
\label{Sec: Conclusion}

Notwithstanding the rich literature of spectral reconstructions for the resolution of the Gibbs phenomenon, studying their applicability in the presence of truncated convolution errors (TCE) remains unprecedented.
In this paper, we have assessed the efficacy of post-processing techniques via product test functions, only to find almost all methods fail to converge correctly in the vicinity of discontinuities, due to the emergence of TCE.
On the other side, we have formulated and proved a significant theorem on the robustness of the classical Gegenbauer method against this computational phenomenon. Eventually, as a case study in photonics, we have proved the existence and indicated the practical resolution of TCE in gating eigenmodes.

\appendix  %changes section numbering to A
\section*{Appendices}\label{Apx}
%\appendix  %changes section numbering to A

\subsection{Spectral Reconstructions}\label{Apx_B}

\renewcommand{\theequation}{B-\arabic{equation}}% redefine the command that creates the equation no.
\setcounter{equation}{0}  % reset counter

Here, we will describe the popular post-processing methods for the resolution of the Gibbs phenomenon utilized throughout the text, in the simplest form.

\subsection*{Vandeven filter}
The Vandeven filter is a method with a proven exponential rate convergence outside a certain neighborhood of discontinuities. Application of this filter on a truncated series is as follows:

\begin{equation}\label{B1}
\Scale[.9]{
\mathcal{V} f_N = \displaystyle\sum\nolimits_{|k|\leqslant N} {\sigma \left({k}/{N}\right){\hat f_k}{e^{ik\pi x}}},\hspace{40pt}
\sigma \left( t \right) = 1 - \frac{{\left( {2p - 1} \right)!}}{{\left( {p - 1} \right){!^2}}}{\displaystyle\int_0^t {{{\left[ {r\left( {1 - r} \right)} \right]}^{p - 1}}dr}}.
}
\end{equation}

\subsection*{Singular Fourier-Pad\'e method}
In this method, the transform $\Scale[.9]{{z = {e^{-i\pi x}}}}$ is used first, and the Fourier expansion is converted to the sum of two Taylor series with variables $\Scale[.9]{z}$ and $\Scale[.9]{z^{-1}}$, as follows:

\begin{eqnarray}\label{B2}
\Scale[.9]{
f( {z = {e^{-i\pi x}}} ) = \displaystyle\sum\nolimits_{n = 0}^\infty  {'{c_n}{z^n}}  + \displaystyle\sum\nolimits_{n = 0}^\infty  {'{c_{ - n}}{z^{ - n}} = {f^ + }( z ) + {f^ - }( {{z^{ - 1}}})}.
}
\end{eqnarray}

In which $'$ in summations means the first summand is halved. Given the discontinuity at $\Scale[.9]{x_1=1}$ and considering $\Scale[.9]{\rho=e^{-i\pi x_1}}$, the following singular representation for the constituent functions is sought, wherein $\Scale[.9]{g_{0,1}^ \pm}$ are rational functions which are analytic in $\Scale[.9]{z=1}$.

\begin{eqnarray}\label{B3}
\Scale[0.9]{
{f^ \pm }\left( z \right) = g_0^ \pm \left( z \right) + g_1^ \pm \left( z \right)\ln \left( {1 - z/\rho^{\pm 1}} \right).
}
\end{eqnarray}

For multiple discontinuities, more singular logarithmic terms appear in the above equation.

\subsection*{Polynomial least-squares}
If $\Scale[.9]{T_M=\left\{ {{\varphi _1},...,{\varphi _M}} \right\}}$ is a set of linearly independent polynomials of the maximum order $\Scale[.9]{M-1}$ with $\Scale[.9]{M = \mathcal{O}(\sqrt{2N+1} )}$, a polynomial reconstruction $\Scale[.9]{\varphi}$ of the function $\Scale[.9]{f}$ to remove the Gibbs phenomenon could be the solution to the following least-squares equation, in which bracket notation simply denotes the $\Scale[.9]{\mathcal{L}^2(-1,1)}$ inner product.

\begin{eqnarray}\label{B4}
\Scale[.9]{
{\min}_{\varphi  \in {T_M}}\left\{ {\displaystyle\sum\nolimits_{|k| \leqslant N} {{{\left| {{\hat f_k}  - \langle \varphi ,e^{ik\pi x}\rangle } \right|}^2}} } \right\}.
}
\end{eqnarray}

For multiple discontinuities, a separate extension for each subdomain is considered and the equations are aggregated. Overall, the reconstruction over the $i$-th smooth subinterval will require a constant $\Scale[.9]{\eta_i=M_i/\sqrt{2N+1}}$, in which $\Scale[.9]{M_i}$ denotes the polynomial order of the reconstruction over the $i$-th smooth subdomain.

\subsection*{Fourier extension}
Let $\Scale[.9]{{G_M}}$ denote the set of $\Scale[.9]{2T}$-periodic functions defined as follows:

\begin{eqnarray}\label{B5}
\Scale[.9]{
g \in {G_M}:g\left( x \right) = \frac{{{\alpha _0}}}{2} + \displaystyle\sum\nolimits_{k = 1}^M {{\alpha _k}\cos \left( {{\pi}kx/{T}} \right) + {\beta _k}\sin \left( {{\pi }kx/{T}} \right)}.
}
\end{eqnarray}

The $\Scale[.9]{M}$-th order Fourier extension of the function $\Scale[.9]{f}$ over $\Scale[.9]{{[-T,T]}}$ is defined to be the solution to the following optimization problem:

\begin{eqnarray}\label{B6}
\Scale[.9]{
{g_n}: = \arg {\min\limits_{g \in {G_M}}}{\left| {\left| {f - g} \right|} \right|_{2}}.
}
\end{eqnarray}

If the Fourier extension is performed over truncated Fourier series with $\Scale[.9]{M = \mathcal{O}(2N+1)}$, it can be shown to eliminate the Gibbs phenomenon and retrieve the function.
For multiple discontinuities, a separate extension for each subdomain is considered and the equations are aggregated. Overall, the Fourier extension over the $i$-th smooth subdomain will require a pair of parameters $\Scale[.9]{(T_i,\eta_i={M_i}/2N+1)}$, in which $\Scale[.9]{T_i}$ and $\Scale[.9]{M_i}$ denote the extension factor and the Fourier extension order of the $i$-th subdomain, respectively.

\subsection*{Freud reconstruction}
The formalism of spectral reprojections was elaborated in section \ref{Sec: Gegen} for Gegenbauer polynomials. The use of Freud orthogonal polynomials is quite similar. These orthogonal polynomials have the weight function $\Scale[.9]{w(x)=\exp\left(-c|x|^{2p}\right)}$, in which $\Scale[.9]{c=-\ln(\varepsilon \ll 1)}$ and $\Scale[.9]{p=\lfloor {\sqrt{N({b - a})/2}} - 2\sqrt{2} \rfloor}$, for the reconstruction interval $\Scale[.9]{[a,b]}$. Moreover, the reconstruction order would be $\Scale[.9]{M = N(b - a)/8}$. Note that Freud polynomials are defined on the whole domain of real numbers $\Scale[.9]{\mathbb{R}}$, though in the reprojection framework we confine them to the $\Scale[.9]{[-1,1]}$ interval.

\subsection{Expansion of the Fourier Modal Matrix}\label{Apx_A}

\renewcommand{\theequation}{A-\arabic{equation}}% redefine the command that creates the equation no.
\setcounter{equation}{0}  % reset counter

The corresponding eigenvalue equation of a lamellar/crossed grating is conventionally derived from the Fourier expansion of Maxwell equations \cite{Moharam1995,Li1997}, however, it can also be derived from the wave equation \cite{Faghihifar2020}, as follows, wherein $\Scale[.9]{k}$ represents the vacuum wavenumber.

\begin{equation}\label{A1}
\Scale[.9]{
\bm\nabla  \times \bm\nabla  \times \bm E(\bm r) - k^2 \bm\epsilon(\bm r)\bm E(\bm r) = \bm{0}.
}
\end{equation}

Let us assume the dielectric constant has a monoclinic form, i.e. $\Scale[.9]{\epsilon_{zx}=\epsilon_{zy}=\epsilon_{yz}=\epsilon_{xz}=0}$. The dielectric function of a lamellar or crossed grating can be expressed in terms of a Fourier series as follows, in which $\Scale[0.9]{{\bm \epsilon }^{m}}$ and $\Scale[0.9]{{\bm \epsilon }^{m,n}}$ are the $\Scale[.9]{3\times 3}$ matrix Fourier coefficients of the one and two-dimensional dielectric functions:

\begin{eqnarray}\label{A2}
\Scale[.9]{
\bm\epsilon(x) = {\displaystyle \sum\limits_{m} {{{{\bm\epsilon }^{m}}{e^{ imx({2\pi }/{\lambda T_x})}}}}},\hspace{40pt}
\bm\epsilon(x,y) = {\displaystyle {\sum\limits_{m,n} { {{{\bm\epsilon }^{m,n}}{e^{ imx({2\pi}/{\lambda _0 T_x})}}{e^{ iny({2\pi}/{\lambda T_y})}}}}}}.
}
\end{eqnarray}

If $\Scale[.9]{M}$ and $\Scale[.9]{N}$ denote Fourier truncation orders along $\Scale[.9]{x}$ and $\Scale[.9]{y}$ axes, the total number of Fourier orders will be $\Scale[.9]{S=2M+1}$ and $\Scale[.9]{S=(2M+1)(2N+1)}$, for one or two-dimensional problems respectively. Let us define $\Scale[0.9]{\mathcal{J}_n = \{-n,...,0,...,n\} }$ and $\Scale[0.9]{\mathcal{K}_n=\{1,2,...,n\} }$ for any positive integer $n$. Now, in order to include/exclude Fourier products and arrange the equations in the matrix form, one needs to define a bijective relation $\Scale[0.9]{\rho:{\mathcal{J}_M} \rightarrow  {\mathcal{K}_S}}$ or $\Scale[0.9]{\rho:{\mathcal{J}_M} \times {\mathcal{J}_N} \rightarrow  {\mathcal{K}_S}}$ so that every one or two-dimensional index with $\Scale[0.9]{m \in \mathcal{J}_M}$ and $\Scale[0.9]{n \in \mathcal{J}_N}$, corresponds to some unique ordinal number $\Scale[0.9]{s \in \mathcal{K}_S}$. The most trivial and common choice of order functions are as following:

\begin{eqnarray}\label{A3}
\Scale[.9]{
s = \rho_{1d}\left( {m} \right) =  m+M+1,
\hspace{40pt}
s = \rho_{2d}\left( {m,n} \right) = \left( {2N + 1} \right)\left( {m + M} \right) +  {n + N + 1}.
}
\end{eqnarray}

If we denote $\Scale[.9]{( {m,n}) = {{\rho}^{ - 1}}( i )}$ and $\Scale[.9]{( {p,q}) = {{\rho}^{ - 1}}( j )}$ for any $\Scale[0.9]{i,j \in \mathcal{K}_S}$ and consider that $\Scale[0.9]{u,v \in \{x,y,z\}}$, the building blocks of the matrix eigenvalue equation could be constructed as follows:

\begin{eqnarray}\label{A4}
\Scale[.9]{
\begin{array}{l}
[\bm{F_{uv}}]_{i,j} =  {\epsilon^{m-p,n-q}},\hspace{40pt}
\left[\bm{N_x}\right]_{i,j} = {\delta _{ij}}({k_{x0}}/{k} + {m}/{T_x}),\\
\left[{\bm{e_u}}\right]_i = {{E_u^{m,n}}},\hspace{70pt}\left[\bm{N_y}\right]_{i,j} = {\delta _{ij}}({k_{y0}/k+ n/T_y}).
\end{array}
}
\end{eqnarray}

In which $\Scale[.9]{\bm{F_{uv}}}$ are (block) Toeplitz matrices of the Fourier coefficients, $\Scale[.9]{\bm{N_x}}$ and $\Scale[.9]{\bm{N_y}}$ are diagonal matrices, and $\Scale[.9]{\bm{e_u}}$ are vertical vectors. Let us define $\Scale[.9]{\bm{e_t} = [\bm{e_x}, \bm{e_y}]}$, $\Scale[.9]{\bm{N} = [\bm{N_x}, \bm{N_y}]}$, $\Scale[.9]{\bm{N_t} = [-\bm{N_y}, \bm{N_x}]}$, $\Scale[.9]{\bm{Q} = \bm{F_{zz}} - \bm{N'N}}$, $\Scale[.9]{\bm{F_t} = [{\bm{F_{xx}}}{\hspace{5pt}}{\bm{F_{xy}}}, {\bm{F_{yx}}{\hspace{5pt}}{\bm{F_{yy}}}}]}$, and $\Scale[.9]{\bm{Q_t} = \bm{F_t} - [{\bm{N'N}}{\hspace{5pt}}{\bm{0}}, {\bm{0}}{\hspace{5pt}}{\bm{N'N}}]}$, where $\Scale[.9]{[\cdot , \cdot]}$ represents vertical concatenation. Note that for a one-dimensional (lamellar) grating, the relationship $\Scale[.9]{\bm{N_y} = ({k_{y0}}/{k})\bm{I}}$ simply holds.
Eventually, the Fourier modal matrix $\Scale[.9]{\bm{P}}$ can be elaborated as follows, in which $\Scale[.9]{{}'}$ denotes the conjugate transpose:

\begin{eqnarray}\label{A5}
\Scale[0.9]{
\bm{P} = \left(\bm{I} + \bm{N}\bm{Q}^{-1}\bm{N}'\right)^{-1}\left(\bm{F_t} - \bm{N_t}\bm{N_t}' \right) = \bm{Q_t} - \bm{N}\left(\bm{N'}-\bm{F_{zz}}^{-1}\bm{N'}\bm{F_t}\right).
}
\end{eqnarray}

%\section*{References}

%\bibliography{Ref}

\end{document}